\documentclass[useAMS,usenatbib,fleqn]{mn2e}

\usepackage{amssymb}
\usepackage{times}
\usepackage{graphicx}
\usepackage{epstopdf}
\usepackage{subfigure}
\usepackage[T1]{fontenc}
\usepackage{aecompl} 
\usepackage{multirow}
\usepackage{pdflscape}
\usepackage{rotating}

\usepackage[fleqn]{amsmath}
\usepackage{mathrsfs}
\usepackage{xcolor}

\usepackage{anyfontsize}

\usepackage[breaklinks=true]{hyperref}
\hypersetup{colorlinks=true,citecolor = blue}

\newcommand{\beq}{\begin{equation}}
\newcommand{\eeq}{\end{equation}}

\def\gs{\mathrel{\lower0.6ex\hbox{$\buildrel {\textstyle >}\over{\scriptstyle \sim}$}}}
\def\ls{\mathrel{\lower0.6ex\hbox{$\buildrel {\textstyle <}\over{\scriptstyle \sim}$}}}
\newcommand{\simgt}{\lower.5ex\hbox{$\; \buildrel > \over \sim \;$}}
\newcommand{\simlt}{\lower.5ex\hbox{$\; \buildrel < \over \sim \;$}}

\newcommand{\aap}{A\&A}

\newcommand{\apj}{ApJ}
\newcommand{\apjl}{ApJ}
\newcommand{\apjs}{ApJS}
\newcommand{\aj}{AJ}

\newcommand{\jcap}{J. Cosmol. Astropart. Phys.}
\newcommand{\pasj}{PASJ}

\newcommand{\prd}{Phys. Rev. D}
\newcommand{\mnras}{MNRAS}
\newcommand{\physrep}{Phisycs Rep.}

\newcommand{\ssr}{Space Science Reviews}

\defcitealias{se+et15_comalit_I}{CoMaLit-I}
\defcitealias{ser+al15_comalit_II}{CoMaLit-II}
\defcitealias{ser15_comalit_III}{CoMaLit-III}
\defcitealias{se+et15_comalit_IV}{CoMaLit-IV}
\defcitealias{se+et17_comalit_V}{CoMaLit-V}
\defcitealias{xxl_I_pie+al15}{XXL-I}
\defcitealias{xxl_II_pac+al15}{XXL-II}
\defcitealias{xxl_III_gil+al15}{XXL-III}
\defcitealias{xxl_IV_lie+al16}{XXL-IV}
\defcitealias{xxl_XIII_eck+al15}{XXL-XIII}

\begin{document}

\title[PSZ2LenS]{PSZ2LenS. Weak lensing analysis of the {\it Planck} clusters in the CFHTLenS and in the RCSLenS}
\author[Sereno et al.]{Mauro Sereno$^{1,2}$\thanks{E-mail: mauro.sereno@oabo.inaf.it (MS)}, Giovanni Covone$^{3,4}$, Luca Izzo$^{5}$, Stefano Ettori$^{1,6}$,  Jean Coupon$^{7}$, \newauthor 
Maggie Lieu$^{8}$\\
$^1$INAF - Osservatorio Astronomico di Bologna, via Piero Gobetti 93/3, I-40129 Bologna, Italy\\
$^2$Dipartimento di Fisica e Astronomia, Universit\`a di Bologna, via Piero Gobetti 93/2, I-40129 Bologna, Italy\\
$^3$Dipartimento di Fisica, Universit\`{a} di Napoli `Federico II', Compl. Univers. di Monte S. Angelo, via Cinthia, I-80126 Napoli, Italia\\
$^4$INFN, Sezione di Napoli, Compl. Univers. di Monte S. Angelo, via Cinthia, I-80126 Napoli, Italia\\
$^5$Instituto de Astrof{\`i}sica de Andaluc{\`i}a (IAA-CSIC), Glorieta de la Astronom{\`i}a s/n, E-18008 Granada, Spain\\
$^6$INFN, Sezione di Bologna, viale Berti Pichat 6/2, I-40127 Bologna, Italia\\
$^7$Department of Astronomy, University of Geneva, ch. d'Ecogia 16, CH-1290 Versoix, Switzerland \\
$^8$European Space Astronomy Centre (ESA/ESAC), Science Operations Department, E-28691 Villanueva de la Caada, Madrid, Spain
}


\maketitle

\begin{abstract}
The possibly unbiased selection process in surveys of the Sunyaev Zel'dovich effect can unveil new populations of galaxy clusters. We performed a weak lensing analysis of the PSZ2LenS sample, i.e. the PSZ2 galaxy clusters detected by the {\it Planck} mission in the sky portion covered by the lensing surveys CFHTLenS and RCSLenS. PSZ2LenS consists of 35 clusters and it is a statistically complete and homogeneous subsample of the PSZ2 catalogue. The {\it Planck} selected clusters appear to be unbiased tracers of the massive end of the cosmological haloes. The mass concentration relation of the sample is in excellent agreement with predictions from the $\Lambda$ cold dark matter model. The stacked lensing signal is detected at $14\sigma$ significance over the radial range $0.1<R<3.2~\text{Mpc}/h$, and is well described by the cuspy dark halo models predicted by numerical simulations. We confirmed that {\it Planck} estimated masses are biased low by $b_\text{SZ}= 27\pm 11\text{(stat)}\pm8\text{(sys)}$ per cent with respect to weak lensing masses. The bias is higher for the cosmological subsample, $b_\text{SZ}= 40\pm 14\text{(stat)}\pm8\text{(sys)}$ per cent.
\end{abstract}


\section{Introduction}
\label{sec_intro}

The prominent role of clusters of galaxies in cosmology and astrophysics demands for a very accurate knowledge of their properties and history. Galaxy clusters are laboratories to study the physics of  baryons and dark matter in the largest gravitationally nearly virialized regions \citep{voi05,pra+al09,arn+al10,gio+al13}. Cosmological parameters can be determined with cluster abundances and the observed growth of massive haloes \citep{man+al10,planck_2015_XXIV}, gas fractions \citep{ett+al09}, or lensing analyses \citep{ser02,jul+al10,lub+al14}. 

Ongoing and future large surveys will provide invaluable information on the multi-wavelength sky \citep{eucl_lau_11,xxl_I_pie+al16}. Large surveys of the Sunyaev Zel'dovich (SZ) sky can find galaxy clusters up to high redshifts. Successful programs have been carried out by the {\it Planck} Satellite \citep{planck_2015_XXVII}, the South Pole Telescope \citep[SPT]{ble+al15} and the Atacama Cosmology Telescope \citep[ACT]{has+al13}. SZ surveys should in principle detect clusters regardless of their distance. Even though the finite spatial resolution can hamper the detection of the most distant objects, SZ selected clusters should be nearly mass limited. The selection function of SZ selected clusters can be well determined.

Furthermore, SZ quantities are quite stable and not significantly affected by dynamical state or mergers \citep{mot+al05,kra+al12,bat+al12}. The relation between mass and SZ flux is expected to have small intrinsic scatter \citep{kay+al12,bat+al12}.  These properties make the determination of cosmological parameters using number counts of SZ detected clusters very appealing \citep{planck_2015_XXVII}. 

If confirmed, the mass limited but otherwise egalitarian selection could make the SZ clusters an unbiased sample of the whole massive haloes in the universe. \citet{ros+al16} characterized the dynamical state of 132 {\it Planck} clusters with high signal to noise ratio using as indicator the projected offset between the peak of the X-ray emission and the position of the brightest cluster galaxy (BCG). They showed that the fraction of dynamically relaxed objects is smaller than in X-ray selected samples and confirmed the early impression that many {\it Planck} selected objects are dynamically disturbed systems. \citet{ros+al17} found that the fraction of cool core clusters is $29\pm4$ per cent and does not show significant time evolution. They found that SZ selected samples are nearly unbiased towards cool cores, one of the main selection effects affecting clusters selected in X-ray surveys.

A crucial ingredient to study cluster physics is the mass determination. Weak lensing (WL) analyses can provide accurate and precise estimates. The physics behind gravitational lensing is very well understood  \citep{ba+sc01} and mass measurements can be provided up to high redshifts \citep{hoe+al12,wtg_I_14,ume+al14,ser15_comalit_III}. 

The main sources of uncertainty and scatter in WL mass estimates are due to triaxiality, substructures and projection effects \citep{ogu+al05,se+um11,men+al10,be+kr11,bah+al12,gio+al14}. Theoretical predictions based on numerical simulations \citep{ras+al12,be+kr11} and recent measurements \citep{man+al15,se+et15_comalit_I} agree on an intrinsic scatter of $\sim$15 per cent. 

More than five hundred clusters with known WL mass are today available \citep{ser15_comalit_III} and this number will explode with future large photometric surveys, e.g., Hyper Suprime-Cam Subaru Strategic Program \citep[HSC-SSP]{hsc_I_17} or Euclid \citep{eucl_lau_11}. However, direct mass measurements are usually available only for the most massive clusters. Mass estimates of lesser clusters have to rely on calibrated mass--observable relations \citep{se+et17_comalit_V}. Due to the low scatter, mass proxies based on SZ observables are among the most promising.

The above considerations motivate the analysis of SZ selected clusters of galaxies with homogeneous WL data. The relation between WL masses and SZ flux of {\it Planck} selected clusters has been investigated by several groups \citep{gru+al14,lin+al14,ser+al15_comalit_II,smi+al16}. The scaling relation between WL mass and integrated spherical Compton parameter $Y_{500}$ of the 115 {\it Planck} selected clusters with known WL mass was studied in \citet{ser+al15_comalit_II} and \citet{se+et15_comalit_IV}, which retrieved a $Y_{500}$-$M_{500}$ in agreement with self-similar predictions, with an intrinsic scatter of $10\pm5$ per cent on the SZ mass proxy. 

The tension between the lower values of the power spectrum amplitude $\sigma_8$ inferred from clusters counts \citep[$\sigma_8\sim0.71$-$0.78$ and references therein]{planck_2015_XXIV} and higher estimates from measurements of the primary Cosmic Microwave Background (CMB) temperature anisotropies \citep[$\sigma_8=0.83\pm0.02$]{planck_2015_XIII} may be due to the $Y_{500}$-$M_{500}$ relation used to estimate cluster masses. Consistency can be achieved if {\it Planck} masses, which are based on SZ/X-ray proxies \citep{planck_2013_XX,planck_2013_XXIX}, are biased low by $\sim$ 40 per cent \citep{planck_2015_XXIV}. 

The level of bias has to be assessed but it is still debated. \citet{gru+al14} presented the WL analysis of 12 SZ selected clusters, including 5 {\it Planck} clusters. 
The comparison of WL masses and Compton parameters showed significant discrepancies correlating with cluster mass or redshift. Comparing the {\it Planck} masses to the WL masses of the WtG clusters \citep[Weighing the Giants,][]{wtg_III_14}, \citet{lin+al14} found evidence for a significant mass bias and a mass dependence of the calibration ratio. The analysis of the CCCP clusters \citep[Canadian Cluster Comparison Project,][]{hoe+al15} confirmed that the bias in the hydrostatic masses used by the {\it Planck} team depends on the cluster mass, but with normalization 9 per cent higher than what found in \citet{lin+al14}. \citet{smi+al16} found that the mean ratio of the {\it Planck} mass estimate to LoCuSS (Local Cluster Substructure Survey) lensing mass is $0.95\pm0.04$.
 
An unambiguous interpretation of the bias dependence in terms of either redshift or masses can be hampered by the small sample size. Exploiting a large collection of WL masses, \citet{ser+al15_comalit_II} and \citet{se+et17_comalit_V} found the bias to be redshift rather than mass dependent. 

Even though some of the disagreement among competing analyses can de due to statistical methodologies not properly accounting for Eddington/Malmquist biases and evolutionary effects, see discussion in \citet{ser+al15_comalit_II,se+et15_comalit_IV,se+et17_comalit_V}, the mass biases found for different cluster samples do not necessarily have to agree. Different samples cover different redshift and mass ranges, where the bias can differ. Furthermore, WL masses are usually available for the most massive clusters only.

In this paper, we perform a WL analysis of a statistically complete and homogeneous subsample of the {\it Planck} detected clusters, the PSZ2LenS. We analyze all the {\it Planck} candidate clusters in the fields of two public lensing surveys, the CFHTLenS  \citep[Canada France Hawaii Telescope Lensing Survey,][]{hey+al12} and the RCSLenS \citep[Red Cluster Sequence Lensing Survey,][]{hil+al16}, which shared the same observational instrumentation and the same data-analysis tools. PSZ2LenS is homogeneous in terms of selection, observational set up, data reduction, and data analysis.

The paper is structured as follows. In Section~\ref{sec_data} we present the main properties of the lensing surveys and the available data. In Section~\ref{sec_planck}, we introduce the second {\it Planck} Catalogue of SZ Sources \citep[PSZ2,][]{planck_2015_XXVII} and the PSZ2LenS sample. In Section~\ref{sec_weak}, we cover the basics of the WL theory. Section~\ref{sec_back_sel} is devoted to the selection of the lensed source galaxies. In Section~\ref{sec_lens}, we detail how we modelled the lenses. The strength of the WL signal of the PSZ2Lens clusters is discussed in Section~\ref{sec_signal}. The Bayesian method used to analyze the lensing shear profiles is illustrated in Section~\ref{sec_infe}. The recovered cluster masses and their consistency with previous results are presented in Section~\ref{sec_masses}. In Section~\ref{sec_c200}, we measure the mass-concentration relation of the PSZ2LenS clusters. Section~\ref{sec_stacking} is devoted to the analysis of the stacked signal. In Section~\ref{sec_planck_bias}, we estimate the bias of the {\it Planck} masses. A discussion of potential systematics effects and residual statistical uncertainties is presented in Section~\ref{sec_systematics}. Candidate clusters which were not visually confirmed are discussed in Section~\ref{sec_not}. Section~\ref{sec_conc} is devoted to some final considerations. In Appendix~\ref{app_radi}, we discuss the optimal radius to be associated to the recovered shear signal.  Appendinx~\ref{sec_lens_weig} details the lensing weighted average of cluster properties. Appendix~\ref{sec_robu_esti} discusses pros and cons of some statistical estimators used for the WL mass.

\subsection{Notations and conventions}

As reference cosmological model, we assumed the concordance flat $\Lambda$CDM ($\Lambda$ and Cold Dark Matter) universe with density parameter $\Omega_\text{M}=0.3$, Hubble constant $H_0=70~\text{km~s}^{-1}\text{Mpc}^{-1}$, and power spectrum amplitude $\sigma_8=0.8$.  When $H_0$ is not specified, $h$ is the Hubble constant in units of $100~\text{km~s}^{-1}\text{Mpc}^{-1}$. 

Throughout the paper, $O_{\Delta}$ denotes a global property of the cluster measured within the radius $r_\Delta$ which encloses a mean over-density of $\Delta$ times the critical density at the cluster redshift, $\rho_\text{cr}=3H(z)^2/(8\pi G)$, where $H(z)$ is the redshift dependent Hubble parameter and $G$ is the gravitational constant. We also define $E_z\equiv H(z)/H_0$. 

The notation `$\log$' is the logarithm to base 10 and `$\ln$' is the natural logarithm. Scatters in natural logarithm are quoted as percents.

Typical values and dispersions of the parameter distributions are usually computed as bi-weighted estimators \citep{bee+al90} of the marginalized posterior distributions.

\section{Lensing data}
\label{sec_data}

We exploited the public lensing surveys CFHTLenS and RCSLenS. In the following, we introduce the data sets.


\subsection{The CFHTLenS}

The CFHTLS (Canada France Hawaii Telescope Legacy Survey) is a photometric survey performed with MegaCam. The wide survey covers four independent fields for a total of $\sim154\deg^2$ in five optical bands $u^*$, $g$, $r$, $i$, $z$ \citep{hey+al12}.

The survey was specifically designed for weak lensing analysis, with the deep $i$-band data taken in sub-arcsecond seeing conditions \citep{erb+al13}. The total unmasked area suitable for lensing analysis covers $125.7\deg^2$. The raw number density of lensing sources, including all objects that a shape was measured for, is 17.8 galaxies per arcmin$^2$ \citep{hil+al16}. The weighted density is 15.1 galaxies per arcmin$^2$.

The CFHTLenS team provided\footnote{The public archive is available through the Canadian Astronomy Data Centre at \url{http://www.cfht.hawaii.edu/Science/CFHLS}.} weak lensing data processed with \texttt{THELI} \citep{erb+al13} and shear measurements obtained with \texttt{lensfit} \citep{mil+al13}. The photometric redshifts were measured with accuracy $\sigma_{z_\text{phot}} \sim 0.04(1+z)$ and a catastrophic outlier rate of about 4 per cent \citep{hil+al12,ben+al13}.

\subsection{The RCSLenS}

The RCSLenS is the largest public multi-band imaging survey to date which is suitable for weak gravitational lensing measurements\footnote{The data products are publicly available at \url{http://www.cadc-ccda.hia-iha.nrc-cnrc.gc.ca/en/community/rcslens/query.html}.}  \citep{hil+al16}. 

The parent survey, i.e., the Red-sequence Cluster Survey 2 \citep[RCS2]{gil+al11} is a sub-arcsecond seeing, multi-band imaging survey in the $griz$ bands initially designed to optically select galaxy cluster. The RCSLenS project later applied methods and tools already developed by CFHTLenS for lensing studies.

The survey covers a total unmasked area of $571.7\deg^2$ down to a magnitude limit of $r\sim24.3$ (for a point source at $7\sigma$).  Photometric redshifts based on four bands ($g$, $r$, $i$, $z$) data are available for an unmasked area covering $383.5\deg^2$, where the raw (weighted) number density of lensing sources is 7.2 (4.9) galaxies per arcmin$^2$. The survey area is divided into 14 patches, the largest being $10\times10\deg^2$ and the smallest $6\times6\deg^2$. 

Full details on imaging data, data reduction, masking, multi-colour photometry, photometric redshifts, shape measurements, tests for systematic errors, and the blinding scheme to allow for objective measurements can be found in \citet{hil+al16}.

The RCSLenS was observed with the same telescope and camera as CFHTLS and the project applied the same methods and tools developed for CFHTLenS. The two surveys share the same observational instrumentation and the same data-analysis tools, which make the shear and the photo-$z$ catalogues highly homogeneous, but some differences can be found in the two data sets. 

CFHTLenS features the additional $u$ band and the co-added data are deeper by $\sim1~\text{mag}$. The CFHTLenS measured shapes of galaxies in the $i$ band. On the other side, since the $i$ band only covers $\sim 70$ per cent of the RCS2 area, the $r$ band was used in RCSLenS for shape measurements because of the longest exposure time and the complete coverage.


\subsection{Ancillary data}

When available, we exploited ancillary data sets to strengthen the measurement of photometric redshifts and secure the selection of background galaxies. For some fields partially covering CFHTLS-W1 and CFHTLS-W4, we complemented the CFHTLenS data with deep near-UV and near-IR observations, supplemented by secure spectroscopic redshifts. The full data set of complementary observations was presented and fully detailed in \citet{cou+al15}, who analyzed the relationship between galaxies and their host dark matter haloes through galaxy clustering, galaxy--galaxy lensing and the stellar mass function. We refer to \citet{cou+al15} for further details.

\subsubsection{Spectroscopic data}
\label{sec_spec}

When available, we used the spectroscopic redshifts collected from public surveys by \citet{cou+al15} instead of the photometric redshift value. \citet{cou+al15} exploited four main spectroscopic surveys to collect 62220 unique galaxy spectroscopic redshifts with the highest confidence flag.

The largest spectroscopic sample within the W1 area comes from the VIMOS (VIsible MultiObject Spectrograph) Public Extragalactic Survey \citep[VIPERS,][]{gar+al14}, designed to study galaxies at $0.5 \la z\la1.2$. The designed survey covers a total area of $16 \deg^2$ in the W1 field and $8 \deg^2$ in the W4 field. The first public data release (PDR1) includes redshifts for 54204 objects (30523 in VIPERS-W1). \citet{cou+al15} only considered the galaxies with the highest confidence flags between 2.0 and 9.5.

The VIMOS-VLT (Very Large telescope) Deep Survey \citep[VVDS,][]{lef+al05} and the Ultra-Deep Survey \citep{lef+al15} cover a total area of $0.75\deg^2$ in the VIPERS-W1 field. \citet{cou+al15} also used the VIMOS-VLT F22 Wide Survey with 12995 galaxies over $4\deg^2$ down to $i < 22.5$ in the southern part of the VIPERS-W4 field \citep{gar+al08}. In total, \citet{cou+al15} collected 5122 galaxies with secure flag 3 or 4.

The PRIsm MUlti-object Survey \citep[PRIMUS,][]{coi+al11} consists of low resolution spectra. \citet{cou+al15} retained the 21365 galaxies with secure flag 3 or 4.

The SDSS-BOSS spectroscopic survey based on data release DR10 \citep{ahn+al14} totals 4675 galaxies with \textsc{zWarning=0} within the WIRCam area, see below.

\subsubsection{The Near-IR observations}

\citet{cou+al15} conducted a $K_s$-band follow-up of the VIPERS fields with the WIRCam instrument at CFHT. Noise correlation introduced by image resampling was corrected exploiting data from the deeper UKIDSS Ultra Deep Survey \citep[$K < 24.5$][]{law+al07}. Sample completeness reaches 80 per cent at $K_s = 22$. 

\citet{cou+al15} also used the additional data set from the WIRCam Deep Survey data \citep{bie+al12}, a deep patch of 0.49 deg$^2$ observed with WIRCam $J$, $H$ and $K_s$ bands.

The corresponding effective area in the CFHTLS after rejection for poor WIRCam photometry and masked CFHTLenS areas covers $\sim23.1\deg^2$, divided into 15 and $\sim8.1\deg^2$ in the VIPERS-W1 and VIPERS-W4 fields, respectively. WIRCAM sources were matched to the optical counterparts based on position.

\subsubsection{The UV-GALEX observations}

UV deep imaging photometry from the GALEX satellite \citep{galex_mar+al05} is also available for some partial area. \citet{cou+al15} considered the observations from the Deep Imaging Survey (DIS). All the GALEX pointings were observed with the NUV channel and cover $\sim10.8\deg^2$ and $\sim1.9\deg^2$ of the WIRCam area in VIPERS-W1 and VIPERS-W4, respectively. FUV observations are available for 10 pointings in the central part of W1.

\section{The PSZ2LenS}
\label{sec_planck}

\begin{table}
\caption{The PSZ2LenS sample. Column~1: cluster name. Column 2: index in the PSZ2-Union catalogue. Columns~3 and 4: right ascension and declination in degrees (J2000) of the associated BCG. Column~5: redshift. A star indicates a photometric redshift. Column~7: lensing survey. Column~8: survey patch. The suffix NIR means that ancillary data were available.}
\label{tab_WL_sample}
\centering
\resizebox{\hsize}{!} {
\begin{tabular}[c]{l rrrl ll}
\hline
	\noalign{\smallskip}  
	 PSZ2  & index   &  \multicolumn{1}{c}{RA}     &  \multicolumn{1}{c}{DEC}   & \multicolumn{1}{c}{$z$}    & survey & field  \\ 
	 	\hline
	\noalign{\smallskip}      
G006.49+50.56	&	21  	&	227.733767	&	5.744914  	&	0.078	&	RCSLenS 	&	1514  	\\
G011.36+49.42	&	38  	&	230.466125	&	7.708881  	&	0.044	&	RCSLenS 	&	1514  	\\
G012.81+49.68	&	43  	&	230.772096	&	8.609181  	&	0.034	&	RCSLenS 	&	1514  	\\
G053.44-36.25	&	212 	&	323.800386	&	$-$1.049615 	&	0.327	&	RCSLenS 	&	2143  	\\
G053.63-41.84	&	215 	&	328.554600	&	$-$3.998100 	&	0.151	&	RCSLenS 	&	2143  	\\
G053.64-34.48	&	216 	&	322.416475	&	0.089100  	&	0.234	&	RCSLenS 	&	2143  	\\
G058.42-33.50	&	243 	&	323.883680	&	3.867200  	&	0.400$^*$	&	RCSLenS 	&	2143  	\\
G059.81-39.09	&	251 	&	329.035737	&	1.390939  	&	0.222	&	RCSLenS 	&	2143  	\\
G065.32-64.84	&	268 	&	351.332080	&	$-$12.124360	&	0.082	&	RCSLenS 	&	2338  	\\
G065.79+41.80	&	271 	&	249.715656	&	41.626982 	&	0.336	&	RCSLenS 	&	1645  	\\
G077.20-65.45	&	329 	&	355.320925	&	$-$9.019929 	&	0.251	&	RCSLenS 	&	2338  	\\
G083.85-55.43	&	360 	&	351.882792	&	0.942811  	&	0.279	&	RCSLenS 	&	2329  	\\
G084.69+42.28	&	370 	&	246.745833	&	55.474961 	&	0.140	&	RCSLenS 	&	1613  	\\
G087.03-57.37	&	391 	&	354.415650	&	0.271253  	&	0.277	&	RCSLenS 	&	2329  	\\
G096.14+56.24	&	446 	&	218.868592	&	55.131111 	&	0.140	&	CFHTLenS	&	W3    	\\
G098.44+56.59	&	464 	&	216.852112	&	55.750253 	&	0.141	&	CFHTLenS	&	W3    	\\
G099.48+55.60	&	473 	&	217.159762	&	56.860909 	&	0.106	&	CFHTLenS	&	W3    	\\
G099.86+58.45	&	478 	&	213.696611	&	54.784321 	&	0.630$^*$	&	CFHTLenS	&	W3    	\\
G113.02-64.68	&	547 	&	8.632500  	&	$-$2.115100 	&	0.081	&	RCSLenS 	&	0047  	\\
G114.39-60.16	&	554 	&	8.617292  	&	2.423011  	&	0.384	&	RCSLenS 	&	0047  	\\
G119.30-64.68	&	586 	&	11.302080 	&	$-$1.875440 	&	0.545	&	RCSLenS 	&	0047  	\\
G125.68-64.12	&	618 	&	14.067088 	&	$-$1.255492 	&	0.045	&	RCSLenS 	&	0047  	\\
G147.88+53.24	&	721 	&	164.379271	&	57.995912 	&	0.528	&	RCSLenS 	&	1040  	\\
G149.22+54.18	&	724 	&	164.598600	&	56.794931 	&	0.135	&	RCSLenS 	&	1040  	\\
G150.24+48.72	&	729 	&	155.836579	&	59.810944 	&	0.199	&	RCSLenS 	&	1040  	\\
G151.62+54.78	&	735 	&	163.722100	&	55.350600 	&	0.470	&	RCSLenS 	&	1040  	\\
G167.98-59.95	&	804 	&	33.671129 	&	$-$4.567300 	&	0.140	&	CFHTLenS	&	W1-NIR	\\
G174.40-57.33	&	822 	&	37.920500 	&	$-$4.882580 	&	0.185	&	CFHTLenS	&	W1-NIR	\\
G198.80-57.57	&	902 	&	45.527500 	&	$-$15.561800	&	0.350$^*$	&	RCSLenS 	&	0310  	\\
G211.31-60.28	&	955 	&	45.302332 	&	$-$22.549510	&	0.400$^*$	&	RCSLenS 	&	0310  	\\
G212.25-53.20	&	956 	&	52.774492 	&	$-$21.009075	&	0.188	&	RCSLenS 	&	0310  	\\
G212.93-54.04	&	961 	&	52.057408 	&	$-$21.672833	&	0.600	&	RCSLenS 	&	0310  	\\
G230.73+27.70	&	1046	&	135.377830	&	$-$1.654880 	&	0.294	&	CFHTLenS	&	W2    	\\
G233.05+23.67	&	1057	&	133.065400	&	$-$5.567200 	&	0.192	&	CFHTLenS	&	W2    	\\
G262.95+45.74	&	1212	&	165.881133	&	$-$8.586525 	&	0.154	&	RCSLenS 	&	1111  	\\
	\hline
	\end{tabular}
	}
\end{table}

\begin{figure}
\resizebox{\hsize}{!}{\includegraphics{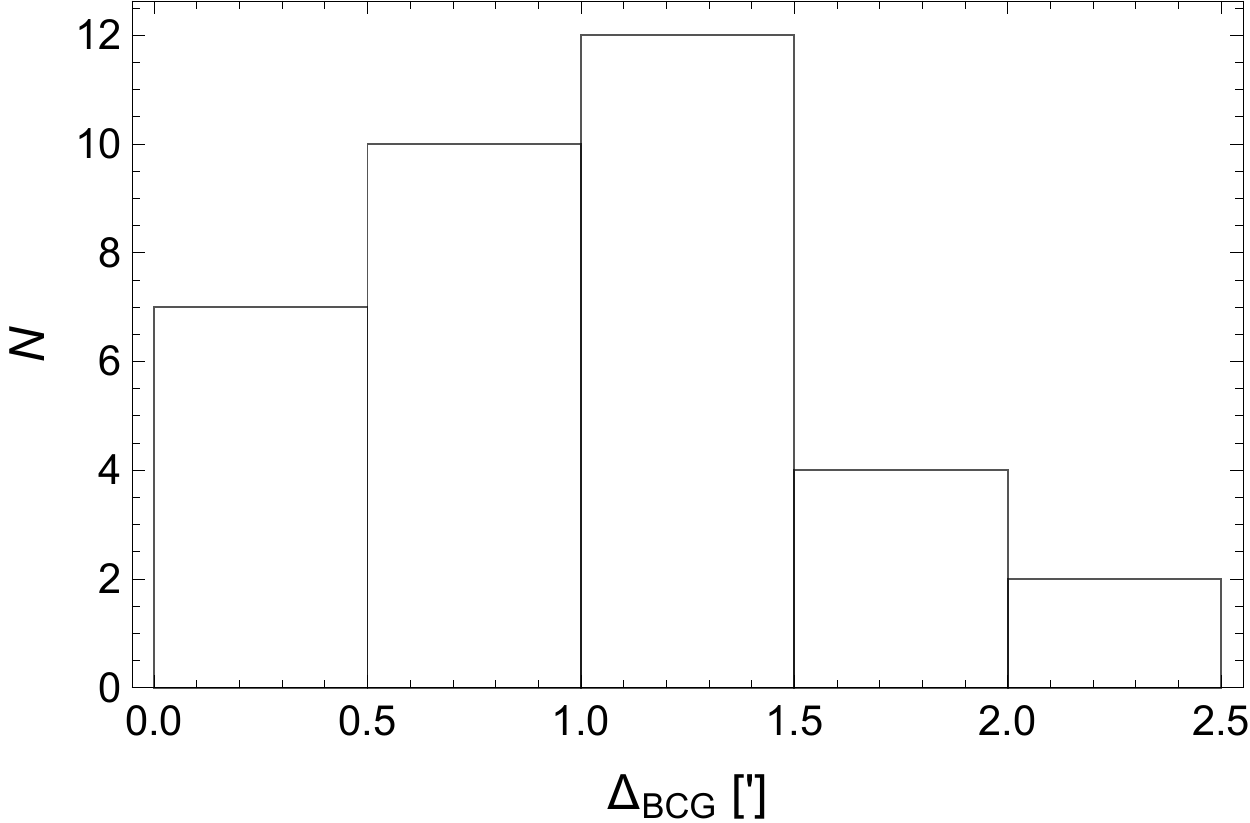}} 
\caption{Distribution of displacements between SZ centroid and BCG in the PSZ2LenS sample. Displacements are in units of arcminutes.}
\label{fig_Delta_BCG_histo}
\end{figure}

\begin{figure}
\resizebox{\hsize}{!}{\includegraphics{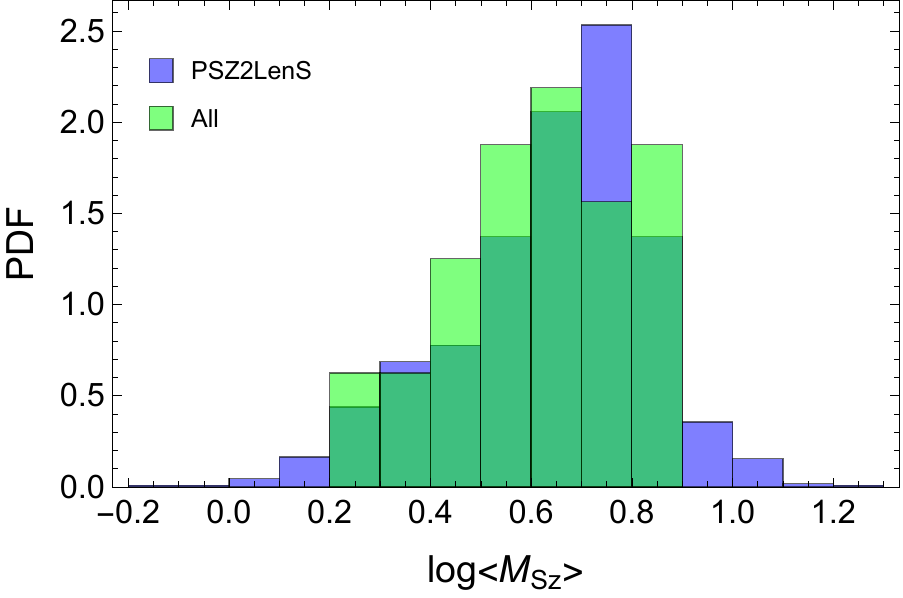}} 
\caption{The mass distribution of the {\it Planck} clusters. The histograms are rescaled to unitary area and show the distribution in mass of all the PSZ2 clusters with identified counterpart (green) and the PSZ2LenS subsample in the fields of the CFHTLenS/RCSLenS (blue). The masses are in units of $10^{14} M_\odot$.}
\label{fig_logMSZ_histo}
\end{figure}

\begin{figure}
\resizebox{\hsize}{!}{\includegraphics{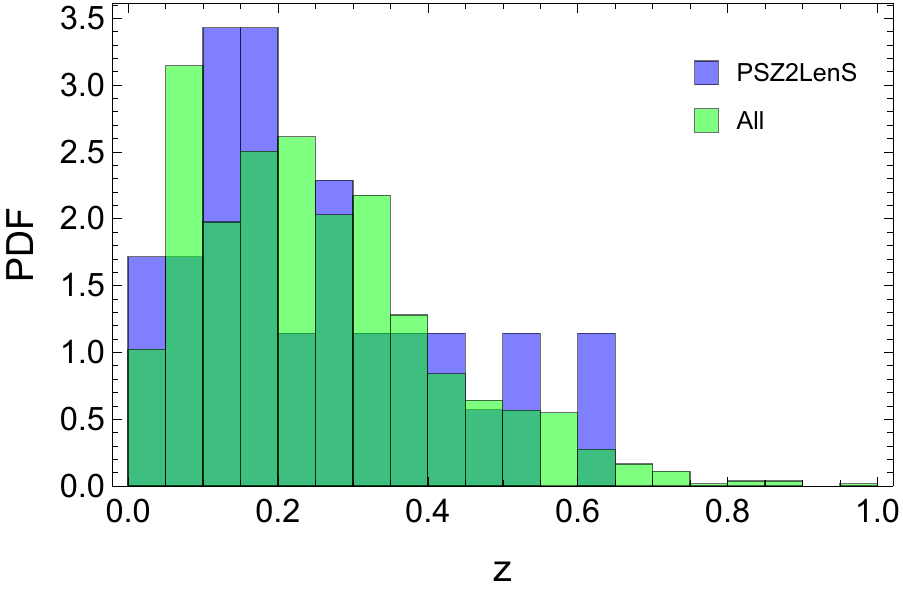}} 
\caption{The redshift distribution of the {\it Planck} clusters. The histograms are rescaled to unitary area and show the distribution in redshift of all the PSZ2 clusters with identified counterpart (green) and the PSZ2LenS subsample in the fields of the CFHTLenS/RCSLenS (blue).}
\label{fig_z_histo}
\end{figure}

\begin{figure}
\resizebox{\hsize}{!}{\includegraphics{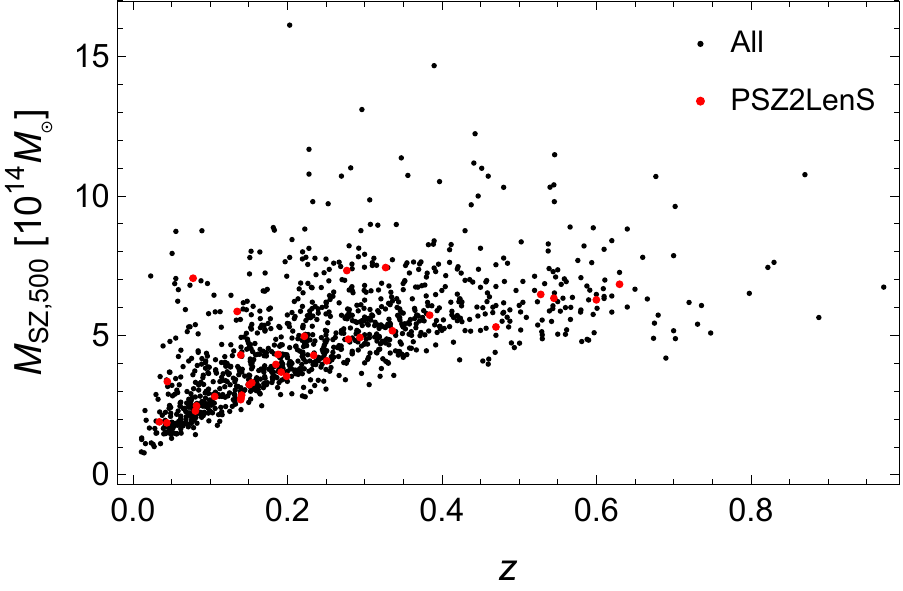}} 
\caption{Distribution of the PSZ2 clusters with known redshift in the $M_\text{SZ}$-$z$ plane. The black and red points denote all the PSZ2 clusters with identified counterpart and the PSZ2LenS subsample in the fields of the CFHTLenS/RCSLenS, respectively.}
\label{fig_z_vs_MSZ}
\end{figure}

The second {\it Planck} Catalogue of Sunyaev-Zel'dovich Sources \citep[PSZ2,][]{planck_2015_XXVII} exploits the 29 month full-mission data. The catalogue contains 1653 candidate clusters and it is the largest, all-sky, SZ selected sample of galaxy clusters yet produced\footnote{The union catalogue \textsc{HFI\_PCCS\_SZ-union\_R2.08.fits} is available from the {\it Planck} Legacy Archive at \url{http://pla.esac.esa.int/pla/}.}.

Only candidates with an SNR (signal-to-noise ratio) above 4.5 detected outside the highest-emitting Galactic regions, the Small and Large Magellanic Clouds, and the point source masks were included. Out of the total, 1203 clusters are confirmed with counterparts identified in external optical or X-ray samples or by dedicated follow-ups. The mean redshift is $z\sim 0.25$ and the farthest clusters were found at $z\ls 1$, which makes PSZ2 the deepest all-sky catalogue of galaxy clusters.

The {\it Planck} team calibrated the masses of the detected clusters with known redshift assuming a best fitting scaling relation between $M_{500}$ and $Y_{500}$, i.e. the spherically integrated Compton parameter within a sphere of radius $r_{500}$ \citep{planck_2013_XX}. These masses are denoted as $M_\text{SZ}$ or $M_{500}^{Y_z}$. The catalogue spans a nominal mass range from $M_\text{SZ}\sim0.8$ to $16\times10^{14}M_\odot$. 

We performed the WL analysis of the clusters centred in the CFHTLenS and RCSLenS fields. Out of the 47 PSZ2 sources within the survey fields, we confirmed 40 clusters by visually inspecting the optical images and identifying the BCGs. Five of these candidate galaxy clusters are located in regions of the RCSLenS where photometric redshifts are not available. Even though these galaxy clusters were clearly identified in the optical images, we could not measure the WL signal since we need photometric redshifts for the selection of background galaxies, see Section~\ref{sec_back_sel}. These clusters are: PSZ2~G054.95-33.39 (PSZ2 index: 221), G055.95-34.89 (225), G081.31-68.56 (349), G082.31-67.00 (354), and G255.51+49.96 (1177).

The final catalogue, PSZ2LenS, includes the confirmed 35 galaxy clusters (out of a total of 41 candidates) located in regions where photometric redshifts are available and is presented in Table~\ref{tab_WL_sample}. The cluster coordinates and redshifts correspond to the BCG. We did not confirm 6 candidates. Spectroscopic redshifts were recovered via the SIMBAD Astronomical Database\footnote{\url{http://simbad.u-strasbg.fr/simbad/}.} for 30 out of the 35 BCGs. Additional updated redshifts for PSZ2~G053.44-36.25 (212) and G114.39-60.16 (554) were found in \citet{car+al17}. For the remaining three clusters, we exploited photometric redshifts. The displacements of the SZ centroid from the BCG are pictured in Fig.~\ref{fig_Delta_BCG_histo}. 

Fifteen clusters out of 35 in PSZ2LenS are part of the cosmological subsample used by the {\it Planck} team for the analysis of the cosmological parameters with number counts.


We could confirm $\sim 85$ per cent of the candidate clusters, in very good agreement with the nominal statistical reliability assessed by the {\it Planck} team \citep{planck_2015_XXVII}, that placed a lower limit of 83 per cent on the purity. 

The results of our identification process are consistent with the the validation process by the {\it Planck} team \citep{planck_2015_XXVII}, who performed a multi-wavelength search for counterparts in ancillary radio, microwave, infra-red, optical, and X-ray data sets. 33 out of the 41 candidates were validated by the {\it Planck} team. This subset shares 32 clusters with PSZ2LenS. There are only a few different assessments by the independent selection processes. We did not include PSZ2~G006.84+50.69 (25), which we identified as a substructure of PSZ2~G006.49+50.56 (21), i.e. Abell~2029, see Section~\ref{sec_not}. On the other hand, we included PSZ2~G058.42-33.50 (243), PSZ2~G198.80-57.57 (902), and PSZ2~G211.31-60.28 (955), which were not validated by the {\it Planck} team.

Since we took all the {\it Planck} clusters without any further restriction, the lensing clusters constitute an unbiased subsample of the full catalogue. This is a strength of our sample with respect to other WL selected collections, which usually sample only the massive end of the full population, see discussion in \citet{ser+al15_comalit_II}.

The mass and redshift distribution of PSZ2LenS is representative of the full population of {\it Planck} clusters, see Figs.~\ref{fig_logMSZ_histo}, \ref{fig_z_histo} and \ref{fig_z_vs_MSZ}. According to the Kolmogorov-Smirnov test, there is a 53 per cent probability that the masses of our WL subsample and of the full sample are drawn from the same distribution. The redshift distributions are compatible at the 96 per cent level. 

The cluster catalogue and the shape measurements are extracted from completely different data sets, the PSZ2-Survey and CFHTLenS/RCSLenS data respectively. The distribution of lenses is then uncorrelated with residual systematics in the shape measurements \citep{miy+al15}.

\section{Weak lensing shear}
\label{sec_weak}


The reduced tangential shear $g_+$ is related to the differential projected surface density $\Delta\Sigma_+$ of the lenses \citep{man+al13,vel+al14,vio+al15}. For a single source redshift, 
\beq
\label{eq_Delta_Sigma_1}
\Delta \Sigma_+(R) = \gamma_+ \Sigma_\text{cr} = \bar{\Sigma}(<R)-\Sigma(R),  
\eeq 
where $\Sigma$ is the projected surface density and $\Sigma_\text{cr}$ is the critical density for lensing, 
\beq 
\label{eq_Delta_Sigma_2}
\Sigma_\text{cr}=\frac{c^2}{4\pi G} \frac{D_\text{s}}{D_\text{d} D_\text{ds}}, 
\eeq 
where $c$ is the speed of light in vacuum, $G$ is the gravitational constant, and $D_\text{d}$, $D_\text{s}$ and $D_\text{ds}$ are the angular diameter distances to the lens, to the source, and from the lens to the source, respectively.

The signal behind the clusters can be extracted by stacking in circular annuli as
\beq
\label{eq_Delta_Sigma_3}
\Delta \Sigma_+ (R) = \frac{\sum_i  (w_i \Sigma_{\text{cr},i}^{-2}) \epsilon_{\text{+},i} \Sigma_{\text{cr},i}} {\sum_i ( w_i \Sigma_{\text{cr},i}^{-2}) },
\eeq
where $\epsilon_{\text{+},i}$ is the tangential component of the ellipticity of the $i$-th source galaxy after bias correction and $w_i$ is the \texttt{lensfit} weight assigned to the source ellipticity. The sum runs over the galaxies included in the annulus at projected distance $R$.

If the redshifts are known with an uncertainty, as it is the case for photometric redshifts, the point estimator in Eq.~\ref{eq_Delta_Sigma_3} is biased. Optimal estimators exploiting the full information contained in the probability density distribution of the photometric redshift have been advocated \citep{she+al04}, but these methods can be hampered by the uncertain determination of the shape of the probability distribution, which is very difficult to ascertain \citep{tan+al17}. However, the level of systematics introduced by the estimator in Eq.~\ref{eq_Delta_Sigma_3} for quality photometric redshifts as those of the CFHTLens/RCSLenS is under control and well below the statistical uncertainty, see Sec.~\ref{sec_syst_phot}. We can safely use it in our analysis.

The raw ellipticity components, $e_{\text{m}, 1}$ and $e_{\text{m}, 2}$, were calibrated and corrected by applying a multiplicative and an additive correction,
\begin{equation}
\label{eq_calibration}
e_{\text{true}, i} = \frac{e_{\text{m}, i} - c_i}{1 + {\bar m}}  \, \hspace{1cm} (i=1,2) \, .
\end{equation}
The bias parameters can be estimated either from simulated images or empirically from the data. 

The multiplicative bias $m$ was identified from the simulated images \citep{hey+al12,mil+al13}. The simulation-based estimate mostly depends on the shape measurement technique and is common to both CFHTLenS and RCSLenS. In each sky area, we considered the average $\bar{m}$, which was evaluated taking into account the weight of the associated shear measurement \citep{vio+al15}, 
\beq
\label{eq_Delta_Sigma_5}
\bar{m}(R) = \frac{\sum_i w_i \Sigma_{\text{cr},i}^{-2} m_i}{\sum_i w_i \Sigma_{\text{cr},i}^{-2}}.
\eeq

The two surveys suffer for a small but significant additive bias at the level of a few times $10^{-3}$. This bias depends on the SNR (signal-to-noise ratio) and the size of the galaxy. The empirical estimate of the additive bias is very sensitive to the actual properties of the data \citep{hey+al12,mil+al13} and it differs in the two surveys \citep{hil+al16}. The residual bias in the first component is consistent with zero ($c_1=0$) for CFHTLenS \citep{hey+al12,mil+al13}, which is not the case for RCSLenS  \citep{hil+al16}. Furthermore, RCSLenS had to model the complex behaviour of the additive ellipticity bias with a two-stage process. The first stage is the detector level correction. Once this is corrected for, the residual systematics attributed to noise bias are removed \citep{hil+al16}.


\section{Background selection}
\label{sec_back_sel}

\begin{figure}
\resizebox{\hsize}{!}{\includegraphics{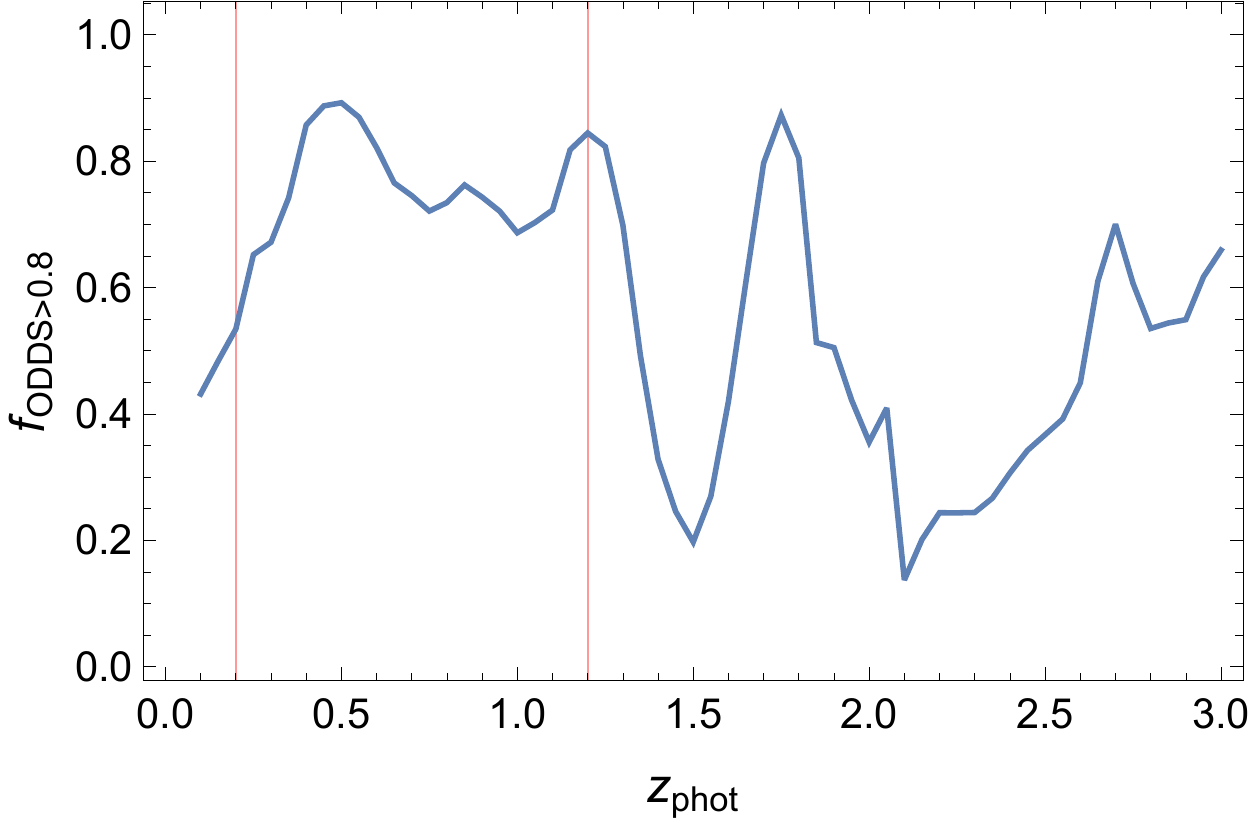}} 
\caption{Fraction of galaxies in the W1 field of the CFHTLenS catalogue with $\texttt{ODDS} \ge 0.8$ as a function of the photometric redshift. Here, redshift estimates exploit the optical $ugriz$ bands. The vertical red lines delimit the allowed redshift range for CFHTLenS sources.}
\label{fig_Delta_ODDS}
\end{figure}

Our source galaxy sample includes all detected galaxies with a non-zero shear weight and a measured photometric redshift \citep{mil+al13}. We did not reject those pointings failing the requirements for cosmic shear but still suitable for galaxy lensing \citep{vel+al14,cou+al15}.

Our selection of background galaxies relies on robust photometric redshifts. Photometric redshifts exploiting the ancillary data sets were computed in \citet{cou+al15} with the template fitting code \texttt{LEPHARE} \citep{ilb+al06}. The spectroscopic sample described in Section~\ref{sec_spec} was used for validation and calibration. These photometric redshifts were retrieved within a dispersion $\sim 0.03$--$0.04(1+z)$ and feature a catastrophic outlier rate of $\sim1$-$4$ per cent. Main improvements with respect to CFHTLenS rely on the choice of isophotal magnitudes and PSF homogenization \citep{hil+al12} at faint magnitude, and the contribution of NIR data above $z \sim 1$. The UV photometry improves the precision of photometric redshifts at low redshifts, $z\la 0.2$. 

As a preliminary step, we identified (as candidate background sources for the WL analysis behind the lens at $z_\text{lens}$) galaxies such that
\beq
\label{eq_zphot_1}
z_\text{s} > z_\text{lens} +\Delta  z_\text{lens},
\eeq
where $z_\text{s}$ is the photometric redshift or, if available, the spectroscopic redshift. For our analysis, we conservatively set $\Delta  z_\text{lens}=0.05$. On top of this minimal criterion, we required that the sources passed more restrictive cuts in either photometric redshift or colour properties, which we discuss in the following.

\subsection{Photometric redshifts}

As a first additional criterion for galaxies with either spectroscopic redshifts or photometric redshift, $z_\text{s} $, we adopted the cuts
\begin{equation} 
\label{eq_zphot_2}
z_\text{2.3\%}  >  z_\text{lens} +\Delta  z_\text{lens} \,\, \text{AND} \,\, z_\text{min}  <  z_\text{s} < z_\text{max},
\end{equation}
where $z_\text{2.3\%}$ is the lower bound of the region including the 2-$\sigma$ (95.4 per cent) of the probability density distribution, i.e. there is a probability of 97.7\% that the galaxy redshift is higher than $z_\text{2.3\%}$.

The redshifts $z_\text{min}$  and $z_\text{max}$ are the lower and upper limits of the allowed redshift range, respectively.

For the galaxies with spectroscopic redshift, $z_\text{min}=0$ whereas  $z_\text{max}$ is arbitrarily large. For the sample with only photometric redshifts, the allowed redshift range was determined according to the available bands. For the galaxies exploiting only the CFHTLenS photometry ($ugriz$), we restricted the selection to $0.2<z_\text{phot}<1.2$; for the RCSLenS photometric redshifts, which lack for the $u$ band, we restricted the selection to $0.4<z_\text{phot}<1.2$; for galaxies with additional NIR data, we relaxed the upper limit, i.e. we set $z_\text{max}$ to be arbitrarily large; for galaxies with ancillary UV data, we relaxed the lower limit, i.e. we set $z_\text{min}=0$.
 
In case of only optical filters without NIR data, we required that the posterior probability distribution of the photometric redshift is well behaved by selecting galaxies whose fraction of the integrated probability included in the primary peak exceeds 80 per cent, 
\beq
\label{eq_zphot_3}
\texttt{ODDS}  \ge 0.8.
\eeq
The $\texttt{ODDS}$ parameter quantifies the relative importance of the most likely redshift \citep{hil+al12}. The additional selection criterion based on the $\texttt{ODDS}$ parameter guarantees for a clean selection but it is somewhat redundant. In fact, most of the galaxies with $\texttt{ODDS} <0.8$ were already cut by retaining only galaxies in the redshift range $z_\text{min}<z<z_\text{max}$, see Fig.~\ref{fig_Delta_ODDS}. For sources in the CFHTLenS without ancillary information, a fraction of $\sim 76$ per cent of the sources in the redshift range $0.2<z<1.2$ meet the $\texttt{ODDS}$ requirement.

By definition, the constraint $z_\text{2.3\%}> z_\text{lens}$ guarantees that the contamination is at the $2.3$ per cent level. The additional $\Delta z_\text{lens}$ requirement in Eq.~(\ref{eq_zphot_2}) makes the contamination even lower. Since $\Delta z_\text{lens}=0.05$ is $\sim 1(0.5)\sigma_{z_\text{phot}}$ at $z_\text{s} =0.2 (1.2)$, we are practically requiring that the contamination is $\sim 0.1$ (0.6) per cent for galaxies at $z_\text{phot} =0.2 (1.2)$.



When available, the impact of ancillary UV and mainly NIR data is significant. Thanks to the increased accuracy in the redshift estimates, we can include in the background sample more numerous and more distant galaxies. In particular, when we could rely on improved photometric redshift estimates based on the NIR additional data set, we did not have to restrict our redshift sample to $z_\text{phot}< 1.2$, increasing the full background source sample by $\sim30$ per cent compared to other CFHTLenS lensing studies, without introducing any systematic bias \citep{cou+al15}. 



\subsection{Colour-colour space}

\begin{figure}
\resizebox{\hsize}{!}{\includegraphics{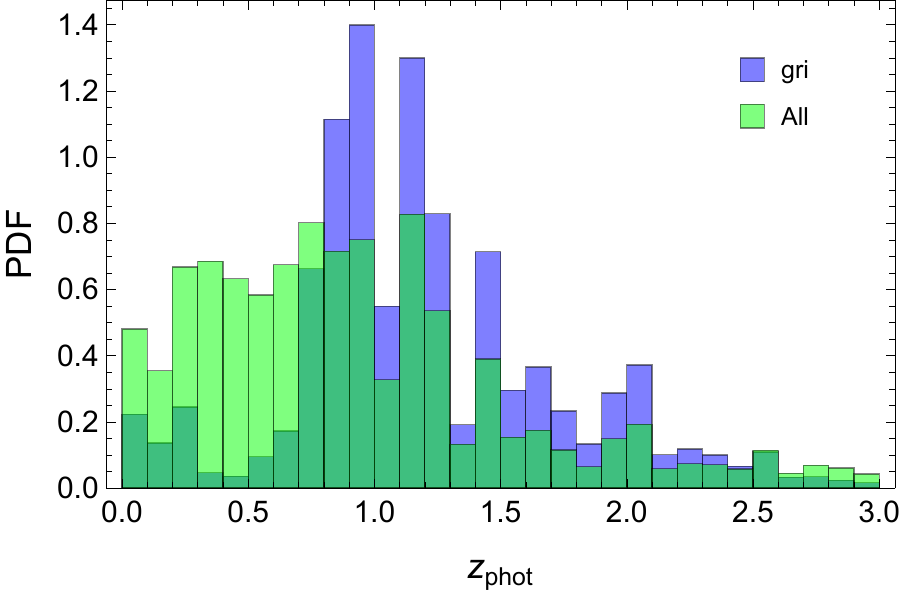}} 
\caption{Photometric redshift distributions of galaxies in the COSMOS catalogue, before (green) and after (blue) the $gri$ colour-colour cut.}
\label{fig_histo_zphot_COSMOS}
\end{figure}

\begin{figure}
\resizebox{\hsize}{!}{\includegraphics{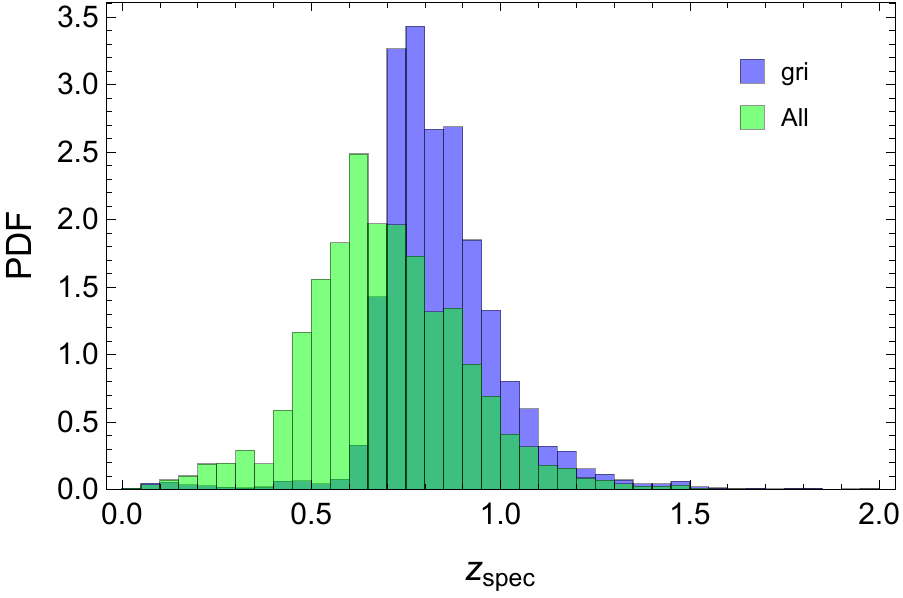}} 
\caption{Spectroscopic redshift distributions of VIPERS/VVDS galaxies in CFHTLS-W1, W4, before (green) and after (blue) the $gri$ colour-colour cut.}
\label{fig_histo_zspec_VIPERS}
\end{figure}

The population of source galaxies can be identified with a colour-colour selection \citep{med+al10,for+al16}. For clusters at $z_\text{lens}<0.65$, we adopted the following criterion exploiting the $gri$ bands, which efficiently select galaxies at $z_\text{s}\ga0.7$ \citep{ogu+al12,cov+al14}:
\begin{equation} 
\label{eq_col_1}
(g-r < 0.3) \\ 
\, \text{OR} \\
\, (r - i >  1.3) \\ 
\, \text{OR} \\
\, (r - i > g-r). 
\end{equation}
To pass this cut, lensing sources have to be detected in the $r$ band and in at least one of the filters $g$ or $i$.

Since we use photometric redshifts to estimate the lensing depth, we required
\beq
\label{eq_col_2}
z_\text{s}  > z_\text{min}, 
\eeq
as for the $z_\text{phot}$ selection. The two-colours method may select as background sources an overdensity of sources at low photometric redshifts \citep{cov+al14}. Most of these sources are characterized by a low value of the $\texttt{ODDS}$ parameter, and $z_\text{phot}$ is not well constrained, hinting to possible degeneracies in the photometric redshift determination based only on optical colours. Since $z_\text{phot}$ still enters in the estimate of the lensing depth, we conservatively excluded these galaxies through Eq.~(\ref{eq_col_2}).

The colour cuts in equation~(\ref{eq_col_1}) were originally proposed by \citet{ogu+al12} based on the properties of the galaxies in the COSMOS photometric catalogue \citep{ilb+al09}, which provides very accurate photometric redshifts down to $i\sim25$. They determined the cuts after inspection of the photometric redshift distributions in the $g$-$r$ versus $r$-$i$ colour space. The criteria are effective, see Fig.~\ref{fig_histo_zphot_COSMOS}. When we analyze the distribution of photometric redshifts, 64.4 per cent of the 385044 galaxies in the COSMOS survey with measured photometric redshift have $z_\text{phot}> 0.63$, i.e. the highest cluster redshift in our sample. After the colour-colour cut, 92.0 per cent of the selected galaxies have $z_\text{phot}> 0.63$. If we limit the galaxy sample to $z_\text{s}> 0.2(0.4)$, as required in Eq.~(\ref{eq_col_2}), 95.4 (98.3) per cent of the selected galaxies have $z_\text{phot}> 0.63$. In fact, a very high fraction of the not entitled galaxies which pass the colour test ($44.4$ per cent) forms an overdensity at $z_\text{phot}\la 0.2$.

We can further assess the reliability of the colour-space selection considering the spectroscopic samples in CFHTLS-W1 and W4 fields. We considered the 61525 galaxies from the VIPERS and VVDS samples with high quality spectroscopic redshifts and good CFHTLS $gri$ photometry. Before the cut, 61.6 per cent of the sources have $z_\text{spec}> 0.63$. After the cut, 97.0 per cent of the 26711 selected galaxies have $z_\text{spec}> 0.63$, see Fig.~\ref{fig_histo_zspec_VIPERS}. If we only consider galaxies with $z_\text{s}> 0.2(0.4)$, as required in Eq.~(\ref{eq_col_2}), 97.7 (98.1) per cent of the selected galaxies have $z_\text{phot}> 0.63$.

Based on the above results, we can roughly estimate that a galaxy passing the $gri$ cuts has a $\ga95$ per cent probability of being at $z>0.63$. When combined with the constraint $z_\text{phot}>z_\text{lens}$, the combined probability of the galaxy of being behind the highest redshift PSZ2LenS cluster goes up to $\ga 98$ per cent.

\section{Lens model}
\label{sec_lens}

The lensing signal is generated by all the matter between the observer and the source. For a single line of sight, we can break the signal down in three main components: the main halo, the correlated matter around the halo, and the uncorrelated matter along the line of sight.

The profile of the differential projected surface density of the lens can then be modelled as
\beq
\Delta \Sigma_\text{tot}=\Delta \Sigma_\text{1h}+\Delta \Sigma_\text{2h} \pm \Delta \Sigma_\text{LSS}.
\eeq
The dominant contribution up to $\sim 3~\text{Mpc}/h$, $\Delta \Sigma_\text{1h}$, comes from the cluster; the second contribution is the 2-halo term, $\Delta \Sigma_\text{2h}$, which describes the effects of the correlated matter distribution around the location of the main halo. The 2-halo term is mainly effective at scales $\ga 10$ Mpc. $\Delta \Sigma_\text{LSS}$ is the noise contributed by the uncorrelated matter.

The cluster can be modelled as a Navarro Frenk White  (NFW) density profile \citep{nav+al97},
\beq 
\rho_\text{NFW} = \frac{\rho_\text{s}}{(r/r_\text{s}) (1 +  r/r_\text{s})^2} ,
\eeq 
where $r_\text{s}$ is the inner scale length and $\rho_\text{s}$ is the characteristic density. In the following, as reference halo mass, we consider $M_{200}$, i.e., the mass in a sphere of radius $r_{200}$. The concentration is defined as $c_{200}=r_{200}/r_\text{s}$. 

The NFW profile may be inaccurate in the very inner or in the outer regions. The action of baryons, the presence of a dominant BCG, and deviations from the NFW predictions \citep{man+al08,du+ma14,ser+al16_einasto} can play a role. However, for CFHTLenS/RCSLenS quality data, systematics caused by poor modelling are subdominant with respect to the statistical noise. Furthermore, in the radial range of our consideration, $0.1<R<3~\text{Mpc}/h$, the previous effects are subdominant.

To better describe the transition region between the infalling and the collapsed matter at large radii, the NFW density profile can be smoothly truncated as \citep[BMO]{bal+al09},
\beq 
\rho_\text{BMO} = \rho_\text{NFW}(r) \left(\frac{r_\text{t}^2}{r^2 +  r_\text{t}^2} \right)^2,
\eeq 
where $r_\text{t}$ is the truncation radius.  For our analysis, we set $r_\text{t} = 3\,r_{200} $ \citep{og+ha11,cov+al14}. 

The 2-halo term $\Delta \Sigma_\text{2h}$ arises from the correlated matter distribution around the location of the galaxy cluster \citep{cov+al14,ser+al15_bias}. The 2-halo shear around a single lens of mass $M$ at redshift $z$ for a single source redshift can be modelled as
\citep{og+ta11,og+ha11}
\begin{equation}
\gamma_{+, \text{2h}} (\theta; M, z) = \int \frac{l d l}{2 \pi} J_2(l \theta) \frac { \bar{\rho}_\text{m} (z) b_\text{h}(M; z)}{ (1+z)^3 \Sigma_\text{cr}  D_\text{d}^2(z)} P_\text{m}(k_l; z),
\label{eq:gamma_t2}
\end{equation}
where $ \theta$ is the angular radius, $J_n$ is the Bessel function of $n$-th order, and $k_l \equiv l / [ (1+z) D_\text{d}(z) ]$. $b_\text{h}$ is the bias of the haloes with respect to the underlying matter distribution \citep{sh+to99,tin+al10,bha+al13}. $P_\text{m}(k_l; z)$ is the linear power spectrum. We computed $P_\text{m}$ following \citet{ei+hu99}, which is fully adequate given the precision needed in our analysis. 

The 2-halo term boosts the shear signal at $\sim10~\text{Mpc}/h$ but its effect is negligible at $R \la3~\text{Mpc}/h$ even in low mass groups \citep{cov+al14,ser+al15_bias}. 
In order to favour a lens modelling as simple as possible but to still account for the correlated matter, we expressed the halo bias $b_\text{h}$ as a known function of the peak eight, i.e. in terms of the halo mass and redshift, as prescribed in \citet{tin+al10}.

The final contribution to the shear signal comes from the uncorrelated large scale structure projected along the line of sight. We modelled it as a cosmic noise which we added to the uncertainty covariance matrix \citep{hoe03}. The noise, $\sigma_\text{LSS}$, in the measurement of the average tangential shear in a angular bin ranging from $\theta_1$ to $\theta_2$ caused by large scale structure can be expressed as \citep{sch+al98b,hoe03}
\beq
\sigma_\text{LSS}^2(\theta_1, \theta_2) = 2 \pi \int_0^{\infty} P_k(l)g^2(l,\theta_1, \theta_2)\ dl \ ,
\eeq
where $P_k(l)$ is the effective projected power spectrum of lensing and the function $g(l,\theta_1, \theta_2)$ 
embodies the filter function $U$ as
\beq
g(l,\theta_1, \theta_2)= \int_0^{\theta_2} \phi U(\phi) J_0(l\phi)\ d \phi \ .
\eeq
The filter of the convergence power spectrum is specified by our choice to consider the azimuthally averaged tangential shear \citep{hoe+al11}. The effects of non-linear evolution on the relatively small scales of our interest were accounted for in the power spectrum following the prescription of \citet{smi+al03}. We computed $\sigma_\text{LSS}$ at the weighted redshift of the source distribution.

The cosmic-noise contributions to the total uncertainty covariance matrix can be significant at very large scales or for very deep observations \citep{ume+al14}. In our analysis, the source density is relatively low and errors are dominated by the source galaxy shape noise. For completeness, we nevertheless considered the cosmic noise in the total uncertainty budget.


\section{Lensing signal}
\label{sec_signal}

\begin{table}
\caption{Background galaxy samples for weak-lensing shear measurements. The signal was collected between $0.1$ and $3.16~\text{Mpc}/h$. Column~1: PSZ2 index of the cluster. Column 2: cluster redshift. Column~3: effective source redshift. Column~4: total number of background galaxies. Column~5: raw number density of background lensing sources per square arc minute, including all objects with measured shape. Column~6: WL signal-to-noise ratio.}
\label{tab_WL_back}
\centering
\begin{tabular}[c]{r l l rrr}
\hline
	\noalign{\smallskip}  
	Index  & $z_\text{lens}$  & $z_\text{back}$    &    \multicolumn{1}{c}{$N_\text{g}$}  &    \multicolumn{1}{c}{$n_\text{g}$}  &  SNR  \\ 
	 	\hline
	\noalign{\smallskip}      
21  	&	0.078	&	0.712	&	13171 	&	1.61 	&	2.71 	\\
38  	&	0.044	&	0.752	&	71520 	&	3.02 	&	$-$0.22	\\
43  	&	0.034	&	0.728	&	121913	&	3.18 	&	2.26 	\\
212 	&	0.327	&	0.875	&	2956  	&	3.71 	&	5.71 	\\
215 	&	0.151	&	0.687	&	2972  	&	1.15 	&	$-$0.25	\\
216 	&	0.234	&	0.820	&	1801  	&	1.40 	&	0.15 	\\
243 	&	0.400	&	0.908	&	981   	&	1.59 	&	0.09 	\\
251 	&	0.222	&	0.847	&	3529  	&	2.54 	&	1.92 	\\
268 	&	0.082	&	0.670	&	22135 	&	2.99 	&	1.75 	\\
271 	&	0.336	&	0.891	&	1741  	&	2.26 	&	1.73 	\\
329 	&	0.251	&	0.857	&	1799  	&	1.56 	&	3.01 	\\
360 	&	0.279	&	0.803	&	2025  	&	2.04 	&	1.17 	\\
370 	&	0.140	&	0.691	&	4242  	&	1.45 	&	1.41 	\\
391 	&	0.277	&	0.874	&	4218  	&	4.21 	&	5.35 	\\
446 	&	0.140	&	0.769	&	24930 	&	8.53 	&	4.48 	\\
464 	&	0.141	&	0.755	&	27632 	&	9.55 	&	5.45 	\\
473 	&	0.106	&	0.760	&	53052 	&	11.24	&	2.25 	\\
478 	&	0.630	&	0.967	&	2831  	&	7.43 	&	2.85 	\\
547 	&	0.081	&	0.688	&	18122 	&	2.38 	&	2.94 	\\
554 	&	0.384	&	0.909	&	1303  	&	2.01 	&	2.40 	\\
586 	&	0.545	&	1.156	&	478   	&	1.09 	&	0.46 	\\
618 	&	0.045	&	0.678	&	43093 	&	1.87 	&	4.24 	\\
721 	&	0.528	&	0.880	&	681   	&	1.51 	&	0.86 	\\
724 	&	0.135	&	0.745	&	11376 	&	3.66 	&	6.51 	\\
729 	&	0.199	&	0.791	&	3530  	&	2.14 	&	1.39 	\\
735 	&	0.470	&	0.921	&	1247  	&	2.45 	&	1.05 	\\
804 	&	0.140	&	0.758	&	41113 	&	14.01	&	2.57 	\\
822 	&	0.185	&	0.789	&	19936 	&	10.80	&	3.08 	\\
902 	&	0.350	&	0.897	&	1713  	&	2.35 	&	1.28 	\\
955 	&	0.400	&	1.036	&	800   	&	1.30 	&	$-$0.11	\\
956 	&	0.188	&	0.695	&	1626  	&	0.90 	&	1.70 	\\
961 	&	0.600	&	1.164	&	123   	&	0.31 	&	0.26 	\\
1046	&	0.294	&	0.863	&	4731  	&	5.13 	&	2.16 	\\
1057	&	0.192	&	0.777	&	8209  	&	4.71 	&	5.07 	\\
1212	&	0.154	&	0.717	&	2999  	&	1.20 	&	0.93 	\\
\hline
	\end{tabular}
\end{table}

\begin{figure}
\resizebox{\hsize}{!}{\includegraphics{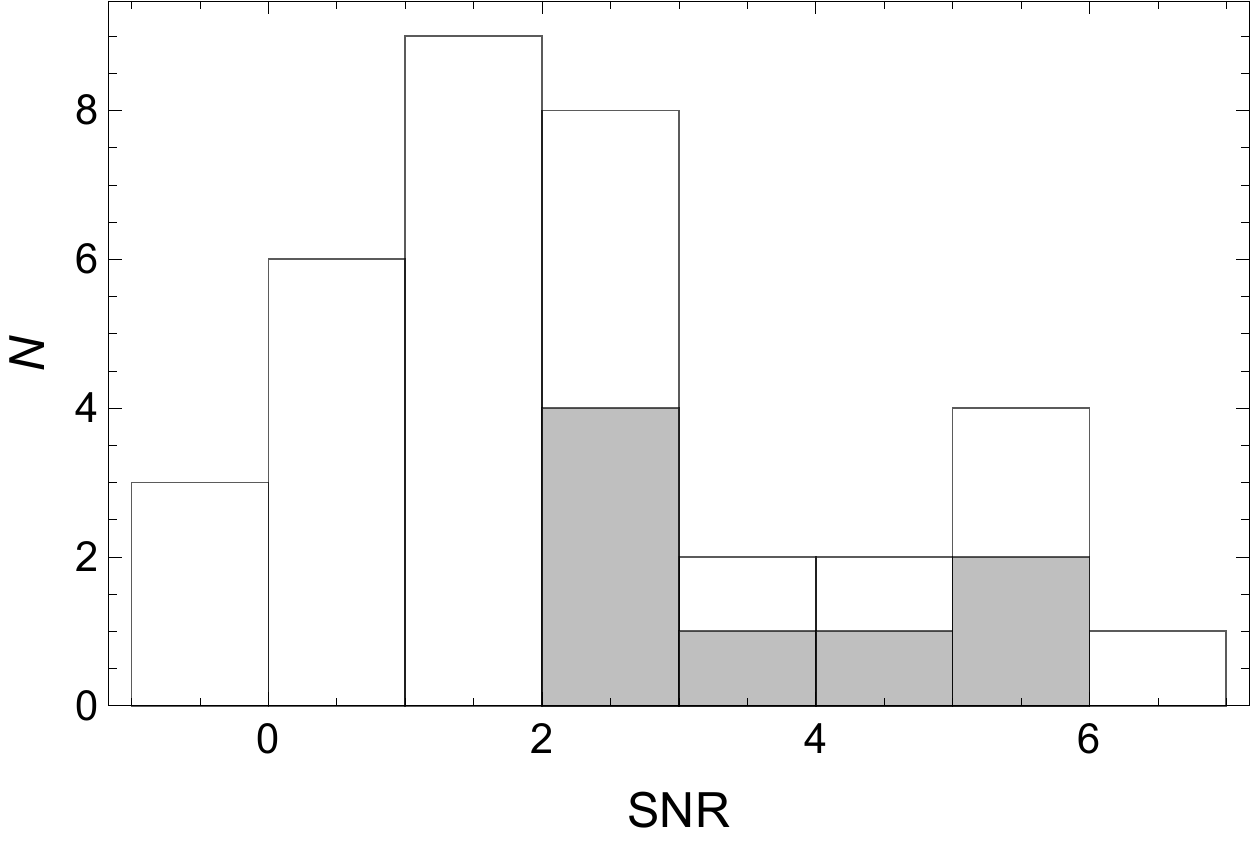}} 
\caption{Distribution of the signal-to-noise ratio of the shear signal of the PSZ2LenS clusters. The white and grey histograms show the combined RCSLenS plus CFHTLenS or the CFHTLenS sample only, respectively.}
\label{fig_SNR_histo}
\end{figure}

\begin{figure}
\resizebox{\hsize}{!}{\includegraphics{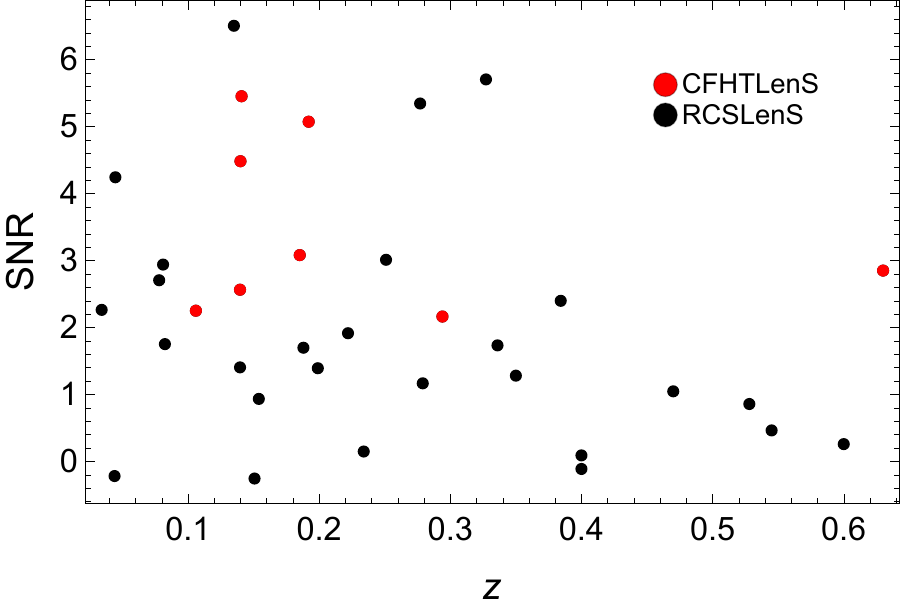}} 
\caption{Signal-to-noise ratio of the shear signal versus the cluster redshift of the PSZ2LenS clusters. The red and black points show the RCSLenS and the CFHTLenS sample, respectively.}
\label{fig_SNR_z}
\end{figure}

Our lensing sample consists of all the PSZ2 confirmed clusters centred in the CFHTLenS and RCSLenS fields with photometric redshift coverage. 
This leaves us with 35 clusters, see Table~\ref{tab_WL_sample}. 

The lensing properties of the background galaxy samples used for the weak-lensing shear measurements are listed in Table~\ref{tab_WL_back}. The effective redshift $z_\text{back}$ of the background population is defined as
\beq
\label{eq_eff_z}
\eta (z_\text{back})= \frac{\sum_i w_i \eta_i}{\sum w_i},
\eeq
where $\eta =D_\text{ds}D_\text{d}/D_\text{s}$. The effective source redshift characterizes the background population. We did not use it in the fitting procedure, where we analyzed the differential surface density derived by considering the individual redshifts of the selected background galaxies, see Eq.~(\ref{eq_Delta_Sigma_3}).

We define the total signal of the detection as the weighted differential density between 0.1 and 3.16~$\text{Mpc}/h$, $\langle \Delta \Sigma+ \rangle_{0.1 <R< 3}$. The signal-to-noise ratio (SNR) is then
\beq
\label{eq_SNR_1}
\text{SNR}=\frac{\langle \Delta \Sigma_+ \rangle_{0.1 <R< 3}}{\delta_+},
\eeq
where $\delta_+$ is the statistical uncertainty.

The distribution of SNR is shown in Fig.~\ref{fig_SNR_histo}. Nine (17) clusters out of 35 sport a SNR in excess of 3 (2). Three clusters exhibit a negative signal. Since we measured the SNR in a fixed physical size, low redshift clusters, which cover a larger area of the sky, were detected with a higher precision, see Fig.~\ref{fig_SNR_z}.

Due to the deeper observations, clusters in the fields of the CFHTLenS have larger SNRs at a given mass and redshift. The median SNR for the CFHTLenS is 3.0, whereas for the RCSLenS clusters it is 1.4. This does not bias our analysis since the subsample of PSZ2LenS in the fields of the CFHTLenS is an unbiased sample of the full PSZ2 catalogue by itself. On the other hand, the survey area of the RCSLenS is three times larger than the CFHTLenS, which counterbalances the smaller number density of background sources as far as the total signal is concerned.

\section{Inference}
\label{sec_infe}

In our reference scheme, the lens is characterized by two free parameters, the mass and the concentration, which we determined by fitting the shear profile. We performed a standard Bayesian analysis \citep{ser+al15_cM}. The posterior probability density function of mass $M_{200}$ and concentration $c_{200}$ given the data $\{{\Delta\Sigma_{+}} \}$ is
\beq
p(M_{200},c_{200}| \{{\Delta\Sigma_{+}} \})  \propto {\cal L}(M_{200,c_{200}}) p_\text{prior}(M_{200})p_\text{prior}(c_{200}),
\eeq
where $\cal L$ is the likelihood and $p_\text{prior}$ represents a prior.

\subsection{Likelihood}
The likelihood can be expressed as ${\cal L}\propto \exp (-\chi^2)$, where the $\chi^2$ function can be written as,
\beq
\label{eq_chi_WL}
\chi^2 =\sum_i \left[ \frac{\Delta\Sigma_{+}(R_i)-\Delta\Sigma_{+}(R_i; M_{200},c_{200})}{\delta_{+}(R_i)}\right]^2;
\eeq
the sum extends over the radial annuli and the effective radius $R_i$ of the $i$-th bin is estimated as a shear-weighted radius, see Appendix~\ref{app_radi}; $\Delta\Sigma_{+}(R_i)$ is the differential surface density in the annulus and $\delta_{+}(R_i)$ is the corresponding uncertainty also accounting for cosmic noise. 

The differential surface density $\Delta\Sigma_{+}$ was measured between 0.1 and $\sim 3.16~\text{Mpc}/h$ from the cluster centre in 15 radial circular annuli equally distanced in logarithmic space. The binning is such that there are 10 bins per decade, i.e. 10 bins between 0.1 and $1~\text{Mpc}/h$. The use of the shear-weighted radius makes the fitting procedure stabler with respect to radial binning, see Appendix~\ref{app_radi}.

The tangential and cross component of the shear were computed from the weighted ellipticity of the background sources as described in Section~\ref{sec_weak}. 

In our reference fitting scheme, we modelled the lens with a BMO profile; alternatively we adopted the simpler NFW profile.


\subsection{Priors}

The probabilities $p_\text{prior}(M_{200})$ and $p_\text{prior}(c_{200})$ are the priors on mass and concentration, respectively. Mass and concentration of massive haloes are expected to be related. $N$-body simulations and theoretical models based on the mass accretion history show that concentrations are higher for lower mass haloes and are smaller at early times \citep{bul+al01,duf+al08,zha+al09,gio+al12c}. A flattening of the $c$-$M$ relation is expected to occur at higher masses and redshifts \citep{kly+al11,pra+al12,lud+al14,me+ra13,du+ma14,di+kr15}.

Selection effects can preferentially include over-concentrated clusters which deviate from the mean relation. This effect is very significant in lensing selected samples but can survive to some extent even in X-ray selected samples \citep{men+al14,ser+al15_cM}. Orientation effects hamper the lensing analysis. As an example, the concentration measured under the assumption of spherical symmetry can be strongly over-estimated for triaxial clusters aligned with the line of sight.

In our reference inference scheme, we then considered both mass and concentration as uncorrelated a priori. As prior for mass and concentration, we considered uniform probability distributions in the ranges $0.05 \le M_{200}/(10^{14}h^{-1}M_\odot) \le 100$ and $1 \le c_{200} \le 20$, respectively, with the distributions being null otherwise.

There are some main advantages with this non-informative approach: (i) the flexibility associated to the concentration can accommodate to deviations of real clusters from the simple NFW modelling; (ii) we can deal with selection effects and apparent very large values of $c_{200}$; (iii) lensing estimates of mass and concentration are strongly anti-correlated and a misleading strong prior on the concentration can bias the mass estimate; (iv) the mass-concentration relation is cosmology dependent with over-concentrated clusters preferred in universes with high values of $\sigma_8$. Since the value of $\sigma_8$ is still debated \citep{planck_2015_XXIV}, it can be convenient to relax the assumption on $\sigma_8$ and on the $c$-$M$ relation.

As an alternative set of priors, we adopted uniform distributions in logarithmically spaced intervals, as suitable for positive parameters \citep{se+co13}: $p_\text{prior}(M_{200})\propto 1/M_{200}$ and $p_\text{prior}(c_{200})\propto 1/c_{200}$ in the allowed ranges and null otherwise. These priors avoid the bias of the concentration towards large values that can plague lensing analysis of good-quality data \citep{se+co13}. On the contrary, in shallow surveys such as the RCSLenS, these priors can bias low the estimates of mass and concentration.

As a third prior for the concentration, we considered a lognormal distribution with median value $c_{200}=4$ and scatter of $0.7$ in natural logarithms. As before, we considered hard limits $1<c_{200}<20$. The median value of the prior is approximately what found for massive clusters in numerical simulations. The scatter is nearly two times what found for the mass-concentration relation \citep{bha+al13,men+al14}.

We did not leave the halo bias as a free parameter, i.e. the prior on the bias is a Dirac delta function $\delta$. In the reference scheme, the 1-halo term is described with a BMO profile and the halo bias is computed as a function of the peak height $\nu$, $b_\text{h}=b_\text{h}[\nu(M_{200},z)]$, as described in \citet{tin+al10}. When we alternatively model the main halo as a NFW profile, we set $b_\text{h}=0$.

\section{Weak lensing masses}
\label{sec_masses}

\begin{table*}
\caption{Weak lensing mass measurements. Over-density masses and radii are reported at $\Delta=$ 2500, 500, 200, and at the viral over-density $\Delta_\text{vir}$, computed according to \citet{br+no98}. Spherical masses within fixed physical radii are reported within 0.5, 1.0 and 1.5 Mpc (columns 10, 11, 12). $M_\Delta$ is the mass within the sphere of radius $r_\Delta$. $M_{n\text{Mpc}}$ is the mass within the sphere of radius $n~\text{Mpc}$. Quoted values are the bi-weight estimators of the posterior probability distributions. Masses and radii are in units of $10^{14}~M_\odot$ and $\text{Mpc}$, respectively.}
\label{tab_radii}
\centering
\resizebox{\hsize}{!} {
\begin{tabular}[c]{r r@{$\,\pm\,$}l r@{$\,\pm\,$}l r@{$\,\pm\,$}l r@{$\,\pm\,$}l r@{$\,\pm\,$}l r@{$\,\pm\,$}l r@{$\,\pm\,$}l r@{$\,\pm\,$}l r@{$\,\pm\,$}l r@{$\,\pm\,$}l r@{$\,\pm\,$}l r@{$\,\pm\,$}l r@{$\,\pm\,$}l r@{$\,\pm\,$}l r@{$\,\pm\,$}l r@{$\,\pm\,$}l r@{$\,\pm\,$}l r@{$\,\pm\,$}l r@{$\,\pm\,$}l }
\hline
	\noalign{\smallskip}  
Index &  \multicolumn{2}{c}{$M_{2500}$}    &  \multicolumn{2}{c}{$r_{2500}$}  &  \multicolumn{2}{c}{$M_{500}$}    &  \multicolumn{2}{c}{$r_{500}$}  &  \multicolumn{2}{c}{$M_{200}$}    &  \multicolumn{2}{c}{$r_{200}$}  &  \multicolumn{2}{c}{$M_\text{vir}$}    &  \multicolumn{2}{c}{$r_\text{vir}$}  &   \multicolumn{2}{c}{$M_\text{0.5Mpc}$} &   \multicolumn{2}{c}{$M_\text{1Mpc}$} &   \multicolumn{2}{c}{$M_\text{1.5Mpc}$}  \\
	 	\hline
	\noalign{\smallskip}      
21  	&	3.2	&	1.1	&	0.6	&	0.1	&	7.2 	&	3.0	&	1.3	&	0.2	&	10.0	&	4.8 	&	2.0	&	0.3	&	12.1	&	6.2 	&	2.6	&	0.5	&	2.6	&	0.6	&	5.5	&	1.5	&	7.9 	&	2.5	\\
38  	&	0.5	&	0.4	&	0.3	&	0.1	&	0.9 	&	0.7	&	0.7	&	0.2	&	1.2 	&	0.9 	&	1.0	&	0.3	&	1.4 	&	1.1 	&	1.3	&	0.4	&	0.7	&	0.4	&	1.2	&	0.7	&	1.5 	&	1.0	\\
43  	&	1.6	&	0.7	&	0.5	&	0.1	&	3.1 	&	1.4	&	1.0	&	0.2	&	4.1 	&	2.1 	&	1.5	&	0.3	&	4.8 	&	2.6 	&	2.0	&	0.4	&	1.7	&	0.5	&	3.1	&	1.0	&	4.0 	&	1.6	\\
212 	&	4.7	&	1.2	&	0.6	&	0.1	&	14.8	&	4.0	&	1.5	&	0.1	&	23.5	&	7.8 	&	2.4	&	0.3	&	28.4	&	10.2	&	3.0	&	0.4	&	3.5	&	0.5	&	8.9	&	1.2	&	14.3	&	2.4	\\
215 	&	0.7	&	0.6	&	0.3	&	0.1	&	1.3 	&	1.2	&	0.7	&	0.2	&	1.7 	&	1.6 	&	1.1	&	0.4	&	2.0 	&	1.9 	&	1.4	&	0.5	&	0.9	&	0.6	&	1.6	&	1.2	&	2.1 	&	1.7	\\
216 	&	0.8	&	0.7	&	0.3	&	0.1	&	1.8 	&	1.7	&	0.8	&	0.3	&	2.5 	&	2.4 	&	1.2	&	0.4	&	2.8 	&	2.9 	&	1.5	&	0.5	&	1.2	&	0.7	&	2.2	&	1.5	&	2.9 	&	2.2	\\
243 	&	0.6	&	0.6	&	0.3	&	0.1	&	1.2 	&	1.2	&	0.6	&	0.2	&	1.7 	&	1.7 	&	1.0	&	0.4	&	1.9 	&	1.9 	&	1.2	&	0.4	&	1.0	&	0.7	&	1.7	&	1.3	&	2.2 	&	1.9	\\
251 	&	4.0	&	1.3	&	0.6	&	0.1	&	6.6 	&	2.2	&	1.2	&	0.1	&	8.1 	&	2.9 	&	1.8	&	0.2	&	8.9 	&	3.3 	&	2.2	&	0.3	&	3.4	&	0.8	&	5.7	&	1.5	&	7.4 	&	2.1	\\
268 	&	1.2	&	0.7	&	0.4	&	0.1	&	2.5 	&	1.5	&	0.9	&	0.2	&	3.5 	&	2.3 	&	1.4	&	0.3	&	4.1 	&	2.9 	&	1.8	&	0.4	&	1.4	&	0.5	&	2.7	&	1.2	&	3.6 	&	1.8	\\
271 	&	2.0	&	1.1	&	0.5	&	0.1	&	3.8 	&	2.2	&	1.0	&	0.2	&	4.9 	&	3.1 	&	1.4	&	0.3	&	5.4 	&	3.6 	&	1.7	&	0.4	&	2.2	&	0.8	&	3.8	&	1.7	&	5.0 	&	2.5	\\
329 	&	2.5	&	1.2	&	0.5	&	0.1	&	9.9 	&	5.2	&	1.4	&	0.3	&	17.3	&	10.4	&	2.3	&	0.5	&	22.2	&	14.1	&	2.9	&	0.6	&	2.4	&	0.6	&	6.4	&	1.8	&	10.7	&	3.6	\\
360 	&	2.3	&	1.4	&	0.5	&	0.1	&	4.6 	&	2.8	&	1.1	&	0.2	&	6.1 	&	4.1 	&	1.6	&	0.4	&	6.9 	&	4.8 	&	1.9	&	0.5	&	2.3	&	0.9	&	4.3	&	1.9	&	5.9 	&	2.9	\\
370 	&	1.1	&	0.8	&	0.4	&	0.1	&	2.8 	&	2.4	&	0.9	&	0.3	&	4.0 	&	3.9 	&	1.4	&	0.5	&	4.8 	&	5.0 	&	1.9	&	0.7	&	1.4	&	0.6	&	2.9	&	1.7	&	4.1 	&	2.9	\\
391 	&	2.6	&	0.9	&	0.5	&	0.1	&	11.6	&	3.6	&	1.5	&	0.2	&	21.4	&	7.9 	&	2.4	&	0.3	&	27.6	&	11.0	&	3.1	&	0.4	&	2.5	&	0.4	&	7.1	&	1.1	&	12.1	&	2.2	\\
446 	&	1.6	&	0.6	&	0.5	&	0.1	&	4.9 	&	1.4	&	1.1	&	0.1	&	7.8 	&	2.6 	&	1.8	&	0.2	&	9.9 	&	3.6 	&	2.4	&	0.3	&	1.8	&	0.3	&	4.2	&	0.7	&	6.5 	&	1.3	\\
464 	&	2.1	&	0.6	&	0.5	&	0.1	&	6.4 	&	1.5	&	1.2	&	0.1	&	10.1	&	2.7 	&	2.0	&	0.2	&	12.8	&	3.8 	&	2.6	&	0.3	&	2.1	&	0.4	&	5.0	&	0.7	&	7.7 	&	1.3	\\
473 	&	0.5	&	0.3	&	0.3	&	0.1	&	1.1 	&	0.6	&	0.7	&	0.1	&	1.6 	&	1.0 	&	1.1	&	0.2	&	2.0 	&	1.3 	&	1.4	&	0.3	&	0.8	&	0.3	&	1.5	&	0.6	&	2.1 	&	1.0	\\
478 	&	2.8	&	1.3	&	0.5	&	0.1	&	6.7 	&	3.1	&	1.1	&	0.2	&	9.6 	&	5.2 	&	1.6	&	0.3	&	10.7	&	6.1 	&	1.9	&	0.4	&	3.1	&	0.9	&	6.3	&	1.9	&	9.0 	&	3.4	\\
547 	&	1.3	&	0.7	&	0.4	&	0.1	&	6.6 	&	3.4	&	1.3	&	0.2	&	12.5	&	7.0 	&	2.2	&	0.4	&	17.5	&	10.1	&	3.0	&	0.6	&	1.6	&	0.4	&	4.7	&	1.4	&	8.0 	&	2.6	\\
554 	&	2.6	&	1.5	&	0.5	&	0.1	&	6.6 	&	4.7	&	1.1	&	0.3	&	9.6 	&	8.0 	&	1.8	&	0.5	&	11.1	&	9.8 	&	2.1	&	0.6	&	2.6	&	0.9	&	5.6	&	2.4	&	8.3 	&	4.4	\\
586 	&	2.6	&	1.4	&	0.5	&	0.1	&	5.0 	&	3.3	&	1.0	&	0.2	&	6.4 	&	4.8 	&	1.5	&	0.4	&	7.0 	&	5.4 	&	1.7	&	0.4	&	2.8	&	1.0	&	5.0	&	2.4	&	6.6 	&	3.7	\\
618 	&	2.0	&	0.9	&	0.5	&	0.1	&	9.2 	&	4.3	&	1.5	&	0.2	&	17.3	&	9.3 	&	2.4	&	0.4	&	24.5	&	14.1	&	3.4	&	0.7	&	1.9	&	0.4	&	5.5	&	1.3	&	9.5 	&	2.6	\\
721 	&	1.7	&	1.4	&	0.4	&	0.1	&	4.2 	&	4.3	&	0.9	&	0.3	&	5.9 	&	6.9 	&	1.4	&	0.6	&	6.6 	&	8.0 	&	1.7	&	0.7	&	2.1	&	1.1	&	4.4	&	3.0	&	6.3 	&	5.0	\\
724 	&	4.4	&	1.2	&	0.6	&	0.1	&	13.3	&	3.7	&	1.6	&	0.1	&	20.8	&	7.5 	&	2.5	&	0.3	&	26.1	&	10.6	&	3.3	&	0.4	&	3.1	&	0.5	&	7.8	&	1.1	&	12.4	&	2.1	\\
729 	&	0.6	&	0.5	&	0.3	&	0.1	&	1.5 	&	1.7	&	0.8	&	0.3	&	2.1 	&	2.8 	&	1.2	&	0.5	&	2.5 	&	3.5 	&	1.5	&	0.7	&	1.0	&	0.6	&	2.0	&	1.6	&	2.8 	&	2.8	\\
735 	&	1.2	&	0.9	&	0.4	&	0.1	&	2.6 	&	2.1	&	0.8	&	0.2	&	3.5 	&	3.1 	&	1.2	&	0.4	&	3.9 	&	3.6 	&	1.4	&	0.5	&	1.6	&	0.8	&	3.0	&	1.9	&	4.1 	&	2.8	\\
804 	&	0.7	&	0.4	&	0.4	&	0.1	&	1.7 	&	0.7	&	0.8	&	0.1	&	2.4 	&	1.2 	&	1.2	&	0.2	&	2.9 	&	1.5 	&	1.6	&	0.3	&	1.1	&	0.3	&	2.0	&	0.7	&	2.8 	&	1.1	\\
822 	&	1.4	&	0.5	&	0.4	&	0.1	&	3.4 	&	1.1	&	1.0	&	0.1	&	4.8 	&	1.8 	&	1.5	&	0.2	&	5.8 	&	2.3 	&	1.9	&	0.3	&	1.7	&	0.4	&	3.4	&	0.8	&	4.7 	&	1.3	\\
902 	&	4.2	&	1.4	&	0.6	&	0.1	&	7.5 	&	3.2	&	1.2	&	0.2	&	9.5 	&	4.5 	&	1.8	&	0.3	&	10.6	&	5.2 	&	2.2	&	0.4	&	3.6	&	0.8	&	6.5	&	1.9	&	8.6 	&	3.0	\\
955 	&	1.9	&	1.5	&	0.4	&	0.1	&	3.7 	&	2.9	&	0.9	&	0.3	&	4.9 	&	4.0 	&	1.4	&	0.4	&	5.5 	&	4.6 	&	1.7	&	0.5	&	2.1	&	1.2	&	3.8	&	2.3	&	5.1 	&	3.3	\\
956 	&	3.9	&	2.2	&	0.6	&	0.1	&	8.4 	&	5.5	&	1.3	&	0.3	&	11.3	&	8.6 	&	2.0	&	0.5	&	13.0	&	10.6	&	2.5	&	0.7	&	3.1	&	1.1	&	6.4	&	2.6	&	9.1 	&	4.3	\\
961 	&	0.8	&	0.7	&	0.3	&	0.1	&	2.4 	&	2.6	&	0.7	&	0.3	&	3.8 	&	4.6 	&	1.2	&	0.5	&	4.3 	&	5.4 	&	1.4	&	0.6	&	1.4	&	0.9	&	3.1	&	2.3	&	4.6 	&	3.9	\\
1046	&	3.4	&	0.9	&	0.6	&	0.0	&	6.5 	&	1.9	&	1.2	&	0.1	&	8.5 	&	2.9 	&	1.8	&	0.2	&	9.6 	&	3.4 	&	2.1	&	0.3	&	3.0	&	0.5	&	5.6	&	1.2	&	7.6 	&	1.9	\\
1057	&	2.4	&	0.6	&	0.5	&	0.0	&	6.6 	&	1.8	&	1.2	&	0.1	&	9.8 	&	3.2 	&	1.9	&	0.2	&	12.0	&	4.3 	&	2.5	&	0.3	&	2.3	&	0.4	&	5.3	&	0.9	&	7.9 	&	1.6	\\
1212	&	1.6	&	1.2	&	0.4	&	0.1	&	3.7 	&	2.9	&	1.0	&	0.3	&	5.2 	&	4.4 	&	1.6	&	0.5	&	6.1 	&	5.5 	&	2.0	&	0.6	&	1.7	&	0.8	&	3.6	&	1.9	&	5.0 	&	3.1	\\
\hline
	\end{tabular}
	}
\end{table*}

Results of the regression procedure for the reference settings of priors are listed in Table~\ref{tab_radii}. Virial over-densities, $\Delta_\text{vir}$, are based on the spherical collapse model and are computed as suggested in \citet{br+no98}.

\begin{figure}
\resizebox{\hsize}{!}{\includegraphics{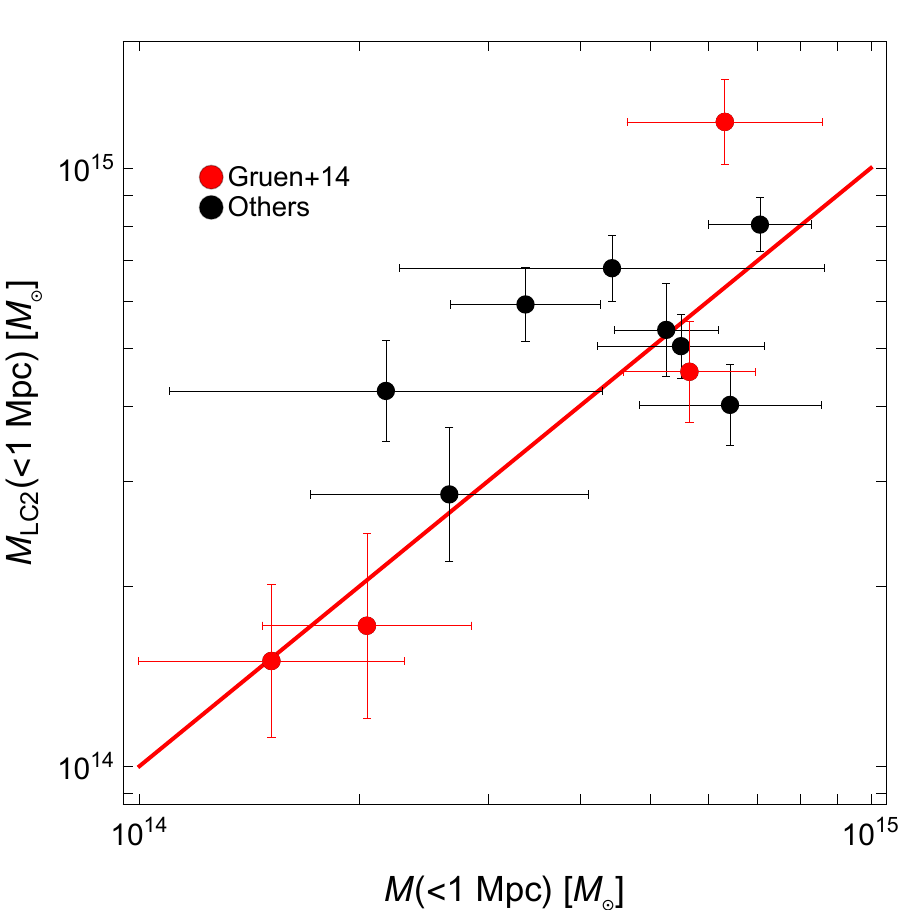}} 
\caption{Comparison between the weak lensing masses within 1~Mpc as measured in this analysis and the masses already available in literature through the LC2-single catalogue, $M_\text{LC2}$. Red points, as detailed in the legend, refer to the analysis in \citet{gru+al14}. The red full line indicates the perfect agreement.}
\label{fig_MWL_MLC2_1Mpc}
\end{figure}

\begin{figure}
\resizebox{\hsize}{!}{\includegraphics{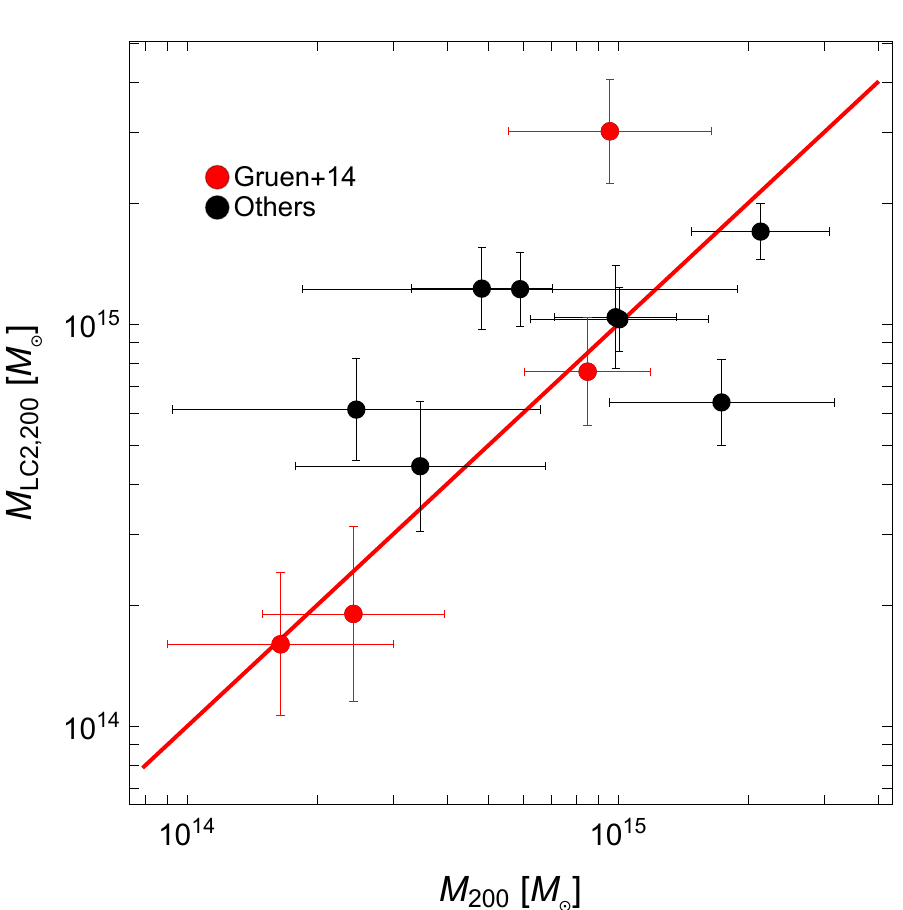}} 
\caption{Comparison between the weak lensing masses $M_{200}$ as measured in this analysis and the masses $M_{\text{LC2},200}$ reported in the LC$^2$-single catalogue. Red points, as detailed in the legend, refer to the analysis in \citet{gru+al14}. The red full line indicates the perfect agreement.}
\label{fig_MWL_MLC2_200}
\end{figure}

\begin{figure}
\resizebox{\hsize}{!}{\includegraphics{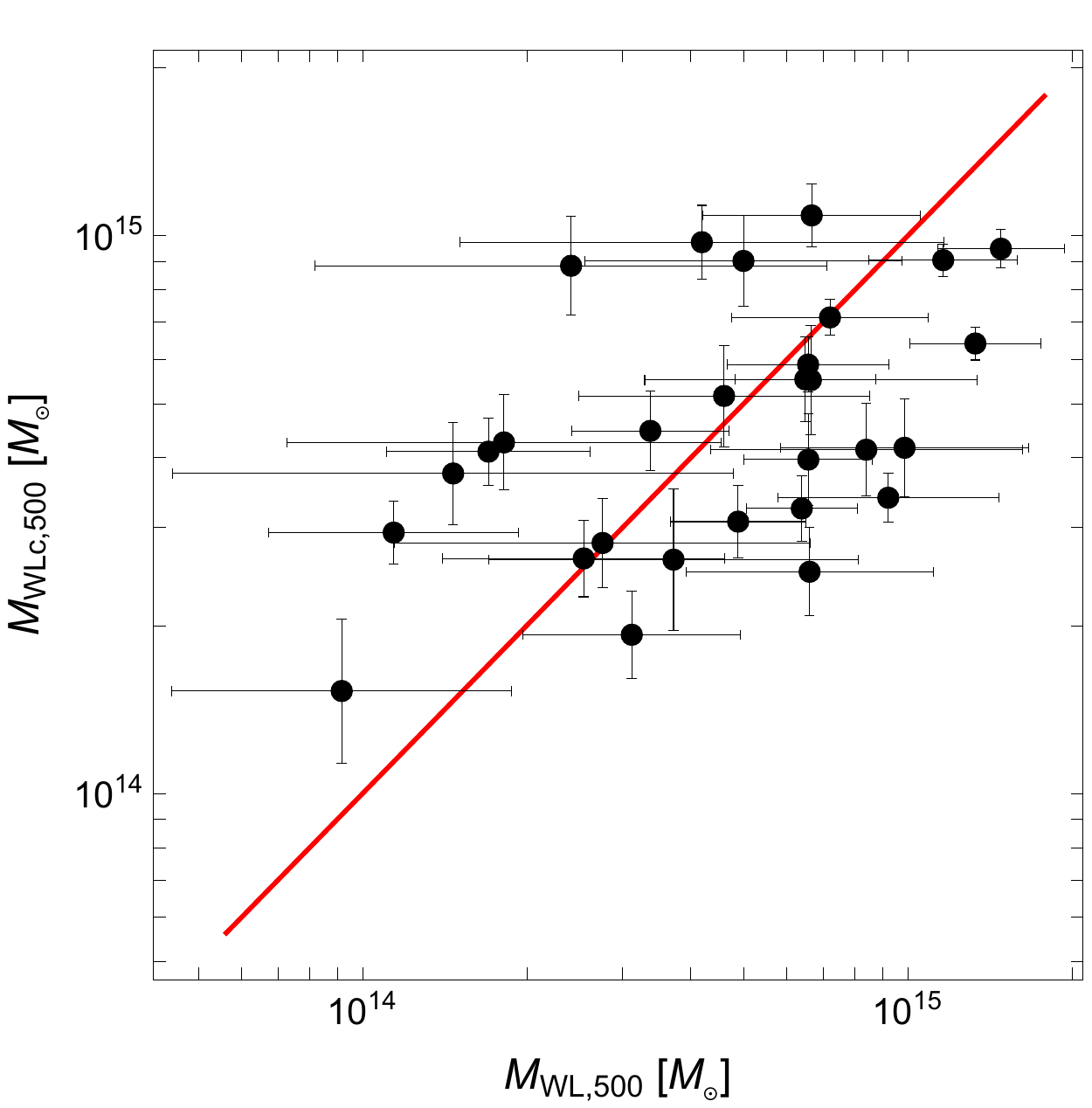}} 
\caption{Comparison between the weak lensing masses $M_{\text{WL},500}$, as measured in this analysis, and the masses $M_{\text{WLc},500}$, based on the Compton parameter $Y_{500}$ and calibrated through a weak lensing subsample by \citet{se+et17_comalit_V}. The red full line indicates the perfect agreement.}
\label{fig_MWL_MWLc_500}
\end{figure}

Some {\it Planck} clusters in CFHTLenS and RCSLensS have been the subject of other WL studies in the past. We collected previous results from the Literature Catalogs of weak Lensing Clusters of galaxies (LC$^2$), the largest compilations of WL masses up to date\footnote{The catalogues are available at \url{http://pico.oabo.inaf.it/\textasciitilde sereno/CoMaLit/LC2/}.} \citep{ser15_comalit_III}. LC$^2$ are standardized catalogues comprising 879 (579 unique) entries with measured WL mass retrieved from 81 bibliographic sources. 

We identified counterparts in the LC$^2$ catalogue by matching cluster pairs whose redshifts differ for less than $\Delta z =0.1$ and whose projected distance in the sky does not exceed $0.5~\text{Mpc}/h$. 

12 PSZ2LenS clusters have already been studied in previous analyses by \citet{dah+al02,dah06,gru+al14,ham+al09,ket+al15,cyp+al04,mer+al15,oka+al10,ume+al14,ume+al16,pe+da07,sha+al12,wtg_III_14,ok+sm16}, for a total of 25 previous mass estimates. For clusters with multiple analyses, we considered the results reported in LC$^2$-single.

We compared spherical WL masses within 1.0 Mpc, see Fig.~\ref{fig_MWL_MLC2_1Mpc}, and within $r_{200}$, see Fig.~\ref{fig_MWL_MLC2_200}. The agreement with previous results is good, $\ln M_\text{LC2}/M_\text{PSZ2LenS} \sim 0.10\pm0.38$ for masses within 1 Mpc and $ \sim 0.12\pm0.67$ for $M_{200}$. The scatter is significant and it is difficult to look for biases, if any.

Four clusters in our sample were investigated in \citet{gru+al14}. The analysis of \citet{gru+al14} was based on the same CFHTLS images but it is independent from ours for methods and tools. They used different pipelines for the determination of galaxy shapes and photometric redshifts; they selected background galaxies based on photometric redshift and they did not exploit colour-colour procedures; they considered a fitting radial range fixed in angular aperture ($2<\theta<15$\arcmin) rather than a range based on a fixed physical length; they measured the shear signal in annuli equally spaced in linear space, which give more weight to the outer regions, rather than intervals equally spaced in logarithmic space; they modelled the lens either as a single NFW profile with a (scattered) mass-concentration relation in line with \citet{duf+al08} or as a multiple component halo. Notwithstanding the very different approaches, the agreement between the two analyses is good, see Figs.~\ref{fig_MWL_MLC2_1Mpc} and ~\ref{fig_MWL_MLC2_200}. 

The most notable difference is in the mass estimate of PSZ2~G099.86+58.45 (478), when they found $M_{500}=18.1^{+5.8}_{-5.3}\times 10^{14}M_\odot$. Part of the difference, which is however not statistically significant, can be ascribed to the cluster redshift $z_\text{lens}=0.69$ assumed in \citet{gru+al14}, which was estimated through the median photometric redshift of 32 visually selected cluster member galaxies and is higher than ours.

\citet{se+et17_comalit_V} estimated the weak lensing calibrated masses $M_{\text{WLc,500}}$ of the 926 {\it Planck} clusters identified through the Matched Multi-Filter method MMF3 with measured redshift\footnote{The catalogue \textsc{HFI\_PCCS\_SZ-MMF3\_R2.08\_MWLc.dat} of {\it Planck} masses is available at \url{http://pico.oabo.inaf.it/\textasciitilde sereno/CoMaLit/forecast/}.}. Masses were estimated based on the spherically integrated Compton parameter $Y_{500}$. They used as calibration sample the LC$^2$-single catalogue and estimated the cluster mass with a forecasting procedure which does not suffer from selection effects, Malmquist/Eddington biases and time or mass evolution.

Weak lensing calibrated masses are available for 29 clusters in the PSZ2LenS sample. The comparison of masses within $r_{500}$ is showed in Fig.~\ref{fig_MWL_MWLc_500}. The agreement is good, $\ln M_\text{WLc}/M_\text{PSZ2LenS} \sim -0.03\pm0.67$.

\section{Concentrations}
\label{sec_c200}

\begin{table}
\caption{Masses and concentrations. Column~1: cluster PSZ2 index; column~2 and 3: bi-weight estimators of $M_{200}$ and concentration $c_{200}$, respectively. Column~4: minimum $\chi^2$. Column~5: number of radial annuli with background galaxies. Masses are in units of $10^{14}~M_\odot/h$.}
\label{tab_c200}
\centering
\begin{tabular}[c]{r r@{$\,\pm\,$}l    r@{$\,\pm\,$}l   r r}
\hline
	\noalign{\smallskip}  
Index &  \multicolumn{2}{c}{$M_{200}$}    &  \multicolumn{2}{c}{$c_{200}$}    & $\chi^2$ &  \multicolumn{1}{c}{$N_\text{bins}$} \\
    	 \hline
	\noalign{\smallskip}      
21  	&	7.0 	&	3.3	&	5.4 	&	3.1	&	17.5	&	15	\\
38  	&	0.8 	&	0.6	&	9.3 	&	5.6	&	13.1	&	15	\\
43  	&	2.9 	&	1.5	&	8.8 	&	5.0	&	10.4	&	15	\\
212 	&	16.5	&	5.5	&	2.9 	&	1.2	&	15.6	&	14	\\
215 	&	1.2 	&	1.1	&	8.6 	&	5.8	&	8.3 	&	15	\\
216 	&	1.7 	&	1.7	&	7.5 	&	5.7	&	22.1	&	14	\\
243 	&	1.2 	&	1.2	&	8.1 	&	5.9	&	14.7	&	14	\\
251 	&	5.7 	&	2.0	&	12.8	&	4.7	&	4.7 	&	14	\\
268 	&	2.4 	&	1.6	&	7.5 	&	5.5	&	13.4	&	15	\\
271 	&	3.4 	&	2.2	&	9.6 	&	5.6	&	17.4	&	14	\\
329 	&	12.1	&	7.3	&	2.0 	&	1.1	&	7.2 	&	15	\\
360 	&	4.3 	&	2.9	&	8.2 	&	5.6	&	2.8 	&	14	\\
370 	&	2.8 	&	2.8	&	5.0 	&	5.2	&	6.6 	&	15	\\
391 	&	15.0	&	5.5	&	1.8 	&	0.7	&	12.4	&	15	\\
446 	&	5.5 	&	1.8	&	2.9 	&	1.3	&	22.2	&	15	\\
464 	&	7.1 	&	1.9	&	2.9 	&	1.1	&	14.8	&	15	\\
473 	&	1.2 	&	0.7	&	5.1 	&	4.5	&	5.9 	&	15	\\
478 	&	6.7 	&	3.6	&	5.4 	&	4.6	&	14.2	&	15	\\
547 	&	8.7 	&	4.9	&	1.6 	&	0.6	&	14.0	&	15	\\
554 	&	6.7 	&	5.6	&	5.1 	&	4.8	&	12.6	&	14	\\
586 	&	4.5 	&	3.4	&	8.9 	&	5.4	&	11.3	&	14	\\
618 	&	12.1	&	6.5	&	1.6 	&	0.7	&	20.5	&	15	\\
721 	&	4.1 	&	4.8	&	4.8 	&	4.8	&	11.8	&	14	\\
724 	&	14.6	&	5.2	&	3.2 	&	1.5	&	15.4	&	15	\\
729 	&	1.5 	&	2.0	&	4.0 	&	5.0	&	13.7	&	15	\\
735 	&	2.5 	&	2.2	&	7.6 	&	5.8	&	19.4	&	13	\\
804 	&	1.7 	&	0.8	&	5.5 	&	4.5	&	10.8	&	15	\\
822 	&	3.4 	&	1.3	&	4.7 	&	2.7	&	10.7	&	15	\\
902 	&	6.7 	&	3.2	&	9.6 	&	4.7	&	4.7 	&	15	\\
955 	&	3.4 	&	2.8	&	9.0 	&	5.8	&	5.6 	&	13	\\
956 	&	7.9 	&	6.1	&	7.3 	&	5.3	&	4.7 	&	14	\\
961 	&	2.6 	&	3.2	&	3.3 	&	3.5	&	5.9 	&	10	\\
1046	&	5.9 	&	2.0	&	7.8 	&	3.8	&	26.0	&	15	\\
1057	&	6.9 	&	2.3	&	3.7 	&	1.5	&	18.6	&	15	\\
1212	&	3.6 	&	3.1	&	6.2 	&	5.4	&	5.2 	&	14	\\
	\hline
	\end{tabular}
\end{table}

\begin{figure}
\centering
\includegraphics[width=\hsize]{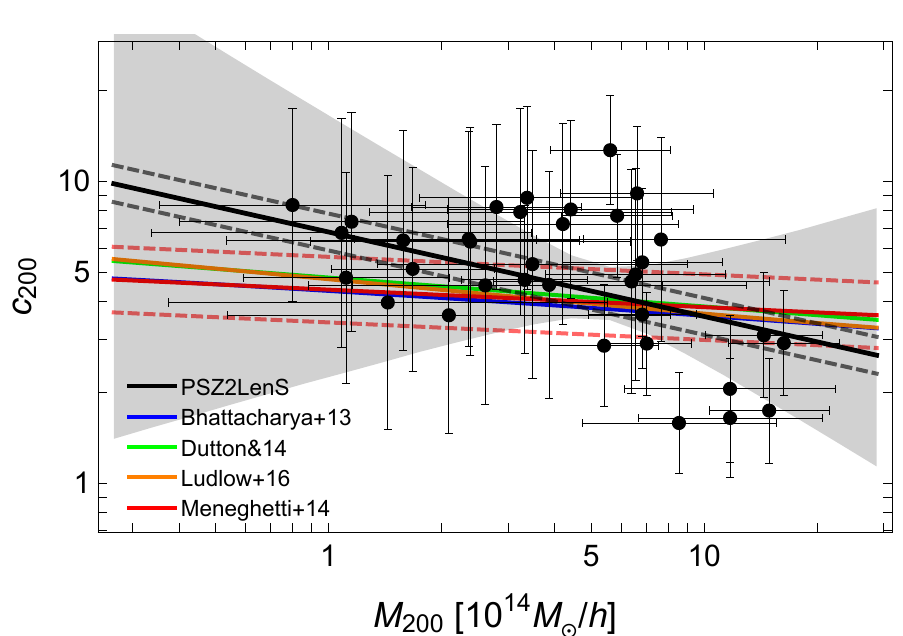} 
\caption{The mass--concentration relation of the PSZ2LenS clusters. The dashed black lines show the median scaling relation (full black line) plus or minus the intrinsic scatter at the median redshift $z=0.20$. The shaded grey region encloses the $68.3$ per cent confidence region around the median relation due to uncertainties on the scaling parameters. The blue, green, orange and red lines plot the mass-concentration relations of \citet{bha+al13}, \citet{du+ma14}, \citet{lud+al16}, and \citet{men+al14}, respectively. The dashed red lines enclose the 1-$\sigma$ scatter region around \citet{men+al14}.}
\label{fig_M200_c200}
\end{figure}

Masses and concentrations at the standard radius $r_{200}$ are reported in Table~\ref{tab_c200}. PSZ2LenS haloes are well fitted by cuspy models. The number of independent data
usually outweighs the $\chi^2$ value.

Due to the low SNR of the observations, concentrations can be tightly constrained only for a few massive haloes. The estimated concentrations can be strongly affected by the assumed priors. Whereas the effect of the priors is negligible in massive clusters with high quality observations \citep{ume+al14,ser+al15_cM}, it can be significant when the SNR is lower \citep{se+co13,ser+al15_cM}. The prior which is uniform in logarithmic space rather than in linear space favours lower concentrations. There is no other way to circumvent this problem than deeper observations.
 
The value of the observed concentrations decreases with mass, see Fig.~\ref{fig_M200_c200}. As customary in analyses of the $c$-$M$, we modelled the relation with a power law,
\beq
\label{eq_cM_1}
c_{200} =10^\alpha  \left( \frac{1+z}{1+z_\text{ref}}\right)^\gamma \left( \frac{M_{200}}{M_\text{pivot}}\right)^\beta;
\eeq
the intrinsic scatter $\sigma_{c|M}$ of the concentration around $c_{200}(M_{200})$ is taken to be lognormal \citep{duf+al08,bha+al13}. 

We performed a linear regression in decimal logarithmic ($\log$) variables using the \textsc{R}-package \texttt{LIRA}\footnote{The package \texttt{LIRA} (LInear Regression in Astronomy) is publicly available from the Comprehensive R Archive Network at \url{https://cran.r-project.org/web/packages/lira/index.html}. For further details, see \citet{ser16_lira}.}. 
\texttt{LIRA} performs a Bayesian hierarchical analysis which can deal with heteroscedastic and correlated measurements uncertainties, intrinsic scatter, scattered mass proxies and time-evolving mass distributions \citep{ser16_lira}. In particular, the anti-correlation between the lensing measured mass and concentration makes the $c$-$M$ relation apparently steeper \citep{aug+al13,du+ma14,du+fa14,ser+al15_cM}. When we correct for this, the observed relation is significantly flatter \citep{ser+al15_cM}. On the other hand, neglecting the intrinsic scatter of the weak lensing mass with respect to the true mass can bias the estimated slope towards flatter values \citep{ras+al12,se+et15_comalit_I}. We accounted for both uncertainty correlations and intrinsic scatter.

A proper modelling of the mass distribution is critical to address Malmquist/Eddington biases \citep{kel07}. Within the \texttt{LIRA} scheme, the distribution of the covariate is modelled as a mixture of time-evolving Gaussian distributions, which can be smoothly truncated at low values to model skewness. The parameters of the distribution are found within the regression procedure. This scheme is fully effective in modelling both selection effects at low masses, where {\it Planck} candidates with SNR<4.5 are excluded, and the steepness of the cosmological halo mass function at large masses. We verified that this approach is appropriate for {\it Planck} selected objects in \citet{ser+al15_comalit_II,se+et15_comalit_IV}. For the analysis of the mass-concentration relation of the PSZ2LenS sample we modelled the mass distribution of the selected objects as a time evolving Gaussian function.

We found $\alpha=0.83\pm0.42$ (for $z_\text{ref}=0.2$), $\beta=-0.27\pm0.57$, $\gamma=0.77\pm0.88$. The relation between mass and concentration is in agreement with theoretical predictions, see Fig.~\ref{fig_M200_c200}, with a very marginal evidence for a slightly steeper relation. There is no evidence for a time-evolution of the relation. The statistical uncertainties make it difficult to distinguish among competing theoretical predictions.

The estimated scatter of the WL masses, $\sigma_{M_\text{WL}|M}=0.11\pm0.08$, is in agreement with the analysis in \citet{se+et15_comalit_I} whereas the scatter of the $c$-$M$ relation, $\sigma_{c|M}=0.06\pm 0.05$, is in line with theoretical predictions \citep[$\sigma_{c|M}\sim0.15$]{bha+al13,men+al14}. 


The observed relation between lensing mass and concentration can differ from the theoretical relation due to selection effects of the sample. Intrinsically over-concentrated clusters or haloes whose measured concentration is boosted due to their orientation along the line of sight may be overrepresented with respect to the global population in a sample of clusters selected according to their large Einstein radii or to the apparent X-ray morphology \citep{men+al14,ser+al15_cM}. 

The Bayesian method implemented in \texttt{LIRA} can correct for evolution effects in the sample, e.g. massive cluster preferentially included at high redshift \citep{ser+al15_cM}. However, if the selected sample consists of a peculiar population of clusters which differ from the global population, we would measure the specific $c$-$M$ relation of this peculiar sample.

Based on theoretical predictions, SZ selected clusters should not be biased, see Section~\ref{sec_intro}. We confirmed this view. We found no evidence for selection effects: the slope, the normalization, the time evolution and the scatter are in line with theoretical predictions based on statistically complete samples of massive clusters. However, the statistical uncertainties are large and we cannot read too much into it.

\section{Stacking}
\label{sec_stacking}

\begin{figure}
\resizebox{\hsize}{!}{\includegraphics{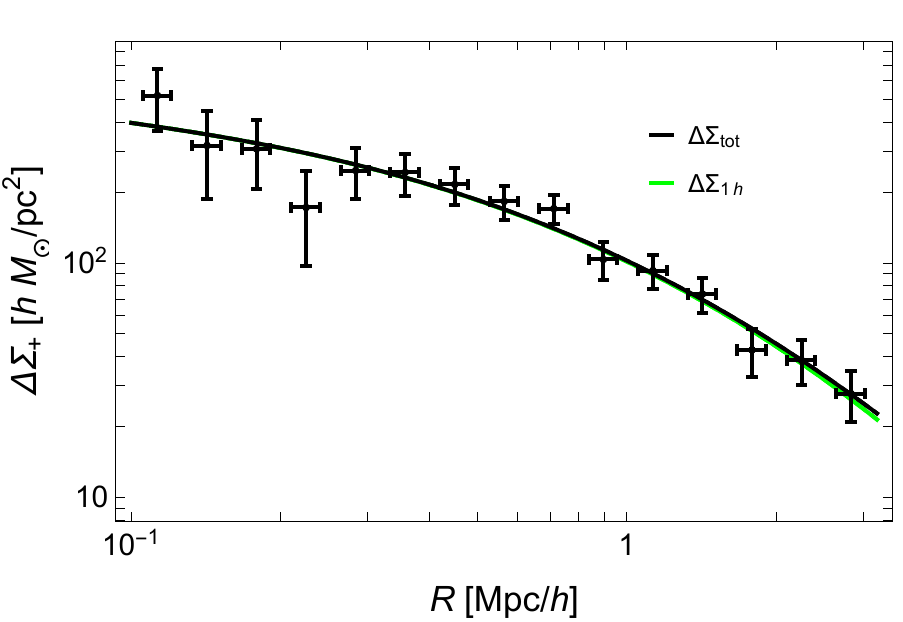}} 
\caption{Stacked differential surface density $\Delta \Sigma_+$ of the PSZ2LenS clusters. Black points are our measurements. The vertical error bars show the square root of the diagonal values of the covariance matrix. The horizontal error bars are the weighted standard deviations of the distribution of radial distances in the annulus. The green curve plots the best fitting contribution by the central halo; the black curve is the overall best fitting radial profile including the 2-halo term.}
\label{fig_Delta_Sigma_t_stacked}
\end{figure}

\begin{figure}
\resizebox{\hsize}{!}{\includegraphics{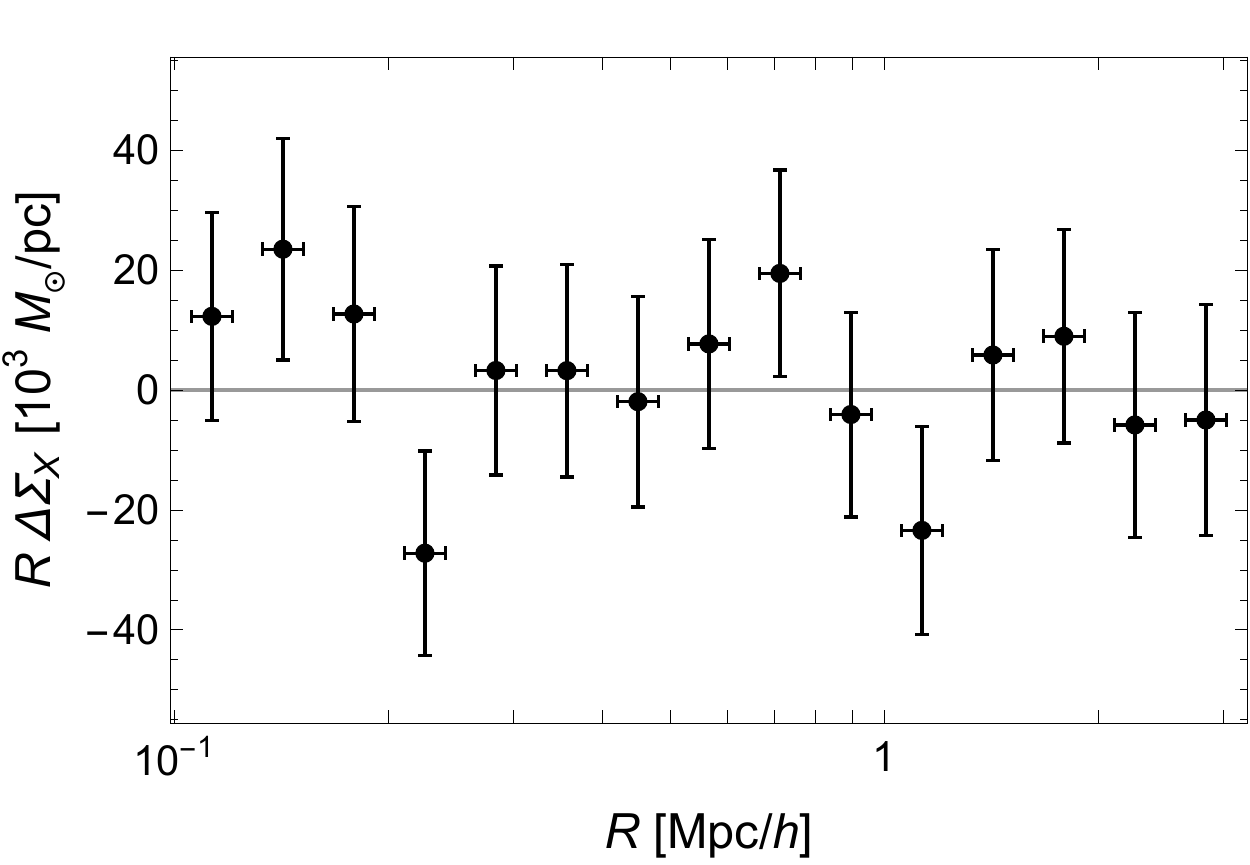}} 
\caption{The renormalized cross-component of the differential shear profile of the stacked sample. Errors bars are as in Fig.~\ref{fig_Delta_Sigma_t_stacked}.}
\label{fig_Delta_Sigma_X_stacked}
\end{figure}

\begin{figure}
\begin{tabular}{c}
\includegraphics[width=6cm]{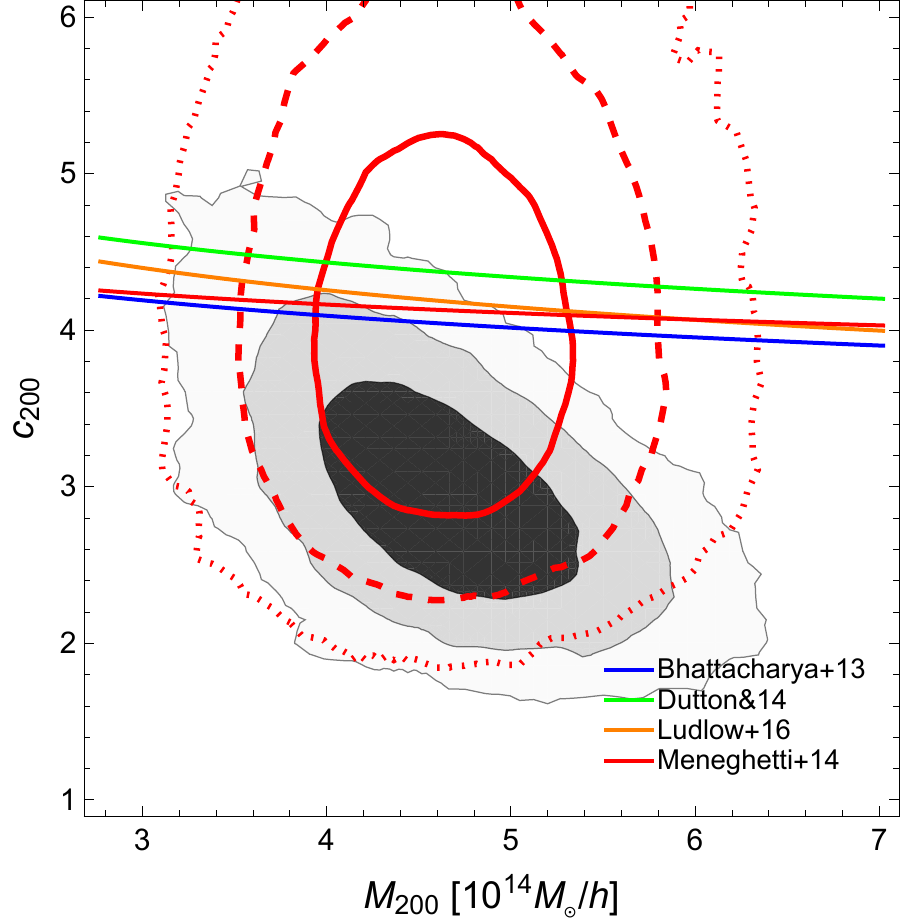}\\ 
\includegraphics[width=6cm]{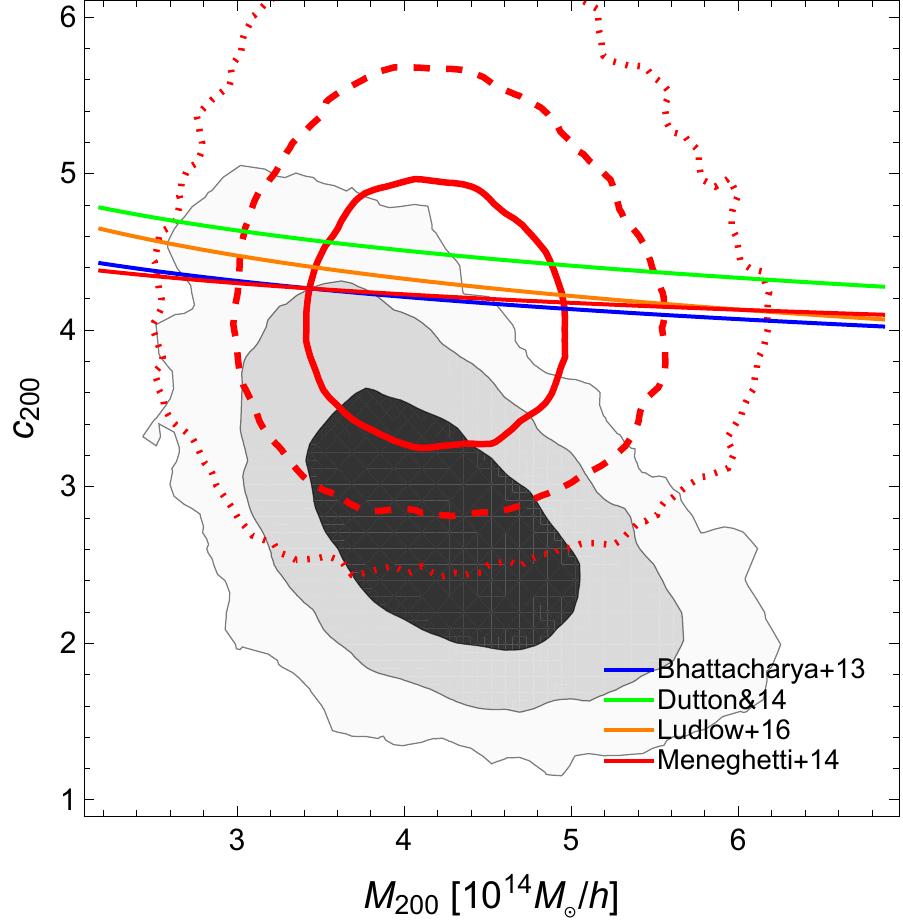}\\ 
\includegraphics[width=6cm]{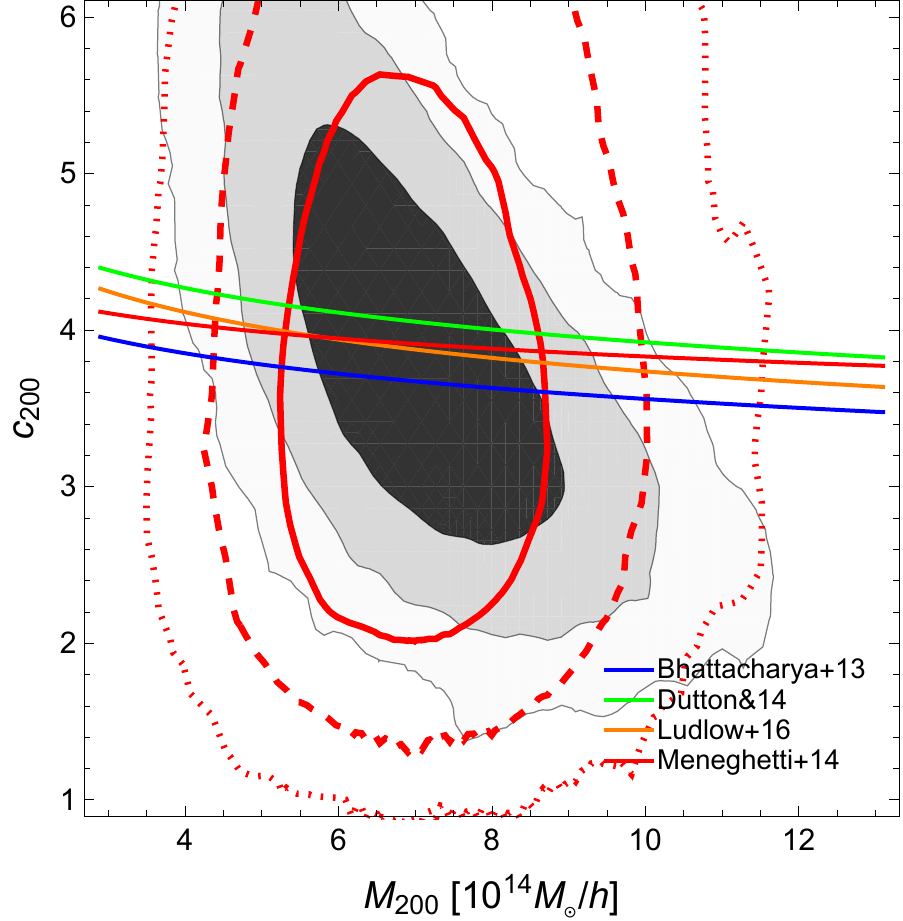}\\ 
\end{tabular}
\caption{Marginalized probability distribution of mass and concentration of the stacked clusters. The grey shadowed regions include the 1-, 2-, 3-$\sigma$ confidence region in two dimensions, here defined as the regions within which the probability density is larger than $\exp[-2.3/2]$, $\exp[-6.17/2]$, and  $\exp[-11.8/2]$ of the maximum, respectively. The blue, green, orange and red lines plot the mass-concentration relations of \citet{bha+al13}, \citet{du+ma14}, \citet{lud+al16}, and \citet{men+al14}, respectively, at the effective redshift. The red contours trace the predicted concentration from \citet{men+al14} given the observed mass distribution and the predicted scatter of the theoretical mass-concentration relation. If needed, published relations were rescaled to our reference cosmology. {\it Top panel.} All clusters were stacked; the effective redshift is $z=0.20$.  {\it Middle panel.} Stacking of the clusters at $z_\text{lens}<0.2$; the effective redshift is $z=0.14$. {\it Bottom panel.} Stacking of the clusters at $z_\text{lens}>0.2$; the effective redshift is $z=0.33$.}
\label{fig_M200_c200_stacked}
\end{figure}


The low signal-to-noise ratio hampers the analysis of single clusters. Some further considerations can be based on the stacked analysis. We followed the usual approach \citep{joh+al07,man+al08,ogu+al12,oka+al13,cov+al14}: we first stacked the shear measurements of the PSZ2LenS clusters and we then fitted a single profile to the stacked signal. 

We combined the lensing signal of multiple clusters in physical proper radii. This procedure does not bias the measurement of mass and concentration since the weight factor is mass-independent for stacking in physical length units \citep{oka+al13,ume+al14}. On the other hand, stacking in radial units after rescaling with the over-density radius can bias the estimates of mass and concentration due to the mass-dependent weight factor \citep{oka+al13}. 

The standard approach we followed is effective in assessing the main properties of the sample. Alternatively, all shear profiles can be fitted at once assuming that all clusters share the same mass and concentration \citep{se+co13}.  More refined Bayesian hierarchical inference models have to be exploited to better study the population properties \citep{lie+al17}.

The stacked signal is showed in Fig.~\ref{fig_Delta_Sigma_t_stacked}. The detection level is of $\text{SNR}=14.3$. As typical redshift of the stacked signal, we weighted the redshifts of the clusters by the lensing factor, see App.~\ref{sec_lens_weig}. The effective lensing weighted redshift is $z_\text{stack}=0.20$,  which is consistent with the median redshift of the sample.

The cross-component of the shear profile, $\Delta \Sigma_\times$ is consistent with zero at all radii, see Fig.~\ref{fig_Delta_Sigma_X_stacked}. This confirms that the main systematics are under control. 

We analyzed the stacked signal as a single lens, see Section~\ref{sec_lens}. Since the cluster centres are well determined and we cut the inner $100~\text{kpc}/h$, we did not model the fraction of miscentred haloes \citep{joh+al07,ser+al15_bias}, which we assumed to be null.

The stacked signal is well fitted by the truncated BMO halo plus the 2-halo term, $\chi^2=6.98$ for 15 bins, see Fig.~\ref{fig_Delta_Sigma_t_stacked}. The contribution by the 2-halo is marginal even at large radii, i.e. $R \sim 3~\text{Mpc}/h$, the radial outer limit of the present analysis. 

Mass, $M_{200}=(4.63\pm0.47)\times10^{14}M_\odot/h$, and concentration, $c_{200}=2.94\pm0.46$, of the stacked signal are in line with theoretical predictions, see Fig.~\ref{fig_M200_c200_stacked}. 

The total stacked signal is mostly driven by very high SNR clusters at low redshifts. We then stacked the signal of the PSZ2LenS clusters in two redshifts bins below or above $z=0.2$. The concentrations of both the low (see Fig.~\ref{fig_M200_c200_stacked}, middle panel) and high (see Fig.~\ref{fig_M200_c200_stacked}, bottom panel) redshift clusters are in line with theoretical predictions.

Recently, the CODEX (COnstrain Dark Energy with X-ray galaxy clusters) team performed a stacked weak lensing analysis of 27 galaxy clusters at $0.40\le z\le0.62$ \citep{cib+al17}. The candidate CODEX clusters were selected in X-ray surface brightness and confirmed in optical richness. They found a stacked signal of $M_{200}\sim6.6\times10^{14}M_\odot/h$ and $c_{200}=3.7$ at a median redshift of $z=0.5$ in agreement with theoretical predictions.

The LoCuSS clusters were instead selected in X-ray luminosity. The analysis of the mass-concentration relation of the sample was found in agreement with numerical simulations and the stacked profile in agreement with the NFW profile \citep{ok+sm16}.

\citet{ume+al16} analyzed the stacked lensing signal of 16 X-ray regular CLASH clusters up to $4~\text{Mpc}/h$. The profile was well fitted by cuspy dark-matter-dominated haloes in gravitational equilibrium, alike the NFW profile. They measured a mean concentration of $c_{200}\sim 3.8$ at $M_{200}\sim 9.9\times10^{14}M_\odot/h$.

Unlike previous samples, PSZ2Lens was SZ selected. Still, our results fit the same pattern and confirm $\Lambda$CDM predictions.

To check for systematics, we compared the stacked lensing mass to the composite halo mass profile $\langle M_{200} \rangle_\text{lw}$ from the sensitivity-weighted average of fits to individual cluster profiles \citep{ume+al14,ume+al16}, see App.~\ref{sec_lens_weig}. From Eq.~(\ref{eq_lens_weig_3}) with $\Gamma=0.65\pm0.10$, we obtain $\langle M_{200} \rangle_\text{lw}=(4.59\pm0.50)\times10^{14}M_\odot/h$, in excellent agreement with the stacked mass, $M_{200}=(4.63\pm0.47)\times10^{14}M_\odot/h$.

\section{The bias of  {\it Planck} masses}
\label{sec_planck_bias}

\begin{figure}
       \resizebox{\hsize}{!}{\includegraphics{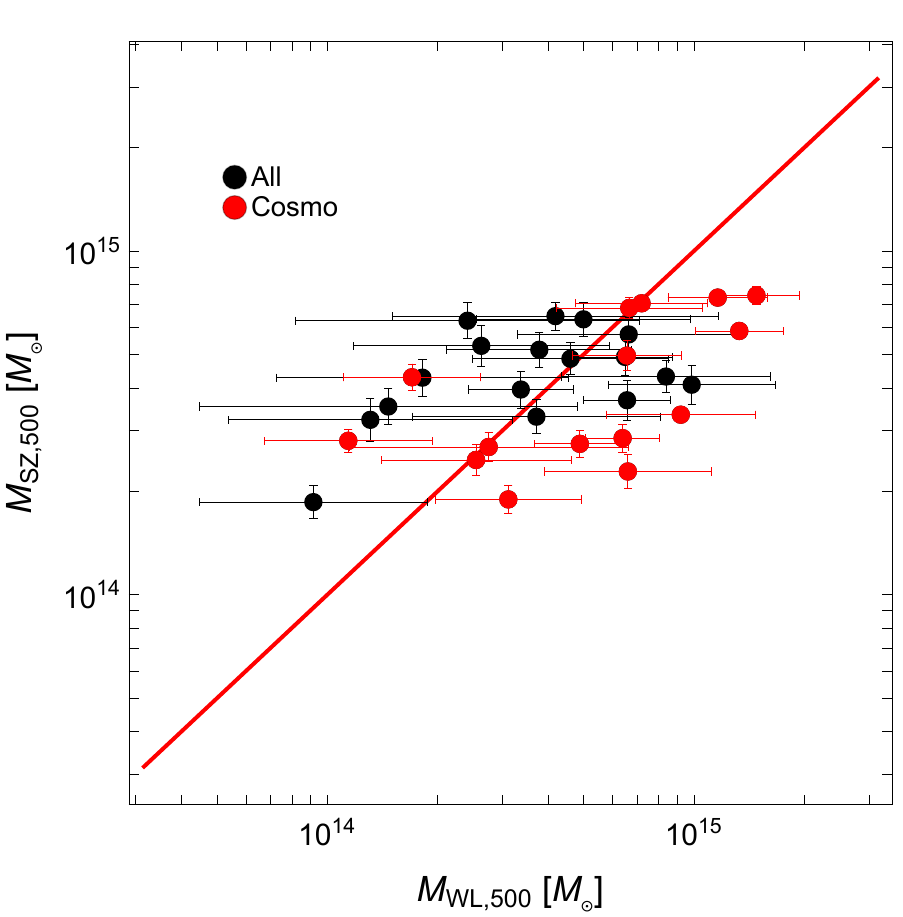}}
       \caption{{\it Planck} SZ masses $M_\text{SZ}$ versus WL masses $M_\text{WL}$ for the PSZ2LenS clusters. Red dots mark the cosmological subsample. Masses are in units of $M_\odot$ and are computed within $r_{500}$. The red line shows the bisection $M_\text{SZ}=M_\text{WL}$.}
	\label{fig_MWL_MSZ}
\end{figure}

\begin{table}
\caption{Bias of the {\it Planck} SZ masses with respect to WL masses. Values for calibration samples other than PSZ2LenS are taken from  \citet{se+et17_comalit_V}. Column~1: sample name. Column~2: number of WL clusters, $N_\text{cl}$. Columns~3 and 4: typical redshift and dispersion. Columns~5 and 6: typical WL mass and dispersion in units of $10^{14}M_\odot$. Column~7: mass bias $b_\text{SZ}=\ln(M_\text{SZ}/M_\text{WL})$. Typical values and dispersions are computed as bi-weighted estimators.}
\label{tab_bias_MSZ}
\resizebox{\hsize}{!} {
\begin{tabular}{ l r c c r c r@{$\,\pm\,$}l}     
Sample		 &	$N_\text{Cl}$	&	$z$	&	$\sigma_z$	&	$M_{500}$	&	$\sigma_{M_{500}}$  &\multicolumn{2}{c}{$b_\text{SZ}$} \\ 
\hline
PSZ2LenS		 &	32	        &	0.20	&	0.15	&	4.8	&	3.4	& $-$0.27 &	0.11 \\
PSZ2LenS Cosmo	 &	15	        &	0.13	&	0.09	&	6.4	&	4.1	& $-$0.40 &	0.14 \\
LC$^2$-single	        &	135	        &	0.24	&	0.14	&	7.8	&	4.8	& $-$0.25 &	0.04 \\
CCCP  		        &	35    		&	0.23	&	0.07	&	8.5	&	3.8	& $-$0.22	&	0.07	\\
CLASH		        &	13		&	0.37	&	0.13	&	11.3	&	3.3	& $-$0.39	&	0.08	\\
LoCuSS  		        &	38    		&	0.23	&	0.04	&	7.5	&	2.8	& $-$0.18 &	0.05	\\
WtG  		        &	37    		&	0.36	&	0.13	&	11.5	&	5.2	& $-$0.43 &	0.06	\\
\hline	
\end{tabular}
}
\end{table}

The bias of the {\it Planck} masses, i.e. the masses reported in the catalogues of the {\it Planck} collaboration, can be assessed by direct comparison with WL masses. For a detailed discussion of recent measurements of the bias, we refer to \citet{ser+al15_comalit_II} and \citet{se+et17_comalit_V}. Most of the previous studies had to identify counterparts of the PSZ2 clusters in previously selected samples of WL clusters. This can make the estimate of the bias strongly dependent on the calibration sample and on selection effects \citep{se+et15_comalit_I,bat+al16}. In fact, WL calibration clusters usually sample the very high mass end of the halo mass function. If the mass comparison is limited to the subsample of SZ detected clusters with WL observations, the estimated bias can be not representative of the full {\it Planck} sample. 

Alternatively, {\it Planck} measurements can be viewed as follow-up observations of a pre-defined weak lensing sample, see discussion in \citet{bat+al16}. Non-detections can be accounted for by setting the SZ signal of non-detected clusters to values corresponding to a multiple of the average noise in SZ measurements. As in the previous case, the calibration sample may be biased by selection effects with respect to the full PSZ2 sample. Here, the inclusion of non-detections makes the sample inconsistent with the {\it Planck} catalogue, which obviously includes only positive detections.

The estimate of the bias through the PSZ2LenS sample does not suffer from selection effects. It is a faithful and unbiased subsample of the whole population of {\it Planck} clusters. We
can estimate the bias by comparing SZ to WL masses, see Fig.~\ref{fig_MWL_MSZ}. To directly compare with the PSZ2 catalogue, we considered $M_{500}$.

We followed \citet{se+et17_comalit_V} and we estimated the bias $b_\text{SZ}$ by fitting the relation\footnote{We define the bias as $b_\text{SZ}=\ln M_\text{SZ} -\ln M_\text{WL}$. This definition slightly differs from that used in the {\it Planck} papers, where the bias is defined as $b_\text{SZ}=M_\text{SZ}/M_{500}-1$. For low values of the bias, the difference is negligible.} 
\beq
\label{eq_bias_13}
\ln \langle M_\text{SZ} \rangle = b_\text{SZ} + \ln \langle M_\text{WL} \rangle.
\eeq
We limited the analysis to the 32 clusters in PSZ2LenS which had a published $M_\text{SZ}$ mass in the {\it Planck} catalogues. 
We performed the regression with \texttt{LIRA}. We modelled the mass distribution of the selected objects as a Gaussian \citep{se+et17_comalit_V}. Corrections for Eddington/Malmquist biases were applied \citep{se+et15_comalit_I,bat+al16} and observational uncertainties and intrinsic scatters in WL and SZ masses were accounted for. We found $b_\text{SZ}=-0.27\pm0.11$. The bias for the 15 clusters in the cosmological subsample is $b_\text{SZ}=-0.40\pm0.14$, which is more prominent but still in good statistical agreement with the result for the full sample.

The intrinsic scatter of the WL masses is $23\pm15\%$, whereas the intrinsic scatter of the SZ masses is $12\pm8\%$. {\it Planck} masses are precise (thanks to the small scatter) but they are not accurate (due to the large bias). 

Based on mock analyses, \citet{shi+al16} found that enhanced scatter in relations confronting WL mass and thermal SZ effect originates from the combination of the projection of correlated structures along the line of sight and the uncertainty in the cluster radius associated with WL mass estimates. Here, we are considering $M_\text{SZ}$ from the {\it Planck catalogue}, which were computed in a X-ray based over-density radius. This makes SZ and WL mass measurements uncorrelated but can increase the relation scatter \citep{ser+al15_comalit_II}.

We determined the bias analyzing the 32 clusters confirmed by both our inspection and the {\it Planck} team. Considering the candidates confirmed by {\it Planck} alone, we should include an additional candidate, PSZ2~G006.84+50.69 (PSZ2 index: 25), which is likely a substructure of the nearby larger PSZ2~G006.49+50.56 (21), see Section~\ref{sec_not}. Taking as lens position and redshift the PSZ2 catalogue entries, we can estimate a mass lens $M_{500}=(0.47\pm0.41) \times 10^{14}M_\odot$. The mass is compatible with a null signal (as expected since we did not find any suitable candidate counter-part) and would slightly reduce the size of the bias to $b_\text{SZ}=-0.24\pm0.11$.

Alternatively, we can assess the level of bias by comparing the effective weak lensing mass $M_\text{WL,stack}$ of the stacked lensing profiles to the sensitivity-weighted average of the {\it Planck} masses $\langle M_{SZ}\rangle_\text{lw}$, see App.~\ref{sec_lens_weig}. By assuming $\Gamma=0.65\pm0.10$, we obtain $\ln (\langle M_{SZ}\rangle_\text{lw}/M_\text{WL,stack})= -0.15\pm0.09$ in good agreement with our reference result.

\citet{bat+al16} argued that if the sample selection preserves the original {\it Planck} selection, as the case for PSZ2LenS, the factor $b_\text{SZ}$ estimated through the {\it Planck} catalog masses can suffer by Eddington bias. By comparison with measurements by ACT, they estimated an Eddington bias correction of order of 15 per cent. In our reference result based on the linear regression, Eddington bias was accounted for by modelling the distribution of WL masses. The distribution of selected mass is quite symmetric. Assuming a log-normal distribution for the mass distribution, the Eddington bias turns out to be negligible when comparing mean values too \citep{se+et17_comalit_V}.

Our result is consistent with previous estimates based on WL comparisons. \citet{lin+al14} found a large bias of $b_\text{SZ}=-0.30\pm0.06$ in the WtG sample \citep{wtg_III_14} . \citet{planck_2015_XXIV} measured $b_\text{SZ}=-0.32\pm0.07$ for the WtG sample, $b_\text{SZ}=-0.22\pm0.09$ for the CCCP \citep{hoe+al15} sample and $b_\text{SZ}\sim 1$ from CMB lensing. The mean bias with respect to the LoCuSS sample is $b_\text{SZ}=-0.05\pm0.04$ \citep{smi+al16}. 

The bias measurements reported in Table~\ref{tab_bias_MSZ} for samples other than PSZ2LenS are taken from \citet{se+et17_comalit_V}, which homogenized the estimates by adopting the same methodology we adopted here. Due to the different methods, the listed values can differ from the values quoted in the original analyses.

\section{Systematics}
\label{sec_systematics}

Weak lensing measurements of masses are very challenging. In fact, masses reported by distinct groups may differ by $\sim$ 20--50 per cent \citep{wtg_III_14,ume+al14,se+et15_comalit_I}. Sources of systematics and residual statistical uncertainties may hinge on calibration errors, the fitting procedure, the selection of background galaxies and their photometric redshift measurements.

The presence of systematics may be tested by comparing results obtained with different methodologies and under different assumptions. Our results are consistent over a variegated sets of circumstances, see Tables~\ref{tab_selections} and \ref{tab_systematics}. Systematic errors on the amplitude of the lensing signal $\Delta \Sigma_+$ are approximated as mass uncertainties through $M \sim \Delta \Sigma_+^{3/2}$, see App.~\ref{sec_lens_weig}.

\subsection{Background selection}

\begin{table*}
\caption{WL analyses exploiting different methods for the selection of background galaxies. Column~1: cluster PSZ2 index. Column 2: cluster redshift. Column~3: raw number density of background lensing sources per square arc minute, including objects with measured shape. Sources were selected with either the colour cuts or the photometric redshift methods. Column~4: effective source redshift. Column~5: WL signal-to-noise ratio. Column~6: $M_{200}$ in units of $10^{14}~M_\odot/h$. Quoted values are the bi-weight estimators of the posterior probability distributions. Columns~7, 8, 9, 10: same as columns 3, 4, 5, and 6 but for sources selected with the colour-colour cuts only. Columns~11, 12, 13, 14: same as columns 3, 4, 5, and 6 but for sources selected with the photometric redshifts only.
}
\label{tab_selections}
\centering
\begin{tabular}[c]{r l r r r r@{$\,\pm\,$}l  r r r r@{$\,\pm\,$}l  r r r r@{$\,\pm\,$}l }
\hline
	\noalign{\smallskip}  
\multicolumn{2}{c}{} &  \multicolumn{5}{c}{$gri$ OR $z_\text{phot}$}  &  \multicolumn{5}{c}{$gri$} &  \multicolumn{5}{c}{$z_\text{phot}$}\\
Index &  $z$    & $n_\text{g}$ &  $z_\text{back}$ & SNR &  \multicolumn{2}{c}{$M_{200}$}  & $n_\text{g}$ &  $z_\text{back}$ & SNR &  \multicolumn{2}{c}{$M_{200}$} & $n_\text{g}$ &  $z_\text{back}$ & SNR &  \multicolumn{2}{c}{$M_{200}$} \\
	 	\hline
	\noalign{\smallskip}      
21   	&	0.078	&	1.61  	&	0.71	&	2.7 	&	7.0 	&	3.3 	&	1.02  	&	0.91	&	3.4 	&	14.1	&	6.7 	&	0.83  	&	0.61	&	1.0 	&	3.3 	&	2.5 	\\
38   	&	0.044	&	3.02  	&	0.75	&	-0.2	&	0.8 	&	0.6 	&	1.99  	&	0.93	&	0.5 	&	1.2 	&	1.1 	&	1.60  	&	0.65	&	-0.7	&	1.2 	&	0.8 	\\
43   	&	0.034	&	3.18  	&	0.73	&	2.3 	&	2.9 	&	1.5 	&	2.02  	&	0.92	&	2.8 	&	4.8 	&	1.9 	&	1.73  	&	0.63	&	1.3 	&	2.3 	&	1.7 	\\
212  	&	0.327	&	3.71  	&	0.88	&	5.7 	&	16.5	&	5.5 	&	3.08  	&	0.91	&	5.4 	&	16.5	&	7.0 	&	1.49  	&	0.79	&	4.5 	&	16.9	&	8.2 	\\
215  	&	0.151	&	1.15  	&	0.69	&	-0.3	&	1.2 	&	1.1 	&	0.66  	&	0.87	&	0.1 	&	2.6 	&	2.3 	&	0.71  	&	0.62	&	-0.2	&	1.4 	&	1.4 	\\
216  	&	0.234	&	1.40  	&	0.82	&	0.1 	&	1.7 	&	1.7 	&	1.02  	&	0.92	&	-0.6	&	2.9 	&	2.1 	&	0.63  	&	0.72	&	0.9 	&	3.3 	&	4.1 	\\
243  	&	0.400	&	1.59  	&	0.91	&	0.1 	&	1.2 	&	1.2 	&	1.38  	&	0.92	&	0.0 	&	1.9 	&	1.6 	&	0.46  	&	0.84	&	0.8 	&	2.1 	&	2.3 	\\
251  	&	0.222	&	2.54  	&	0.85	&	1.9 	&	5.7 	&	2.0 	&	1.96  	&	0.93	&	1.5 	&	5.5 	&	2.3 	&	1.01  	&	0.73	&	1.1 	&	5.4 	&	3.0 	\\
268  	&	0.082	&	2.99  	&	0.67	&	1.8 	&	2.4 	&	1.6 	&	1.70  	&	0.90	&	0.9 	&	2.0 	&	1.5 	&	1.75  	&	0.59	&	1.6 	&	2.5 	&	2.1 	\\
271  	&	0.336	&	2.26  	&	0.89	&	1.7 	&	3.4 	&	2.2 	&	1.80  	&	0.93	&	1.2 	&	2.5 	&	2.1 	&	0.86  	&	0.81	&	1.8 	&	9.3 	&	7.4 	\\
329  	&	0.251	&	1.56  	&	0.86	&	3.0 	&	12.1	&	7.3 	&	0.79  	&	1.24	&	1.0 	&	3.2 	&	5.2 	&	0.84  	&	0.74	&	3.0 	&	15.9	&	8.8 	\\
360  	&	0.279	&	2.04  	&	0.80	&	1.2 	&	4.3 	&	2.9 	&	1.52  	&	0.86	&	0.7 	&	5.0 	&	3.7 	&	0.95  	&	0.74	&	1.0 	&	5.8 	&	4.3 	\\
370  	&	0.140	&	1.45  	&	0.69	&	1.4 	&	2.8 	&	2.8 	&	0.86  	&	0.86	&	2.5 	&	17.6	&	12.4	&	0.87  	&	0.63	&	1.0 	&	3.4 	&	3.4 	\\
391  	&	0.277	&	4.21  	&	0.87	&	5.3 	&	15.0	&	5.5 	&	3.37  	&	0.93	&	5.1 	&	16.0	&	4.5 	&	1.70  	&	0.78	&	2.9 	&	10.6	&	6.6 	\\
446  	&	0.140	&	8.53  	&	0.77	&	4.5 	&	5.5 	&	1.8 	&	6.12  	&	0.91	&	3.6 	&	5.0 	&	2.1 	&	5.44  	&	0.70	&	3.5 	&	5.7 	&	2.0 	\\
464  	&	0.141	&	9.55  	&	0.75	&	5.5 	&	7.1 	&	1.9 	&	6.51  	&	0.91	&	4.5 	&	7.6 	&	2.1 	&	6.38  	&	0.69	&	4.9 	&	6.9 	&	2.2 	\\
473  	&	0.106	&	11.24 	&	0.76	&	2.3 	&	1.2 	&	0.7 	&	8.11  	&	0.90	&	1.8 	&	1.1 	&	0.7 	&	7.42  	&	0.70	&	2.5 	&	1.7 	&	0.9 	\\
478  	&	0.630	&	7.43  	&	0.97	&	2.9 	&	6.7 	&	3.6 	&	7.00  	&	0.96	&	3.1 	&	7.6 	&	3.9 	&	1.65  	&	1.03	&	2.6 	&	15.1	&	10.2	\\
547  	&	0.081	&	2.38  	&	0.69	&	2.9 	&	8.7 	&	4.9 	&	0.96  	&	1.03	&	1.1 	&	4.7 	&	4.4 	&	1.60  	&	0.62	&	2.8 	&	10.1	&	5.6 	\\
554  	&	0.384	&	2.01  	&	0.91	&	2.4 	&	6.7 	&	5.6 	&	1.18  	&	1.03	&	1.1 	&	5.5 	&	3.7 	&	1.04  	&	0.83	&	2.0 	&	8.3 	&	7.1 	\\
586  	&	0.545	&	1.09  	&	1.16	&	0.5 	&	4.5 	&	3.4 	&	0.80  	&	1.25	&	1.0 	&	6.9 	&	5.3 	&	0.34  	&	1.01	&	-0.9	&	4.5 	&	4.4 	\\
618  	&	0.045	&	1.87  	&	0.68	&	4.2 	&	12.1	&	6.5 	&	0.84  	&	0.98	&	1.2 	&	2.8 	&	3.5 	&	1.23  	&	0.61	&	4.1 	&	13.7	&	7.9 	\\
721  	&	0.528	&	1.51  	&	0.88	&	0.9 	&	4.1 	&	4.8 	&	1.36  	&	0.88	&	0.5 	&	4.1 	&	4.8 	&	0.31  	&	0.91	&	0.2 	&	7.9 	&	10.4	\\
724  	&	0.135	&	3.66  	&	0.75	&	6.5 	&	14.6	&	5.2 	&	2.40  	&	0.90	&	5.9 	&	26.2	&	8.0 	&	1.89  	&	0.65	&	4.6 	&	12.9	&	4.9 	\\
729  	&	0.199	&	2.14  	&	0.79	&	1.4 	&	1.5 	&	2.0 	&	1.46  	&	0.90	&	0.9 	&	1.3 	&	1.7 	&	1.03  	&	0.70	&	1.4 	&	2.9 	&	4.3 	\\
735  	&	0.470	&	2.45  	&	0.92	&	1.0 	&	2.5 	&	2.2 	&	2.23  	&	0.92	&	1.4 	&	3.0 	&	2.6 	&	0.59  	&	0.89	&	0.4 	&	5.1 	&	4.8 	\\
804  	&	0.140	&	14.01 	&	0.76	&	2.6 	&	1.7 	&	0.8 	&	9.31  	&	0.95	&	2.2 	&	2.2 	&	1.0 	&	12.81 	&	0.76	&	2.3 	&	1.6 	&	0.8 	\\
822  	&	0.185	&	10.80 	&	0.79	&	3.1 	&	3.4 	&	1.3 	&	7.04  	&	0.99	&	1.9 	&	2.2 	&	1.1 	&	9.96  	&	0.79	&	3.5 	&	4.1 	&	1.5 	\\
902  	&	0.350	&	2.35  	&	0.90	&	1.3 	&	6.7 	&	3.2 	&	1.95  	&	0.91	&	1.1 	&	6.3 	&	3.2 	&	0.78  	&	0.85	&	2.4 	&	14.8	&	7.7 	\\
955  	&	0.400	&	1.30  	&	1.04	&	-0.1	&	3.4 	&	2.8 	&	0.92  	&	1.27	&	0.1 	&	3.6 	&	3.1 	&	0.43  	&	0.83	&	-0.1	&	5.9 	&	5.5 	\\
956  	&	0.188	&	0.90  	&	0.69	&	1.7 	&	7.9 	&	6.1 	&	0.25  	&	1.19	&	1.3 	&	15.6	&	6.5 	&	0.68  	&	0.64	&	1.3 	&	7.9 	&	8.1 	\\
961  	&	0.600	&	0.31  	&	1.16	&	0.3 	&	2.6 	&	3.2 	&	0.21  	&	1.23	&	0.2 	&	1.1 	&	1.1 	&	0.11  	&	1.04	&	0.4 	&	41.2	&	27.1	\\
1046 	&	0.294	&	5.13  	&	0.86	&	2.2 	&	5.9 	&	2.0 	&	4.32  	&	0.91	&	2.6 	&	8.0 	&	2.8 	&	2.92  	&	0.78	&	1.6 	&	5.2 	&	2.1 	\\
1057 	&	0.192	&	4.71  	&	0.78	&	5.1 	&	6.9 	&	2.3 	&	3.22  	&	0.90	&	4.2 	&	6.9 	&	2.4 	&	3.01  	&	0.71	&	4.1 	&	6.0 	&	3.0 	\\
1212 	&	0.154	&	1.20  	&	0.72	&	0.9 	&	3.6 	&	3.1 	&	0.43  	&	1.30	&	1.8 	&	16.1	&	13.4	&	0.82  	&	0.64	&	0.1 	&	2.5 	&	2.2 	\\
Stack	&	0.203	&	127.29	&	0.79	&	14.3	&	4.63	&	0.47	&	89.81	&	0.93	&	12.0	&	4.78	&	0.56	&	75.86	&	0.71	&	11.6	&	4.59	&	0.57	\\
\hline
	\end{tabular}
\end{table*}

\begin{figure}
\resizebox{\hsize}{!}{\includegraphics{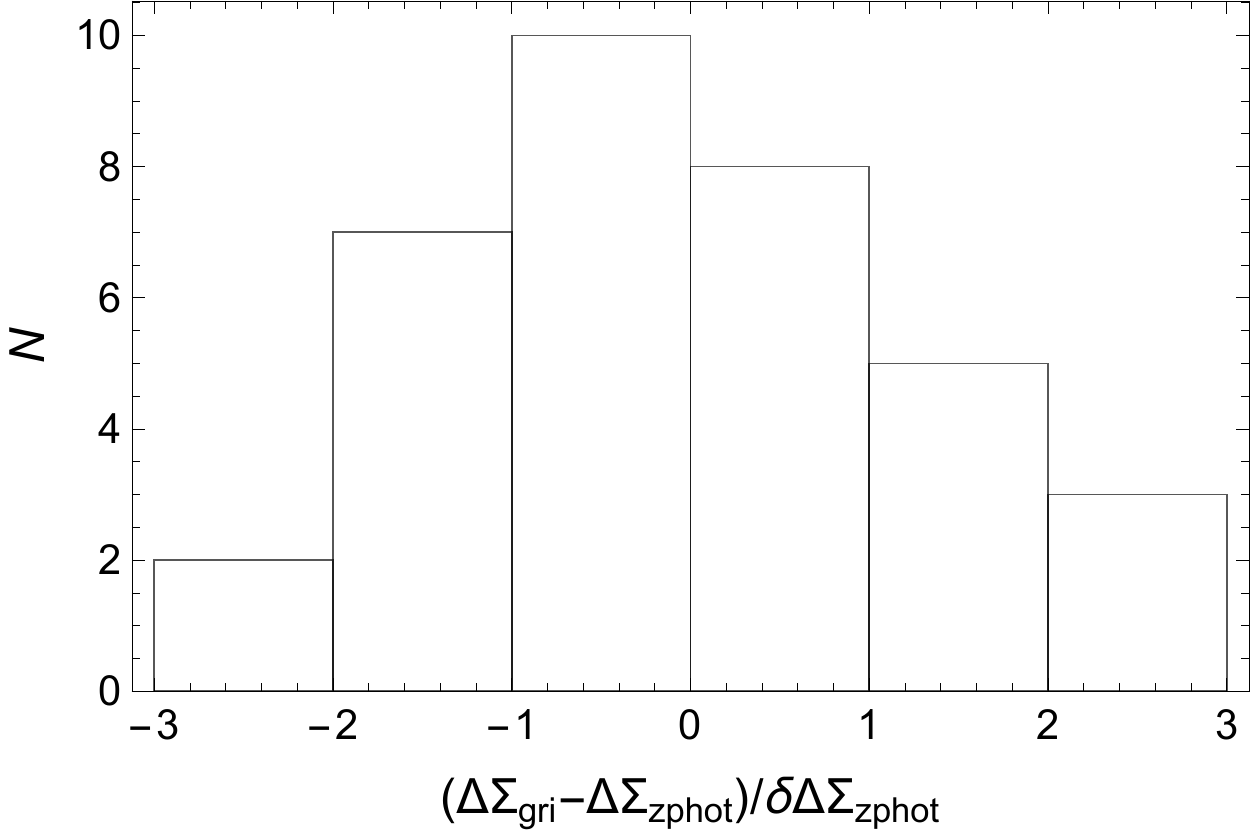}} 
\caption{Distribution of the differences between the differential surface density as measured with a background source population of galaxies selected with the photometric redshifts, $\Delta \Sigma_\text{zphot}$, or with the $g-r-i$ colour-colour method, $\Delta \Sigma_\text{gri}$. The difference is in units of the statistical uncertainties on $\Delta \Sigma_\text{zphot}$. We plotted the shear signals generated by the PSZ2LenS clusters in the outer annulus, $2.51\le R\le 3.16~\text{Mpc}/h$.}
\label{fig_Delta_DeltaSigma_histo}
\end{figure}

The purity of the selected background galaxies is crucial to a proper WL analysis.  Cluster members or foreground galaxies not properly identified can dilute the lensing signal. Contamination by foreground galaxies is most severe in the inner regions. We tried to overcome this by considering conservative selection criteria based on either photometric redshifts or colour-colour cuts. Our selection criteria suffer by a nominal $\ga 1$ per cent contamination. The price for a conservative selection procedure is the low number of retained background galaxies.

\subsubsection{Consistency}

We checked for consistency by redoing the analysis and considering the selection procedures separately, see Table~\ref{tab_selections}. The two selection criteria, i.e. either cuts in $z_\text{phot}$ or in $g-r-i$ colours, are complementary. On average, only $15$ per cent of the total number of retained galaxies is selected by both methods. The percentage is slightly higher ($\sim 18$ per cent) for low redshift clusters ($z_\text{lens}<0.2$). 

The colour-colour cuts are very effective in selecting background galaxies at $z\ga 0.7$ whereas the $z_\text{phot}$ method can also sample lower redshifts sources. As a consequence, the effective source redshift of the galaxies selected by the $g-r-i$ cuts is larger. On one side, these galaxies have a large lensing depth due to the geometrical distance factor. On the other side, the $z_\text{phot}$ method selects nearer and brighter galaxies, whose shape is better determined and which have a larger shear weight. 

The comparison of the estimated differential surface density as obtained with the two different selection methods is showed in Fig.~\ref{fig_Delta_DeltaSigma_histo}. Measurements are very well consistent within the statistical uncertainties. 

We also re-estimated the masses adopting the reference fitting scheme for each selection method. Results are listed in Table~\ref{tab_selections}. For clusters with high SNR, the agreement between the masses estimates is excellent. For lower SNR lenses, the statistical agreement is still good but the uncertainties affecting the mass estimates are very large and the comparison is not so significant. 

The complementarity and the consistent results justify the combined use of the two selection methods.

\subsubsection{Cluster member dilution}

\begin{figure}
\resizebox{\hsize}{!}{\includegraphics{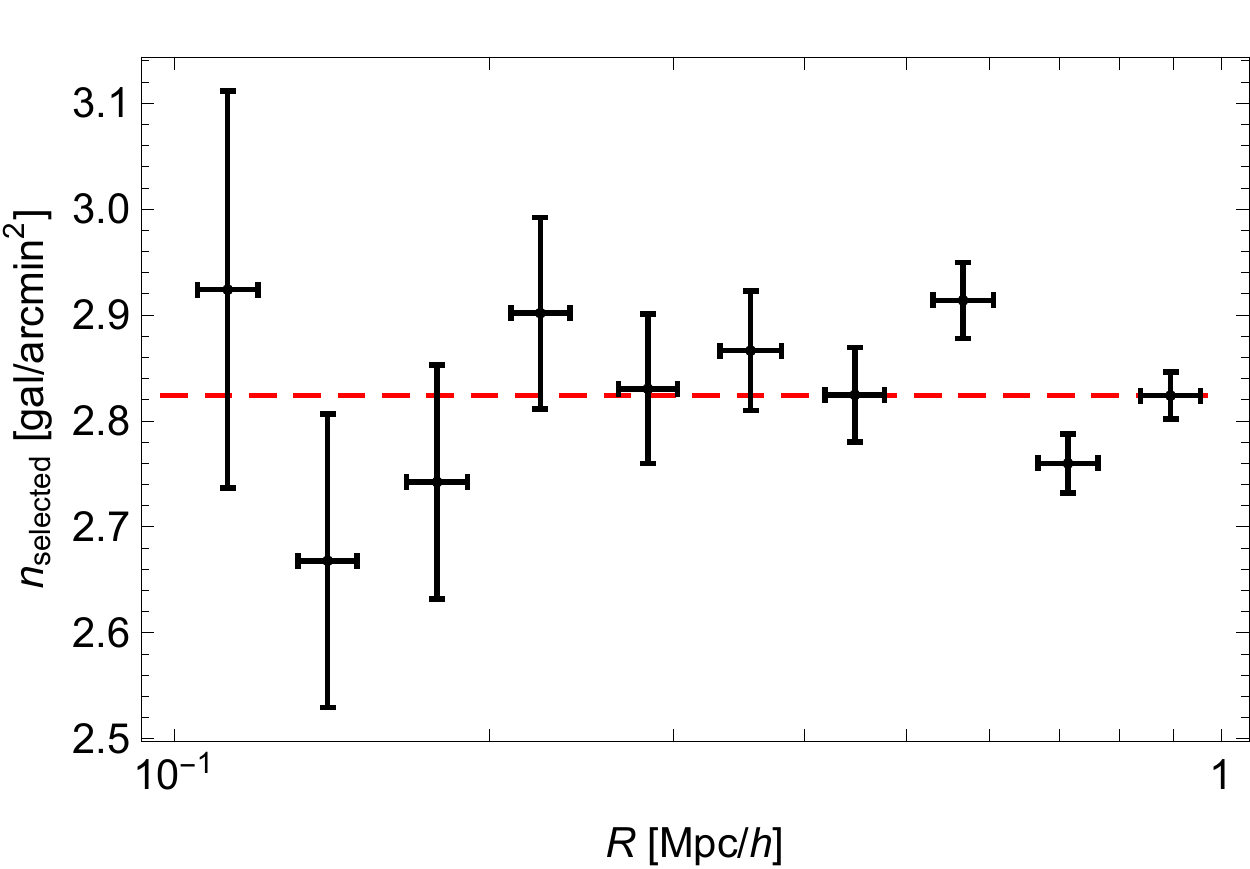}} 
\caption{Mean number density profile of the selected background sources as a function of the radial distance from the cluster centre.}
\label{fig_dilution}
\end{figure}

Cluster members can dilute the lensing signal mostly in central regions. Thanks to our conservative background selection, this effect is not significant in our analysis and we preferred not to introduce corrective boosting factors. We checked the dilution effects in two ways. Firstly, the radial distribution of the number density profile is constant to a good degree, with no bump in the inner regions, see Fig.~\ref{fig_dilution}.

Secondly, the mass measurement do not change significantly if we excise a larger inner region, see columns~2 and 4 of Table~\ref{tab_systematics}. In particular, if we set $R_\text{min}=0.5~\text{Mpc}/h$, the estimated mass $M_{200}$ of the stacked profile changes by $\sim 6$ per cent, well below the uncertainty of $\sim 10$ per cent. The variation is due to the lower statistical power of the data sets (excluding the inner bins, the SNR is 12.1) and the lower capability of breaking the mass-concentration degeneracy rather than being significant of a systematic uncertainty.

\subsubsection{Foreground contamination}

In Sec.~\ref{sec_back_sel}, we showed that the contamination affecting the sample of selected background galaxies is contained to the $\la 2$ per cent level. Since in absence of intrinsic alignments foreground galaxies do not contribute a net shear signal, the contamination depletes the shear signal by the same amount, which causes an under-estimation of the mass by $\la 3$ per cent.

\subsection{Priors}

\begin{table*}
\caption{Masses determined assuming different halo modellings, priors or radial ranges. The setting is specified in the first five rows before the line break, where we list the density profile of the main halo (either NFW or BMO in row 1), the priors for mass (row 2), concentration (row 3) and halo bias (row 4), and the radial range (row 5). The symbols ${\cal U}$, $log{\cal U}$, $log{\cal N}$ and $\delta$ denote the uniform prior in linear space, the uniform prior in log-intervals, the lognormal distribution, and the Dirac delta, respectively. Mass priors are renormalized between $0.05$ and $100\times 10^{14}M_\odot/h$, concentration priors between $c_{200}=1$ and $20$. For the halo bias, the function $b_\text{h}[\nu(M_{200},z)]$ follows \citet{tin+al10}. For the reference case (column~2), we also report the best-fitting value in round brackets. Cluster PSZ2 indexes are listed in Column~1. Masses are in units of $10^{14}~M_\odot/h$, lengths in units of $\text{Mpc}/h$. Bi-weight estimators of central location and scale of the posterior distributions are reported.}
\label{tab_systematics}
\centering
\begin{tabular}[c]{l r@{(}c@{)}@{$\,\pm\,$}l    r@{$\,\pm\,$}l    r@{$\,\pm\,$}l  r@{$\,\pm\,$}l  r@{$\,\pm\,$}l  r@{$\,\pm\,$}l r@{$\,\pm\,$}l }
\hline
	\noalign{\smallskip}  
1-halo 		&  \multicolumn{3}{c}{BMO}    & \multicolumn{2}{c}{BMO}  & \multicolumn{2}{c}{BMO}   & \multicolumn{2}{c}{BMO} &  \multicolumn{2}{c}{NFW}    & \multicolumn{2}{c}{NFW}    & \multicolumn{2}{c}{NFW}\\
$p_\text{prior}(M_{200})$ 	&  \multicolumn{3}{c}{${\cal U}$}    & \multicolumn{2}{c}{${\cal U}$}    & \multicolumn{2}{c}{${\cal U}$}&  \multicolumn{2}{c}{$log{\cal U}$}    & \multicolumn{2}{c}{${\cal U}$}    & \multicolumn{2}{c}{${\cal U}$}  & \multicolumn{2}{c}{${\cal U}$} \\
$p_\text{prior}(c_{200})$		
& \multicolumn{3}{c}{${\cal U}$} & \multicolumn{2}{c}{$log{\cal N}$} & \multicolumn{2}{c}{${\cal U}$} & \multicolumn{2}{c}{$log{\cal U}$}    
& \multicolumn{2}{c}{${\cal U}$} & \multicolumn{2}{c}{$log{\cal N}$} & \multicolumn{2}{c}{${\cal U}$} \\
$p_\text{prior}(b_\text{h})$	
& \multicolumn{3}{c}{$\delta [b_\text{h} (\nu)]$} & \multicolumn{2}{c}{$\delta [b_\text{h} (\nu)]$}   
& \multicolumn{2}{c}{$\delta [b_\text{h} (\nu)]$}  & \multicolumn{2}{c}{$\delta [b_\text{h} (\nu)]$} 
& \multicolumn{2}{c}{$\delta [0]$} & \multicolumn{2}{c}{$\delta [0]$} & \multicolumn{2}{c}{$\delta [0]$}  \\
$R$-range &  \multicolumn{3}{c}{$[0.1, 3]$}    & \multicolumn{2}{c}{$[0.1, 3]$}    & \multicolumn{2}{c}{$[0.5, 3]$} &  \multicolumn{2}{c}{$[0.1, 3]$}    & \multicolumn{2}{c}{$[0.1, 3]$}   &  \multicolumn{2}{c}{$[0.1, 3]$}  & \multicolumn{2}{c}{$[0.1, 2]$} \\
    	 \hline
	\noalign{\smallskip}      
21   	&	7.0 	&	7.8 	&	3.3 	&	9.1 	&	3.7 	&	9.1 	&	2.7 	&	6.3 	&	3.4 	&	6.7 	&	3.2 	&	8.5 	&	3.5 	&	9.4 	&	6.0 	\\
38   	&	0.8 	&	0.5 	&	0.6 	&	1.1 	&	0.9 	&	1.3 	&	0.9 	&	0.4 	&	0.4 	&	0.8 	&	0.6 	&	0.9 	&	0.7 	&	1.2 	&	0.9 	\\
43   	&	2.9 	&	2.8 	&	1.5 	&	3.7 	&	1.7 	&	2.8 	&	1.8 	&	2.5 	&	1.5 	&	2.8 	&	1.5 	&	3.7 	&	1.9 	&	2.5 	&	1.3 	\\
212  	&	16.5	&	16.3	&	5.5 	&	14.4	&	4.4 	&	12.8	&	5.1 	&	15.1	&	5.0 	&	16.3	&	5.7 	&	16.7	&	5.7 	&	12.2	&	4.6 	\\
215  	&	1.2 	&	0.2 	&	1.1 	&	1.6 	&	1.7 	&	2.6 	&	2.0 	&	0.3 	&	0.4 	&	1.5 	&	1.6 	&	1.7 	&	2.1 	&	1.4 	&	1.6 	\\
216  	&	1.7 	&	0.6 	&	1.7 	&	2.5 	&	2.4 	&	3.6 	&	3.2 	&	0.6 	&	0.9 	&	1.7 	&	1.6 	&	2.2 	&	2.0 	&	1.7 	&	1.7 	\\
243  	&	1.2 	&	0.1 	&	1.2 	&	1.7 	&	1.8 	&	1.4 	&	1.4 	&	0.3 	&	0.3 	&	1.2 	&	1.2 	&	1.5 	&	1.6 	&	0.9 	&	0.9 	\\
251  	&	5.7 	&	4.8 	&	2.0 	&	6.4 	&	2.4 	&	5.1 	&	2.5 	&	5.6 	&	2.1 	&	5.6 	&	1.8 	&	7.1 	&	2.8 	&	6.2 	&	2.2 	\\
268  	&	2.4 	&	2.9 	&	1.6 	&	3.5 	&	2.1 	&	4.1 	&	2.0 	&	1.7 	&	1.6 	&	2.2 	&	1.4 	&	2.9 	&	1.8 	&	2.1 	&	1.5 	\\
271  	&	3.4 	&	2.1 	&	2.2 	&	5.3 	&	3.0 	&	4.1 	&	2.6 	&	2.0 	&	1.8 	&	4.2 	&	3.1 	&	5.0 	&	2.7 	&	2.7 	&	1.6 	\\
329  	&	12.1	&	15.4	&	7.3 	&	15.7	&	7.5 	&	11.3	&	4.1 	&	8.8 	&	6.1 	&	9.6 	&	5.1 	&	12.2	&	5.8 	&	9.9 	&	5.2 	\\
360  	&	4.3 	&	4.5 	&	2.9 	&	5.6 	&	3.4 	&	4.9 	&	3.0 	&	2.5 	&	2.5 	&	4.6 	&	3.0 	&	5.6 	&	3.4 	&	5.8 	&	4.4 	\\
370  	&	2.8 	&	4.1 	&	2.8 	&	4.3 	&	3.3 	&	3.5 	&	2.7 	&	1.1 	&	1.6 	&	2.7 	&	2.4 	&	5.8 	&	4.8 	&	2.6 	&	2.7 	\\
391  	&	15.0	&	17.8	&	5.5 	&	14.8	&	5.5 	&	15.1	&	5.5 	&	14.3	&	4.8 	&	13.0	&	4.9 	&	13.9	&	4.9 	&	9.1 	&	4.3 	\\
446  	&	5.5 	&	5.6 	&	1.8 	&	5.6 	&	1.7 	&	5.7 	&	1.5 	&	5.5 	&	2.0 	&	5.2 	&	1.8 	&	5.0 	&	1.7 	&	5.1 	&	2.2 	\\
464  	&	7.1 	&	7.2 	&	1.9 	&	7.1 	&	1.7 	&	6.3 	&	1.4 	&	6.9 	&	1.8 	&	6.5 	&	1.7 	&	6.5 	&	1.6 	&	8.5 	&	2.7 	\\
473  	&	1.2 	&	1.5 	&	0.7 	&	1.5 	&	0.8 	&	1.5 	&	0.7 	&	1.0 	&	0.7 	&	1.1 	&	0.6 	&	1.4 	&	0.7 	&	1.0 	&	0.6 	\\
478  	&	6.7 	&	9.0 	&	3.6 	&	8.6 	&	3.5 	&	7.4 	&	2.9 	&	5.9 	&	3.8 	&	7.2 	&	3.7 	&	7.8 	&	3.5 	&	8.3 	&	7.2 	\\
547  	&	8.7 	&	9.3 	&	4.9 	&	8.2 	&	4.8 	&	8.1 	&	3.6 	&	6.4 	&	4.6 	&	7.2 	&	4.8 	&	8.1 	&	4.5 	&	4.8 	&	4.8 	\\
554  	&	6.7 	&	13.2	&	5.6 	&	7.7 	&	4.5 	&	9.8 	&	6.0 	&	4.2 	&	4.8 	&	7.0 	&	4.9 	&	9.7 	&	6.3 	&	4.1 	&	3.4 	\\
586  	&	4.5 	&	2.5 	&	3.4 	&	6.8 	&	4.8 	&	5.7 	&	4.9 	&	2.2 	&	2.8 	&	4.0 	&	3.6 	&	6.4 	&	3.6 	&	5.6 	&	5.0 	\\
618  	&	12.1	&	15.7	&	6.5 	&	10.3	&	5.0 	&	7.5 	&	4.7 	&	10.3	&	6.2 	&	9.9 	&	6.4 	&	10.1	&	5.3 	&	3.1 	&	2.5 	\\
721  	&	4.1 	&	3.7 	&	4.8 	&	6.1 	&	5.3 	&	7.1 	&	6.0 	&	0.5 	&	0.7 	&	7.0 	&	7.3 	&	6.3 	&	5.3 	&	2.6 	&	3.0 	\\
724  	&	14.6	&	14.4	&	5.2 	&	15.7	&	4.8 	&	15.5	&	5.2 	&	12.7	&	3.9 	&	14.3	&	4.6 	&	15.4	&	4.3 	&	8.7 	&	3.9 	\\
729  	&	1.5 	&	2.5 	&	2.0 	&	2.1 	&	2.1 	&	3.1 	&	2.8 	&	0.3 	&	0.4 	&	1.9 	&	2.5 	&	2.2 	&	2.1 	&	0.8 	&	0.9 	\\
735  	&	2.5 	&	2.8 	&	2.2 	&	3.3 	&	3.1 	&	4.3 	&	3.0 	&	0.7 	&	1.1 	&	2.4 	&	2.2 	&	3.0 	&	2.1 	&	2.3 	&	2.1 	\\
804  	&	1.7 	&	2.0 	&	0.8 	&	2.0 	&	0.9 	&	1.8 	&	0.8 	&	1.5 	&	0.9 	&	1.6 	&	0.7 	&	1.8 	&	0.8 	&	2.0 	&	1.0 	\\
822  	&	3.4 	&	3.7 	&	1.3 	&	3.9 	&	1.5 	&	3.6 	&	1.2 	&	3.2 	&	1.3 	&	3.1 	&	1.2 	&	3.5 	&	1.2 	&	3.6 	&	1.6 	\\
902  	&	6.7 	&	6.6 	&	3.2 	&	9.1 	&	4.4 	&	6.0 	&	3.4 	&	6.0 	&	2.6 	&	6.2 	&	2.9 	&	8.4 	&	3.2 	&	8.0 	&	4.6 	\\
955  	&	3.4 	&	1.8 	&	2.8 	&	5.4 	&	5.1 	&	4.2 	&	3.8 	&	0.5 	&	0.9 	&	3.3 	&	2.6 	&	5.4 	&	5.2 	&	4.0 	&	3.0 	\\
956  	&	7.9 	&	7.6 	&	6.1 	&	9.9 	&	6.4 	&	8.2 	&	6.0 	&	3.0 	&	2.9 	&	6.4 	&	5.4 	&	9.5 	&	7.1 	&	6.2 	&	4.1 	\\
961  	&	2.6 	&	0.1 	&	3.2 	&	6.9 	&	6.8 	&	16.5	&	17.1	&	0.3 	&	0.5 	&	2.9 	&	4.2 	&	4.5 	&	4.2 	&	1.5 	&	1.7 	\\
1046 	&	5.9 	&	6.3 	&	2.0 	&	7.4 	&	2.3 	&	6.3 	&	2.0 	&	5.7 	&	2.0 	&	5.8 	&	2.0 	&	6.8 	&	2.2 	&	11.8	&	4.7 	\\
1057 	&	6.9 	&	7.0 	&	2.3 	&	7.7 	&	2.5 	&	5.4 	&	1.5 	&	6.6 	&	2.2 	&	6.7 	&	2.3 	&	7.0 	&	2.0 	&	6.1 	&	2.2 	\\
1212 	&	3.6 	&	3.8 	&	3.1 	&	5.6 	&	4.8 	&	5.8 	&	3.2 	&	1.4 	&	1.9 	&	3.4 	&	2.6 	&	4.1 	&	3.0 	&	4.0 	&	3.7 	\\
Stack	&	4.63	&	4.65	&	0.47	&	4.65	&	0.45	&	4.34	&	0.47	&	4.64	&	0.46	&	4.35	&	0.43	&	4.33	&	0.42	&	4.43	&	0.56	\\
	\hline
	\end{tabular}
\end{table*}

The effect of priors on mass and concentration is usually negligible but it can play a role when the signal-to-noise ratio of the observations is low. Regression results obtained under different assumptions are summarized in Table~\ref{tab_systematics}. 

Differences among prior schemes are smaller than statistical uncertainties. The only scheme which gives systematically lower masses, mostly at the low mass tail, is that exploiting priors which are uniform in logarithmic units, see column~5 of Table~\ref{tab_systematics}. For low SNR systems, these priors can bias the results towards lower values. This has to be counterbalanced by a careful choice of the lower mass limit , which can make the prior informative again. For this reason, we preferred uniform priors in linear space.

We verified that a more informative prior on the concentration inspired by numerical simulations, see columns~3 and 7 of Table~\ref{tab_systematics}, significantly improves neither the accuracy nor the precision, compare with columns~2 and 6 of Table~\ref{tab_systematics}, which makes the less informative priors preferable.

\subsection{Mass estimator}

A careful choice of the estimator is crucial. The choice has to be tuned to the quality of the data \citep{bee+al90}. In particular, the maximum likelihood method can be less stable in low SNR systems, see Appendix~\ref{sec_robu_esti}. At the low mass end, the best-fitting value can underestimate the mass with respect to the bi-weight estimator, see column~2 of of Table~\ref{tab_systematics}. However, differences are smaller than the statistical uncertainties. For larger mass clusters, differences are negligible.

\subsection{Halo modelling}

Recent $N$-body simulations have showed that the traditional NFW functional form may fail to describe the structural properties of cosmic objects at the percent level required by precision cosmology \citep{du+ma14,kly+al16,men+al14}. 

The Einasto radial profiles can provide a more accurate description of the main halo. \citet{ser+al16_einasto} computed the systematic errors expected for weak lensing analyses of clusters of galaxies if one wrongly models the lens density profile. At the typical mass of the PSZ2LenS clusters, $M_{200}\sim 4.6\times10^{14}M_\odot/h$, the systematic error is below the per cent level whereas the viral masses and concentrations of the most massive halos at $M_{200}\sim 10^{15}M_\odot/h$ can be over- and under-estimated by $\sim5$ per cent, respectively.

The inclusion of the inner regions, $R_\text{min}=0.1~\text{Mpc}/h$ in column~2 of Table~\ref{tab_systematics}, does not significantly improve the statistical accuracy of the results with respect to fitting procedure neglecting them, $R_\text{min}=0.5~\text{Mpc}/h$ in column~4, but it can make the results more accurate thanks to a much better determination of the concentration and the breaking of the related degeneracy

The proper modelling of the outer parts of the shear profile can be crucial in high SNR systems. For analyses that include the outer regions, i.e., $R\ga2~\text{Mpc}/h$, the effect of correlated matter may be significant and the use of the NFW profile can be worrisome  \citep{og+ha11}. The truncation of the profile can remove the unphysical divergence of the total mass of the NFW halo and partially removes systematic errors. However, only accounting for the 2-halo term can accurately describe the transition between the cluster and the correlated matter which occurs beyond the virial radius in the transition region from the infalling to the collapsed material \citep{di+kr14}.

Thanks to our treatment of the 2-halo term, we could fit the shear profile up to large radii, $R=3.16~\text{Mpc}/h$. Even though differences are smaller than statistical uncertainties, some features emerge. The inclusion of the outer regions improves both accuracy, i.e. the size of the systematic error, and precision, i.e. the size of the statistical uncertainty.

If we do not truncate the main halo and we do not consider the 2-halo term, fitting up to large radii can underestimate the halo concentration and bias the mass high \citep{og+ha11}. We found that the NFW fitting out to large radii, see column~6 of of Table~\ref{tab_systematics}, can overestimate masses with respect to the more complete modelling based on the truncated BMO density profile plus the 2-halo term, see column~2 of of Table~\ref{tab_systematics}. By proper modelling the outer regions, we correct a potential systematic error of $\sim 6$ per cent.

Inclusion of outer regions can significantly improve the precision too. As can be seen from the comparison of the case in column~6 of Table~\ref{tab_systematics} where $R_\text{min}=3.16~\text{Mpc}/h$ with the case $R_\text{min}=2.0~\text{Mpc}/h$ in column~8, the statistical uncertainty decreases by $\sim25$ per cent. This feature is crucial in low SNR systems where most of the signal is collected in the outer regions.

In summary, residual systematic bias due to halo modelling is at the per cent level if we properly model the deviations from the NFW profile, mostly at large radii.

\subsection{Centring}

\begin{table}
\caption{Masses and SNR determined assuming different centres, i.e. either the BCG or the SZ centroid. Column~1: cluster PSZ2 index. Columns~2, 3 and 4: displacement between the BCG and the SZ centroid in units of arcminutes, $\text{kpc}/h$, and $r_{200}$, respectively. Columns~5 and 6: mass and SNR assuming that haloes are centred in the BCG. Columns~7 and 8: same as columns~5 and 6 but assuming the SZ centroid as centre. Masses are in units of $10^{14}~M_\odot/h$.}
\label{tab_systematics_centers}
\centering
\resizebox{\hsize}{!} {
\begin{tabular}[c]{r  r r r r@{$\,\pm\,$}l   r  r@{$\,\pm\,$}l  r }
\hline
	\noalign{\smallskip}  
   &  \multicolumn{3}{c}{$\Delta_\text{BCG}$}		 &  \multicolumn{3}{c}{BCG}   &  \multicolumn{3}{c}{SZ}  \\
Index 	& $[\arcmin]$& $[\text{kpc}/h]$	& $[r_{200}]$& \multicolumn{2}{c}{$M_{200}$}    & SNR &  \multicolumn{2}{c}{$M_{200}$}    & SNR\\
    	 \hline
	\noalign{\smallskip}      
21   	&	1.42	&	88 	&	0.06	&	7.0 	&	3.3 	&	2.7 	&	7.3 	&	4.0 	&	2.7 	\\
38   	&	2.20	&	80 	&	0.12	&	0.8 	&	0.6 	&	-0.2	&	0.7 	&	0.6 	&	-0.4	\\
43   	&	0.30	&	9  	&	0.01	&	2.9 	&	1.5 	&	2.3 	&	2.9 	&	1.6 	&	2.2 	\\
212  	&	0.30	&	59 	&	0.03	&	16.5	&	5.5 	&	5.7 	&	15.2	&	4.3 	&	5.7 	\\
215  	&	0.62	&	69 	&	0.09	&	1.2 	&	1.1 	&	-0.3	&	0.8 	&	0.7 	&	-0.2	\\
216  	&	1.02	&	160	&	0.19	&	1.7 	&	1.7 	&	0.1 	&	2.1 	&	1.8 	&	-0.2	\\
243  	&	0.88	&	198	&	0.29	&	1.2 	&	1.2 	&	0.1 	&	1.7 	&	1.6 	&	-0.1	\\
251  	&	1.81	&	272	&	0.22	&	5.7 	&	2.0 	&	1.9 	&	1.6 	&	1.5 	&	0.8 	\\
268  	&	0.78	&	51 	&	0.05	&	2.4 	&	1.6 	&	1.8 	&	2.2 	&	1.4 	&	1.8 	\\
271  	&	1.68	&	339	&	0.34	&	3.4 	&	2.2 	&	1.7 	&	1.4 	&	1.6 	&	1.0 	\\
329  	&	0.80	&	132	&	0.08	&	12.1	&	7.3 	&	3.0 	&	15.1	&	7.4 	&	2.9 	\\
360  	&	1.41	&	250	&	0.23	&	4.3 	&	2.9 	&	1.2 	&	3.0 	&	2.1 	&	1.2 	\\
370  	&	0.75	&	78 	&	0.08	&	2.8 	&	2.8 	&	1.4 	&	3.3 	&	3.2 	&	1.4 	\\
391  	&	0.99	&	175	&	0.10	&	15.0	&	5.5 	&	5.3 	&	13.8	&	4.1 	&	5.0 	\\
446  	&	1.19	&	124	&	0.10	&	5.5 	&	1.8 	&	4.5 	&	5.3 	&	1.8 	&	4.2 	\\
464  	&	2.47	&	257	&	0.19	&	7.1 	&	1.9 	&	5.5 	&	5.8 	&	1.9 	&	4.7 	\\
473  	&	1.01	&	82 	&	0.11	&	1.2 	&	0.7 	&	2.3 	&	1.3 	&	0.7 	&	2.3 	\\
478  	&	0.64	&	183	&	0.16	&	6.7 	&	3.6 	&	2.9 	&	7.9 	&	4.2 	&	2.9 	\\
547  	&	1.43	&	92 	&	0.06	&	8.7 	&	4.9 	&	2.9 	&	10.3	&	4.7 	&	3.3 	\\
554  	&	0.29	&	65 	&	0.05	&	6.7 	&	5.6 	&	2.4 	&	7.8 	&	5.4 	&	2.3 	\\
586  	&	1.17	&	313	&	0.31	&	4.5 	&	3.4 	&	0.5 	&	1.0 	&	1.1 	&	-0.3	\\
618  	&	1.63	&	60 	&	0.04	&	12.1	&	6.5 	&	4.2 	&	15.8	&	7.1 	&	4.3 	\\
721  	&	1.43	&	377	&	0.38	&	4.1 	&	4.8 	&	0.9 	&	5.6 	&	4.5 	&	1.3 	\\
724  	&	0.22	&	22 	&	0.01	&	14.6	&	5.2 	&	6.5 	&	15.6	&	5.6 	&	6.4 	\\
729  	&	0.47	&	65 	&	0.08	&	1.5 	&	2.0 	&	1.4 	&	0.7 	&	0.9 	&	1.4 	\\
735  	&	0.70	&	173	&	0.20	&	2.5 	&	2.2 	&	1.0 	&	3.3 	&	2.5 	&	1.3 	\\
804  	&	1.35	&	140	&	0.16	&	1.7 	&	0.8 	&	2.6 	&	1.4 	&	0.7 	&	2.2 	\\
822  	&	1.41	&	184	&	0.17	&	3.4 	&	1.3 	&	3.1 	&	3.2 	&	1.5 	&	2.9 	\\
902  	&	0.52	&	108	&	0.09	&	6.7 	&	3.2 	&	1.3 	&	6.8 	&	3.2 	&	1.2 	\\
955  	&	0.47	&	105	&	0.11	&	3.4 	&	2.8 	&	-0.1	&	3.7 	&	3.1 	&	-0.3	\\
956  	&	1.11	&	147	&	0.10	&	7.9 	&	6.1 	&	1.7 	&	6.8 	&	4.0 	&	1.8 	\\
961  	&	1.21	&	341	&	0.42	&	2.6 	&	3.2 	&	0.3 	&	6.0 	&	5.6 	&	0.2 	\\
1046 	&	0.37	&	68 	&	0.06	&	5.9 	&	2.0 	&	2.2 	&	6.2 	&	2.2 	&	2.1 	\\
1057 	&	1.63	&	219	&	0.16	&	6.9 	&	2.3 	&	5.1 	&	6.9 	&	3.9 	&	4.3 	\\
1212 	&	0.92	&	104	&	0.09	&	3.6 	&	3.1 	&	0.9 	&	4.2 	&	3.3 	&	1.3 	\\
Stack	&	-   	&	-  	&	-   	&	4.63	&	0.47	&	14.3	&	4.61	&	0.51	&	13.4	\\
	\hline
	\end{tabular}
	}
\end{table}

\begin{figure}
\resizebox{\hsize}{!}{\includegraphics{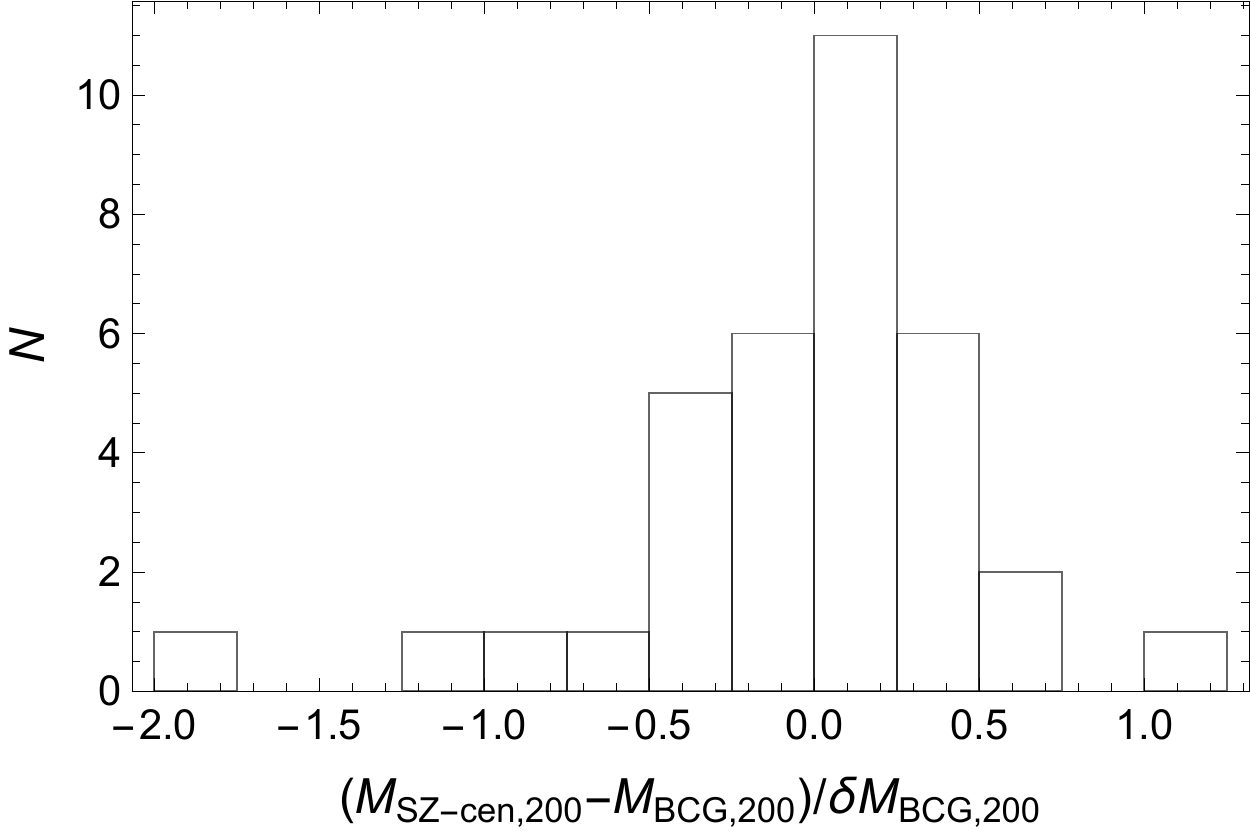}} 
\caption{Distribution of the differences between masses measured around the BCG, $M_{\text{BCG},200}$, or around the SZ centroid, $M_{\text{SZ-cen},200}$. Differences are in units of the statistical uncertainties on $M_{\text{BCG},200}$.}
\label{fig_Delta_M200_centroid_histo}
\end{figure}

Locating the centres of dark matter haloes is critical for the unbiased analysis of mass profiles \citep{geo+al12}. Miscentreing leads to underestimate $\Delta \Sigma_+$ at small scales and to bias low the measurement of the concentration \citep{joh+al07}.

We identified the centre of the cluster as the BCG. Bright galaxies or X-ray emission from hot plasma can be used to trace the halo centre. \citet{geo+al12} investigated the consequences of miscentring on the weak lensing signal from a sample of 129 X-ray-selected galaxy groups in the COSMOS field with redshifts $0 \la z \la 1$ and halo masses in the range $10^{13}$-$10^{14}M_\odot$. By measuring the stacked lensing signal around different candidate centres, they found that massive galaxies near the X-ray centroids trace the centre of mass to $\la75~\text{kpc}$, whereas the X-ray position or the centroids based on the mean position of member galaxies have larger offsets primarily due to the statistical uncertainties in their positions (typically $\sim50$-$150~\text{kpc}$).

In complex clusters, the BCG defining the cluster centre might be misidentified or it might not coincide with the matter centroid, but this second effect is generally small and negligible at the weak lensing scale \citep{geo+al12,zit+al12}. 

Our choice to identify the BCG as the cluster centre and to cut the inner $R<0.1~\text{Mpc}/h$ region makes the effects of miscentring of second order in our analysis. We checked this by re-extracting the shear signal of the clusters around the SZ centroid and by recomputing the masses as described in Section~\ref{sec_infe}. Masses are consistent and differences are well below the statistical uncertainties, see Fig.~\ref{fig_Delta_M200_centroid_histo} and Table~\ref{tab_systematics_centers}. By comparing the stacked profiles, we found that the systematic error in mass due off-centering is negligible ($\sim 0.5$ per cent).

In fact, the typical displacement between the BCG and the SZ centroid is of the order of the arcminute, well below the maximum radius considered in the WL analysis. The displacement is also very small with respect to $r_{200}$. Most of the times, the inner cut encompasses the shift. This makes the estimate of the total SNR within $3~\text{Mpc}/h$ mostly insensitive to the accurate determination of the centre, see Table~\ref{tab_systematics_centers}.

\subsection{Photometric redshift systematics}
\label{sec_syst_phot}

\begin{figure}
\resizebox{\hsize}{!}{\includegraphics{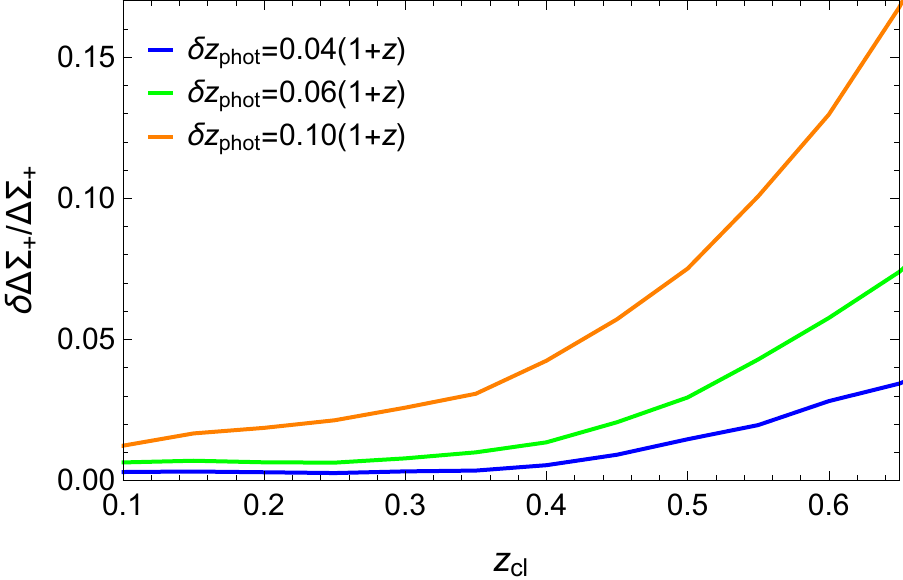}} 
\caption{Systematic relative difference of lensing weighted differential density due to scatter in the photometric redshifts of the background galaxies.}
\label{fig_bias_Sigma_Cr_zphot}
\end{figure}

\begin{figure}
\resizebox{\hsize}{!}{\includegraphics{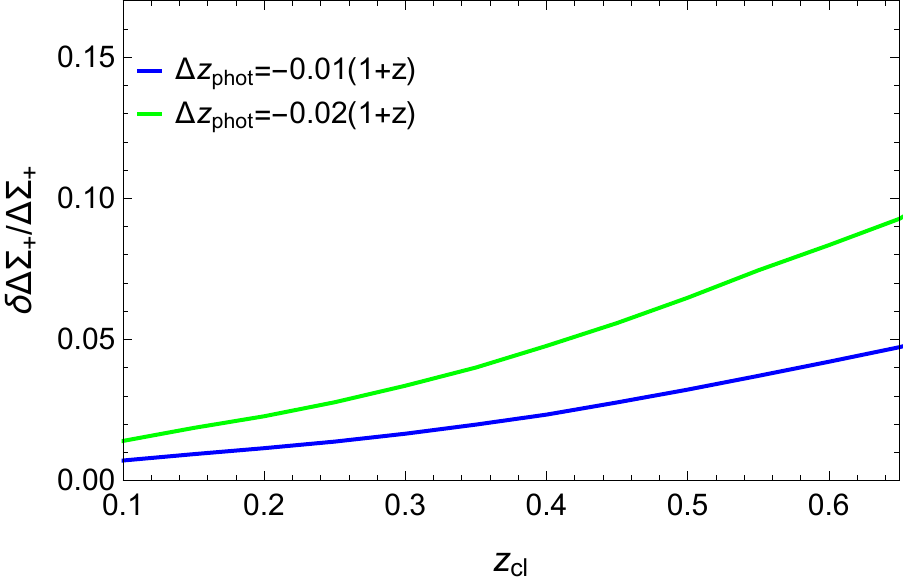}} 
\caption{Systematic relative difference of lensing weighted differential density due to bias in the photometric redshifts of the background galaxies.}
\label{fig_bias_Sigma_Cr_zphot_bias}
\end{figure}

Photometric redshift systematics can impact weak lensing analyses by biasing the estimation of the surface critical density $\Sigma_\text{cr}$. For our estimator of the differential density, we computed the critical density for each source at the peak of the photometric redshift probability density. This is justified since we limited the selection of background galaxies to redshift ranges where the photo-$z$ probability density distribution is mostly well behaved and single peaked, see Fig.~\ref{fig_Delta_ODDS}.

\citet{cou+al15} tested the impact of including high-redshift sources and the reliability of the point estimator for the critical density. They verified that photometric redshifts and shape measurements in CFHTLenS with additional NIR data are robust enough beyond $z_\text{s} > 1.2$. They selected an arbitrary sample of low-redshift lens galaxies with a spectroscopic redshift and they measured the galaxy-galaxy lensing signal using all sources with $0.8 < z_\text{s} < 1.2$ and all sources with $z_\text{s} > 1.2$ and they found no significant difference between the two signals.

To quantitatively estimate the systematic error, we performed a simulation. We approximated the true redshift distribution of the field galaxies as the distribution of the measured photometric redshifts in a CFHTLenS field. The distribution of photometric redshifts was then simulated by considering a Gaussian error $\delta z_\text{phot} \propto 1+z_\text{phot}$. We simulated the lens as a NFW toy model and extracted as background galaxies the sub-sample with $z_\text{phot}>z_\text{lens}+\Delta z_\text{lens}$, with $\Delta z_\text{lens}=0.05$, see Sec.~\ref{sec_back_sel}. We assigned to each source the true shear distortion and its real lensing weight from the shear catalog. We finally computed the $\Delta \Sigma_+$ estimator for the simulated input or the scattered redshifts. Results are summarized in Fig.~\ref{fig_bias_Sigma_Cr_zphot} for different lens redshift and photo-$z$ uncertainties.

For a redshift uncertainty of $\delta z_\text{phot} \la 0.06(1+z_\text{phot})$, as typical of the CHFTLenS survey in the range $0.2 \la z_\text{phot} \la 1.2$ or for the RCSLenS in the range $0.4 \la z_\text{phot} \la 1.2$, the systematic error on the differential density is below the percent level for lenses up to $z_\text{cl} \sim 0.4$. For the highest redshift clusters in our sample  at $z_\text{cl}\ga 0.6$, the uncertainty is $\sim 5$ per cent. 

Together with the scatter, a bias in the estimated $z_\text{phot}$ can affect the mass calibration. The bias, defined as the mean $(z_\text{phot}-z_\text{spes})/ (1+z_\text{spec})$ including the outliers, in RCSLenS for sources with $\texttt{ODDS}>0.8$ (as by our selection) is of order of $\sim 0.01$ for redshifts in the range $0.4\la z_\text{phot}\la 1.0$, and it is stays well below 5 per cent even relaxing the selection criteria \citep{hil+al16}. 

To quantitatively estimate the related systematic error in the mass calibration, we performed a simulation as before but we applied a constant bias rather than scattering the distribution of true redshifts. Results are summarized in Fig.~\ref{fig_bias_Sigma_Cr_zphot_bias}. For a bias of $-0.01$, the systematic error on the shear signal is  $\la 2$ per cent in an ample lens redshift range.

Our treatment did not explicitly consider catastrophic outliers as a secondary population in the source redshift distribution. Outliers are defined as objects with $\Delta z = (z_\text{phot}-z_\text{spes})/ (1+z_\text{spec})$ larger than an arbitrary threshold. In CFHTLenS, less than 4 per cent of estimated redshifts are regarded as outliers \citep[$|\Delta z |>0.15$,][]{hil+al12}. The fraction of outliers is significantly lower if galaxies are selected by the $\texttt{ODDS}$ parameter \citep{hil+al16}.

However, we accounted for outliers in two ways, which can reproduce their main effects. Firstly, the bias estimates includes outliers. Secondly, we considered Gaussian distributions with quite extended tails. For $\delta z_\text{phot} /(1+z_\text{phot}) = 0.1$ (0.06), $\sim13$  (1.2) per cent of the sources are seen as outliers.

The systematic error on the mass, accounting for both scatter and bias, can be derived from the amplitude error of the lensing signal by using $\delta M_{200} \propto \Delta \Sigma_+^{0.65}$, see App.~\ref{sec_lens_weig}. We can then estimate a mass uncertainty of $\sim 5$ per cent. Our result is in good agreement with the analysis in \citet{mel+al17}, who investigated how the estimate of the mean critical density varies as a function of lens redshift among different photometric redshift algorithms.

\subsection{Shear systematics}

A small calibration uncertainty in the shape measurements at the level of a few per cents can severely limit the accuracy on the mass \citep{wtg_I_14,ume+al14}. 

Multiplicative and additive biases in shape measurement for the CFHTLenS \citep{hey+al12,mil+al13} and the RCSLens \citep{hil+al16} were identified on simulated images. The multiplicative bias mostly depends on the shape measurement technique rather than on the actual properties of the data and can be well assessed with a simulation-based estimate. The average calibration correction to the RCSLenS ellipticities is of order of $\sim5$ per cent \citep{hil+al16}.

\citet{liu+al16} proposed a data driven approach to calibrate the multiplicative bias $m$ by cross-correlating CFHTLenS galaxy density maps with CFHTLenS shear maps and {\it Planck} CMB lensing maps. The additional correction for fainter galaxies may be relevant for cosmic shear analysis, but we could neglect it for our analysis.

Whereas simulation-testing shows that the multiplicative bias is well controlled, detailed comparison of separate shape catalogues of actual data can find that the residual systematic is larger. \citet{jar+al16} performed a detailed comparison of two independent shape catalogues from the Dark Energy Survey Science Verification data and found a systematic uncertainty of $\delta m \sim 0.03$. We can conservatively assume that this is the shear systematics affecting our analysis too, which entails a related mass uncertainty of $\sim 4.5$ per cent.

\subsection{Line-of-sight projections}

Two neighbouring clusters that fall along the line of sight may be blended by the SZ cluster finder into a single, apparently larger cluster. Whereas the Compton parameters add approximately linearly, projection effects can severely impact the weak lensing mass. The lensing amplitude $\Delta \Sigma_+$ is a differential measurement and the estimated mass of the blended cluster can be well below the sum of the masses of the aligned halos. Then, the blended object deviates from the mean scaling relation between SZ signal and mass. 

To estimate this effect, we follow \citet{sim+al17a}. The systematic uncertainty due to projection effects can be approximated as
\beq
\delta M/M \sim \frac{p(\epsilon -0.5)}{1+p(\epsilon -0.5)},
\eeq
where $p$ is the fraction of aligned clusters and $\epsilon$ is an effective parameter which characterizes the effective mass contribution of the projected halo. The parameter $\epsilon$ depends on the relative position of the two blended haloes along the line of sight, and on their shape, elongation and concentration. If $\epsilon = 0.5$, we correctly estimate the total mass; if $\epsilon = 0$, the second halo is hidden and contributes no mass.

{\it Planck} objects are rare and the chance to have two or more of them aligned is small, $\ls5$ per cent considering their tendency to be correlated. The systematic error on mass due to projection effects is then negligible ($\la 1$ per cent).

\subsection{Summary}

\begin{table}
\caption{Systematic error budget on the mass calibration of the PSZ2LenS clusters. Sources of systematics (col.~1) are taken as uncorrelated.}
\label{tab_mass_syst}
\centering
\begin{tabular}[c]{l  l}
\hline
	\noalign{\smallskip}  
Source				&  Mass error [\%]  \\
\hline
Shear measurements	& 5\\
Photometric redshifts	& 5 \\
Line-of-sight projections	& 1\\
Contamination	and membership dilution		& 3 \\
Miscentering			& 0.5 \\
Halo modelling			& 1 \\
Total					& 8 \\
   	 \hline
	\noalign{\smallskip}      
	\end{tabular}
\end{table}

Residual systematic and statistical uncertainties on the mass calibration not corrected for in our analysis are listed in Table~\ref{tab_mass_syst}. We assumed that systematics related to priors, mass estimators, and radial range were properly accounted for and eliminated in our analysis. The main contributors to the systematic error budget are the calibration uncertainties of the multiplicative shear bias, the photo-$z$ performance and the selection of the source galaxies. We estimated that the total level of systematic uncertainty affecting our mass calibration and estimate of the {\it Planck} mass bias if $\sim8$ per cent. 

Even though the systematics are specific to the data set and to the analysis, our systematic assessment is comparable to \citet{mel+al17}, who performed a weak lensing mass calibration of redMaPPer galaxy clusters in Dark Energy Survey Science Verification data, and to \citet{sim+al17a}, who measured the weak lensing mass-richness relation of the SDSS (Sloan Digital Sky Survey) redMaPPer clusters.

We did not consider as systematics triaxiality, orientation and substructures. The presence of substructures can dilute or enhance the tangential shear signal \citep{men+al10,be+kr11,gio+al12a,gio+al14}, and lensing effects depend on the cluster orientation \citep{ogu+al05,ser07,se+um11,lim+al13}. For systems whose major axis points toward the observer, WL masses derived under the standard assumption of spherical symmetry are typically over-estimated. The opposite occurs for clusters elongated in the plane of the sky, which are in the majority if the selected sample is randomly oriented. 

We treated these effects as sources of intrinsic scatter, which quantifies the difference between the deprojected WL mass measurement and the true halo mass \citep{se+et15_comalit_I}, rather than as systematic errors. In our regression scheme, we modelled the scatter of the WL mass, which we found to be $25\pm 18$ per cent from the analysis of the mass-concentration relation, see Section~\ref{sec_c200}, and $23\pm 15$ per cent from the analysis of the {\it Planck} mass bias, see Section~\ref{sec_planck_bias}.

\section{Not confirmed clusters}
\label{sec_not}

 



We could not confirm seven out of the 47 candidate PSZ2 clusters in the CFHTLenS and RCSLenS fields. Out the subsample suitable for our WL analysis, i.e. the 41 candidate PSZ2 sources in the fields with photometric redshifts, we could not find evident counterparts for six candidates.

By visual inspection, we could detect neither any evident galaxy overdensity in the optical images nor an extended X-ray signal from archive {\it ROSAT} (R\"ontgensatellit) or {\it XMM}-Newton (X-ray Multi-Mirror) images near the candidates PSZ2~G006.84+50.69 (PSZ2 index: 25), G098.39+57.68 (463), and G233.46+25.46 (1062). We also did not find any galaxy cluster in the SIMBAD Astronomical Database within the uncertainty region associated with the PSZ2 source.

The analysis of the WL shear around them could not support the presence of a counterpart either. In particular, PSZ2~G006.84+50.69 (25) may be a substructure of the nearby larger PSZ2~G006.49+50.56 (21), i.e. Abell~2029. The SNR around the SZ centroid is $~1.06$ and the WL signal is compatible with no mass lens, $M_{200}=(0.4\pm0.4) \times 10^{14}M_\odot/h$. 

For PSZ2~G098.39+57.68 (463), we estimated a $\text{SNR}=-0.83$ by assuming as lens redshift the median redshift of the PSZ2 clusters, i.e $z=0.224$. 

The median redshift of the galaxies nearby PSZ2~G233.46+25.46 (1062) is $z_\text{phot}=0.85$. This supposed lens redshift is too high to perform a reliable WL analysis. We found just 3 source galaxies passing our criteria behind this candidate. 

For two candidates, PSZ2~G084.69-58.60 (371) and G201.20-42.83 (912), a galaxy overdensity is seen in the photometric redshift distribution, but we could not assign a clear-cut BCG
based on visual inspection or available information from the public catalogs. The WL signal around these candidates can be tentatively measured by locating the halo at the SZ centroid and estimating the redshift as the peak of the distribution of measured $z_\text{phot}$ along the line of sight. The SNR of PSZ2~G084.69-58.60 (371), and G201.20-42.83 (912) is $0.20$ and $-0.34$, respectively. There is no indication from WL alone of the presence of a massive halo. 

For two more candidate PSZ2 clusters, the identification was ambiguous because more than one counterpart could be assigned within the uncertainty region: PSZ2~G092.69+59.92 (421) and G317.52+59.94 (1496). For the source PSZ2~G092.69+59.92 (421), the closest candidate BCG, at $z_\text{spec}=0.59$, is located 3.8\arcmin away from the SZ centroid. The weak lensing SNR around this position is 1.72, with a mass of $M_{200}=(3.2\pm2.8)\times 10^{14}M_\odot/h$. For PSZ2~G317.52+59.94 (1496), there is a possible identification with a galaxy cluster at  $z_\text{phot}=0.58$, but photometric redshifts for the background sources in the RCSLenS are not available in that regions, and so it is excluded from our final catalog.

\section{Conclusions}
\label{sec_conc}

Ongoing and future surveys are providing deep and accurate multi-wavelength observations of the sky. SZ selected samples of clusters of galaxies have some very coveted qualities. In principle, they should provide unbiased and mass limited samples representative of the full population of cosmic haloes up to high redshifts.

To date, quality multi-probe coverage is still restricted to limited areas. We performed a WL analysis of the clusters of galaxies which were SZ selected by the {\it Planck} mission in the fields covered by the CFHTLenS and the RCSLenS. The surveys are not deep but the sample, which we named PSZ2LenS, is statistically complete and homogeneous in terms of observing facilities, and data acquisition, reduction, and analysis.

Clusters are selected in SZ nearly independently of their dynamical and merging state. They should sample all kinds of clusters. In fact, we found that the {\it Planck} selected clusters are standard haloes in terms of their density profile, which is well fitted by cuspy halo models, and in terms of their concentrations, which nicely fit the $\Lambda$CDM prediction by numerical simulations. This suggests that the SZ detection does not suffer from over-concentration biases, as also inferred by \citet{ros+al17} based on the comparison of the X-ray properties of the highest SNR {\it Planck} clusters with X-ray selected samples.

Thanks to the statistical completeness of the PSZ2LenS sample, which is a faithful subsample of the whole population of {\it Planck} clusters, we could asses the bias of the SZ  {\it Planck} masses by comparison with the WL masses. We found a mass bias of $-0.27\pm0.11\text{(stat.)}\pm0.08\text{(sys.)}$. We could estimate the effective bias over the full mass and redshift range of the {\it Planck} clusters. Most of the previous analyses considered small mass ranges, i.e. the massive end of the mass function, or they were limited to intermediate redshifts, where the WL signal is optimized. The most sensible comparison is with \citet{se+et17_comalit_V} who extended their analysis to lower masses and higher redshifts by exploiting a heterogeneous data set. Our results are in full agreement. 

By comparison with WL masses, we confirmed that {\it Planck} masses are precise, i.e. the statistical uncertainties and the intrinsic scatter is small, but they are not accurate, i.e. they are systematically biased.

\section*{Acknowledgements}
MS thanks Keiichi Umetsu and Peter Melchior for useful discussions. SE and MS acknowledge the financial contribution from contracts ASI-INAF I/009/10/0, PRIN-INAF 2012 `A unique dataset to address the most compelling open questions about X-Ray Galaxy Clusters', and PRIN-INAF 2014 1.05.01.94.02 `Glittering Kaleidoscopes in the sky: the multifaceted nature and role of galaxy clusters'. SE acknowledges the financial contribution from contracts NARO15 ASI-INAF I/037/12/0 and ASI 2015-046-R.0. 

This research has made use of NASA's Astrophysics Data System (ADS) and of the NASA/IPAC Extragalactic Database (NED), which is operated by the Jet Propulsion Laboratory, California Institute of Technology, under contract with the National Aeronautics and Space Administration. 

This work is based on observations obtained with MegaPrime/MegaCam, a joint project of CFHT and CEA/IRFU, at the Canada-France-Hawaii Telescope (CFHT) which is operated by the National Research Council (NRC) of Canada, the Institut National des Sciences de l'Univers of the Centre National de la Recherche Scientifique (CNRS) of France, and the University of Hawaii. This research used the facilities of the Canadian Astronomy Data Centre operated by the National Research Council of Canada with the support of the Canadian Space Agency. CFHTLenS and RCSLenS data processing was made possible thanks to significant computing support from the NSERC Research Tools and Instruments grant program.


\setlength{\bibhang}{2.0em}

\appendix

\section{Radius}
\label{app_radi}

\begin{figure}
\resizebox{\hsize}{!}{\includegraphics{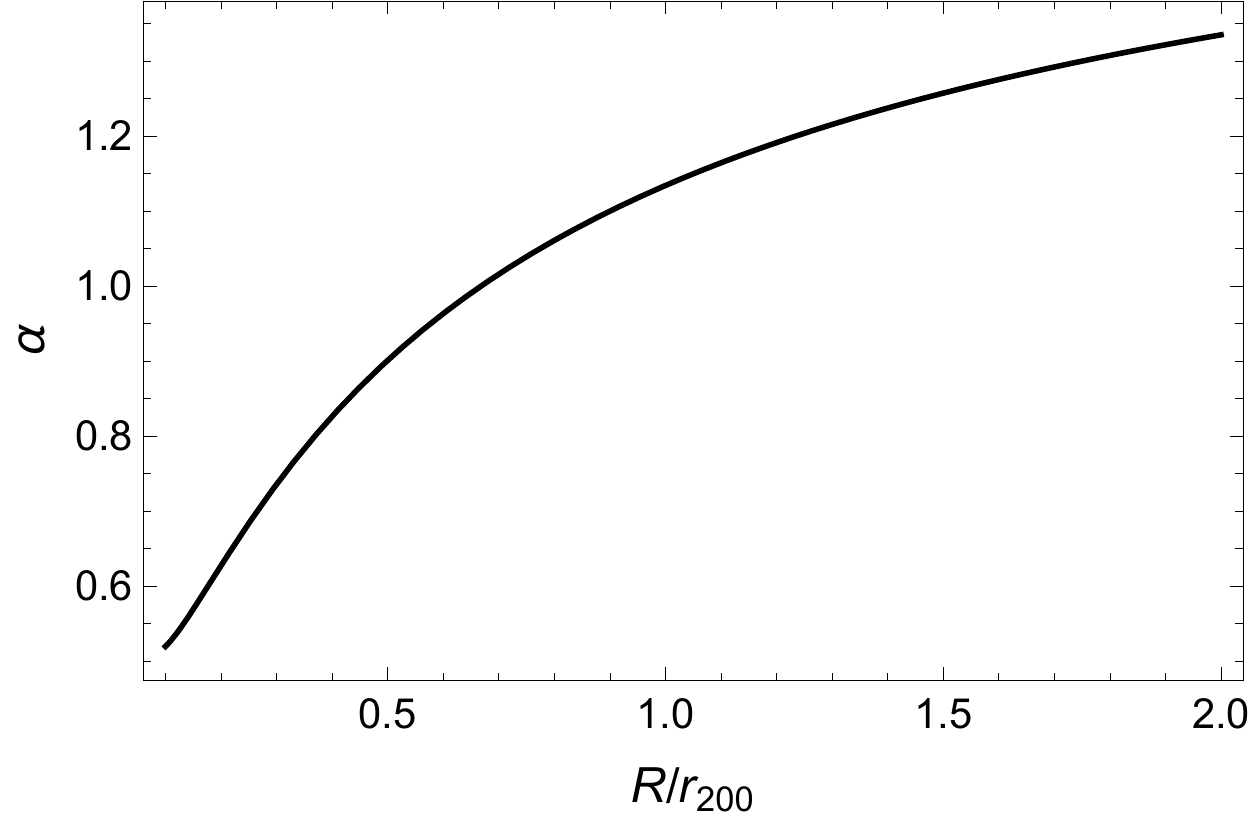}} 
\caption{Logarithmic slope of the reduce shear profile as a function of radius for a NFW lens of mass $M_{200}=5\times10^{14}M_\odot/h$ and concentration in agreement with the relation found in \citet{du+ma14} for the {\it Planck} cosmology, i.e. $c_{200}\sim 3.78$. We placed the lens at $z_\text{d}=0.5$ and the background sources at $z_\text{s}=1.0$. The radius is in units of $r_{200}$.}
\label{fig_alpha_NFW}
\end{figure}

\begin{figure}
\begin{tabular}{c}
\resizebox{8.2cm}{!}{\includegraphics{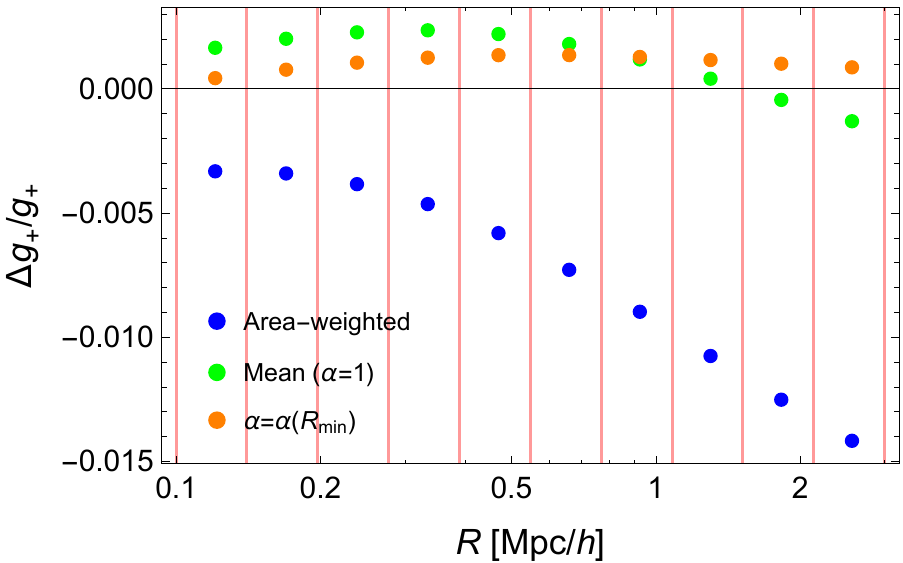}} \\
\resizebox{8.2cm}{!}{\includegraphics{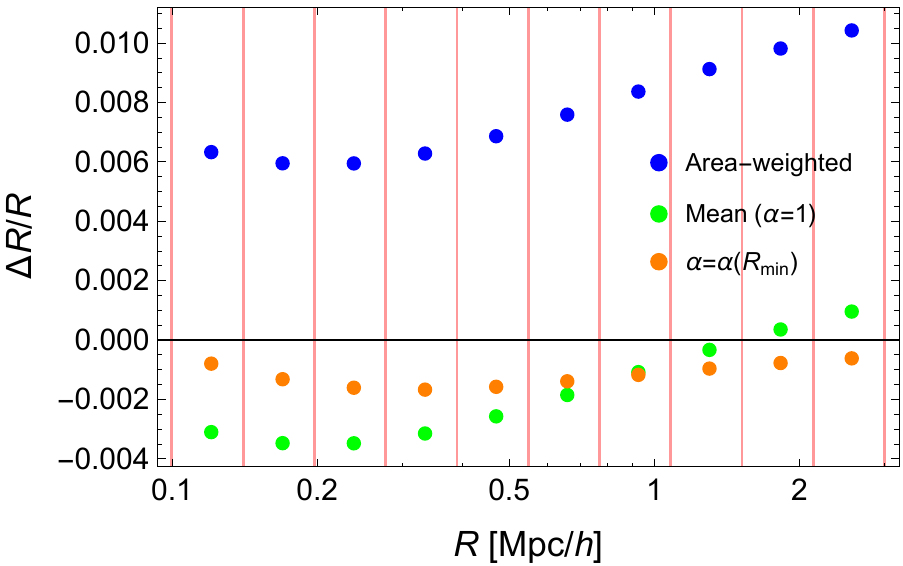}} \\
\end{tabular}
\caption{{\it Top panel}: relative difference between the average reduced shear and the reduced shear computed at different radii. The vertical red lines delimit the radial annuli, which are ten bins equally spaced in logarithmic units between 0.1 and $3~\text{Mpc}/h$. The blue, green and orange points refers to effective radii computed as the mean, $\langle R \rangle_\text{mean}$, the area-weighted mean $\langle R \rangle_\text{aw}$, or the shear-weighted mean $\langle R \rangle_\text{gw}$ with logarithmic slope computed at the inner radius. The lens properties are as in Fig.~\ref{fig_alpha_NFW}. {\it Bottom panel}: same as the top panel but for the relative difference between different estimators of the radius and the true shear-weighted radius.}
\label{fig_DeltaGamma_R}
\end{figure}

Different recipes for the effective radius of a radial annulus have been proposed. A simple estimator is the mean of the inner $R_\text{min}$ and outer  $R_\text{max}$ radii,
\beq
\label{eq_rad_1}
\langle R \rangle_\text{mean}= \frac{R_\text{min}+ R_\text{max}}{2}.
\eeq

Alternatively, for a spatially uniform number density of background galaxies, the effective radius can be estimated as the area-weighted mean,
\beq
\label{eq_rad_2}
\langle R \rangle_\text{aw}= \frac{\int_{R_\text{min}}^{R_\text{max}} R^{2}dR}{\int_{R_\text{min}}^{R_\text{max}} R dR} = \frac{2}{3}\frac{R_\text{min}^2+R_\text{max}^2+R_\text{min}R_\text{max}}{R_\text{min}+R_\text{max}}.
\eeq
The area-weighted mean is higher than the simple mean, since most of the area is near the outer radius.

Here, we define the effective radius $\langle R \rangle_\text{gw}$ as the shear weighted radius,
\beq
\label{eq_rad_3}
g_+(\langle R \rangle_\text{gw}) = \frac{\int_{R_\text{min}}^{R_\text{max}} g_+(R) R dR}{\int_{R_\text{min}}^{R_\text{max}} R dR}.
\eeq
For a power-law shear profile, $g_+ \sim R^{-\alpha}$,
\beq
\label{eq_rad_4}
\langle R \rangle_\text{gw}= \left( \frac{2}{2-\alpha}\frac{R_\text{max}^{2-\alpha}-R_\text{min}^{2-\alpha}}{
R_\text{max}^2-R_\text{min}^2} \right)^{-1/\alpha}.
\eeq
Equations~(\ref{eq_rad_1}) and ~(\ref{eq_rad_2}) are particular cases of equation~(\ref{eq_rad_4}) for $\alpha=1$ and $-1$, respectively. 

In general, the logarithmic slope of the reduced shear profile, $\alpha$, varies with the radius. The notable exception is the singular isothermal sphere with $\alpha_\text{SIS} = 1$. For most profiles the slope is close to one, see Fig.~\ref{fig_alpha_NFW} for the case of the NFW halo.

In Fig.~\ref{fig_DeltaGamma_R}, we show that the simple mean, i.e. the shear-weighted radius with $\alpha=1$, provides a very good approximation of the effective radius. The area weighted radius, whereat the mean shear is under-estimated, is larger than the effective radius. In fact, the area-weighting scheme accounts for most of the galaxies being near the upper radius but does not account for their lower shear.

Since the lens properties are not known when we stack the signal, the shear-weighted radius with $\alpha=1$ is an acceptable choice. More elaborate schemes, as that fixing the slope at its value at the inner radius, which would nevertheless require some knowledge of the profile, do not improve the radius estimates significantly, see Fig.~\ref{fig_DeltaGamma_R}.

The previous discussion relied on the continuous limit where the background galaxy distribution is uniform and lie at  a single redshift. For sparse populations which are redshift distributed, we have to compute the effective radius as
\beq
\langle R \rangle_\text{gw} = \left( \frac{\sum_i w_i \Sigma_{\text{cr},i}^{-2} R_i^{-\alpha}}{\sum_i w_i \Sigma_{\text{cr},i}^{-2}} \right)^{-1/\alpha},
\eeq
where we exploited the power-law approximation for the shear profile. The shear-weighted radius makes the fitting procedure to shear profiles less dependent on the binning scheme.

\section{Lensing weighted average}
\label{sec_lens_weig}

\begin{figure}
\resizebox{\hsize}{!}{\includegraphics{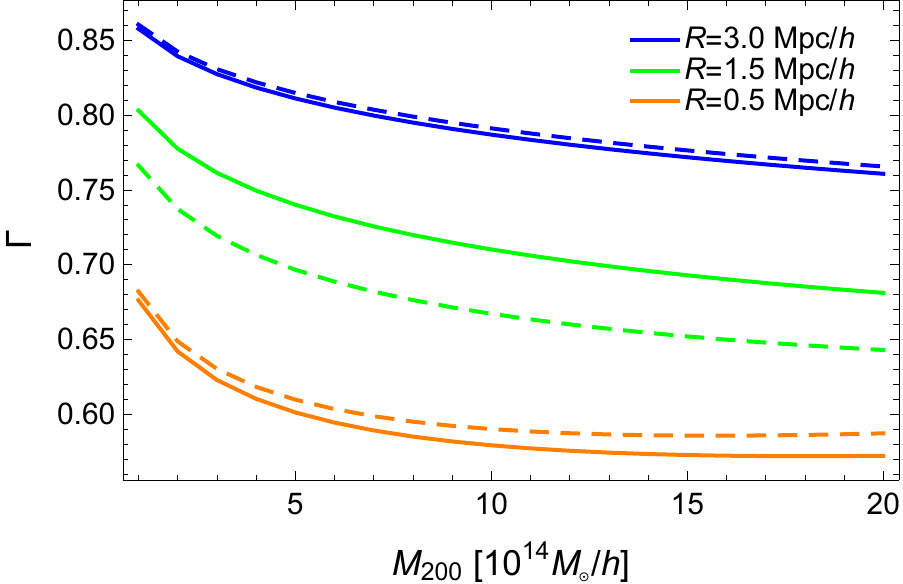}} 
\caption{Logarithmic slope of the differential density profile of a NFW halo, $\Gamma = d \ln \Delta \Sigma_{+} /d \ln M_{200}$, as a function of mass, at different redshifts and radial distances from the cluster centre. Concentrations are assigned through the mass-concentration relation from \citet{men+al14}. The full and dashed lines are for lenses at $z_\text{cl}=0.2$ and $0.4$, respectively. The blue, green, and orange lines (from top to bottom) are for radii $R=0.5$, $1$, and $3~\text{Mpc}/h$, respectively.}
\label{fig_NFW_slope}
\end{figure}

Stacking and combining lensing data or results are a highly non-linear process. A sensible way to define the central estimate of a cluster property ${\cal O}$ for a lensing sample is the lensing weighted average \citep{ume+al14},
\beq
\label{eq_lens_weig_1}
\langle {\cal O} \rangle_\text{lw} = \frac{\sum_i {\cal W}_i {\cal O}_i}{\sum_i {\cal W}_i},
\eeq
where the sums runs over the cluster sample and the weight ${\cal W}$ of the $i$-th cluster is
\beq
\label{eq_lens_weig_2}
{\cal W}_i= \sum_j  w_{i,j} \Sigma_{\text{cr},j}^{-2},
\eeq
where the sum run over the selected background galaxies behind the $i$-th cluster. The weight ${\cal W}$ accounts for the total shear weight of the cluster and accounts for the shear weight $w$ of each lens-source pair, the lens and source redshifts through the critical density, and the angular size of the clusters, since lower redshift clusters subtend a larger angle in the sky for a fixed physical length and hence a larger number of background galaxies. 

We verified, for example, that the definition in Eq.~(\ref{eq_lens_weig_1}) is appropriate to assign a redshift to the stacked profile, i.e. the recovered mass $M_\text{stack}$ of a stacking sample of clusters with the same mass $M_{200}=M_\text{cl}$ but at different redshifts is equal to $\langle M_{200} \rangle_\text{lw} \simeq M_\text{cl}$ if $z_\text{stack} =\langle z_\text{cl} \rangle_\text{lw}$.

The lensing average in Eq.~(\ref{eq_lens_weig_1}) can be modified for some observables to account for the fact that we stack the density profiles $\Delta \Sigma_{+}$. In practice, we have to recover the mean observable from the stacked profile. If $\Delta \Sigma_{+} \propto {\cal O}^\Gamma$, then \citep{mel+al17} 
\beq
\label{eq_lens_weig_3}
\langle {\cal O} \rangle_\text{lw} = \left( \frac{\sum_i {\cal W}_i {\cal O}_i^\Gamma}{\sum_i {\cal W}_i}\right)^{1/\Gamma}.
\eeq

If we consider the mass as the observable, the exponent $\Gamma$ can differ from 1. The dependence of the mass on the density profile can be approximated with a power low $\Delta \Sigma_{+} \propto M_{200}^\Gamma$ with
\beq
\label{eq_lens_weig_4}
\Gamma = \frac{d \ln \Delta \Sigma_{+}}{d \ln M_{200}}; 
\eeq
for an isothermal model, $\Gamma=1$. For a NFW halo, the logarithmic density slope for a range of radii and redshifts is shown in Fig.~\ref{fig_NFW_slope}. The slope is larger at small radii or large redshifts and spans a range from $\sim0.5$ to $1$. Based on some toy model simulations mimicking our stacking analysis, we found that $\Gamma \sim 0.65$ is appropriate for our range of masses and redshifts and for our fitting procedure.

\section{Robust estimator}
\label{sec_robu_esti}

\begin{figure}
\resizebox{\hsize}{!}{\includegraphics{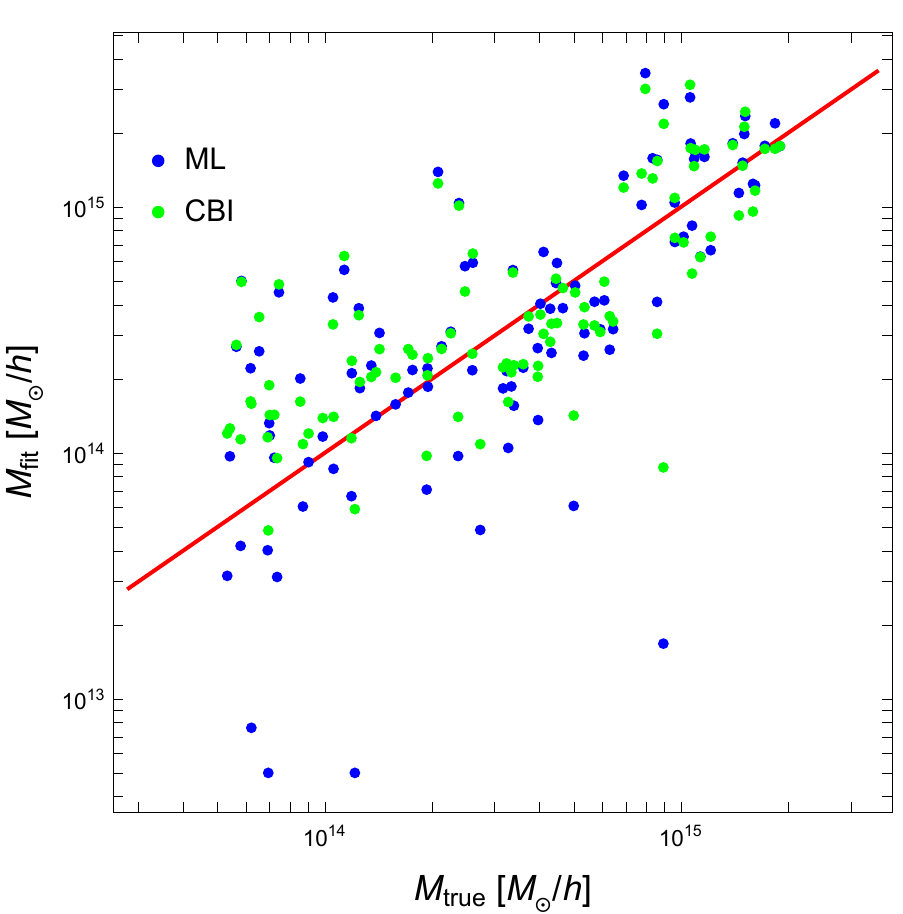}} 
\caption{Fitted mass versus true input mass $M_{200}$ for simulated NFW lenses. The green points and the associated error bars denote the central biweight estimator. The blue points denote the maximum likelihood estimator.}
\label{fig_ML_vs_CBI}
\end{figure}

The posterior probability density function of the mass of low signal-to-noise ratio systems can be asymmetric or peaked near one of the imposed borders. As an extreme example, the more likely mass of a low mass group detected with a negative signal-to-noise ratio will coincide with the lower limit of the allowed parameter range. The problem is then to identify a reliable and stable mass estimator. The median \citep{got+al01} or the bi-weight location estimator $C_\text{BI}$ \citep{bee+al90} are regarded as robust choices for the central location and have been considered in WL analyses \citep{se+um11}. Here, we want to compare the performances of the bi-weight location estimator against the maximum likelihood estimator.

We simulated the shear profile of clusters with shallow quality data. Lenses were modeled as NFW haloes at redshift $z_\text{d}=0.3$. We assumed a shape noise error dominated by the intrinsic distribution of ellipticities, with a dispersion of $\sigma_\text{e}=0.3$, and we also considered the noise from the large scale structure. We considered a background population at $z_\text{s}=0.8$ with a source density of $n_\text{g}=2$ background galaxies per square arc minute. 

We simulated 100 lens masses with a constant logarithmic spacing from $M_{200}=5\times10^{13}$ to $2\times10^{15}~M_\odot/h$. Concentration were associated assuming the scattered relation from \citet{du+ma14}. The shear profiles were finally simulated in 10 equally spaced logarithmic radial annuli between 0.1 and $3~\text{Mpc}/h$.

We fitted the simulated profiles as in Section~\ref{sec_infe}. Results are summarized in Fig.~\ref{fig_ML_vs_CBI}. At the high mass end ($M_{200}\ga 10^{14}~M_\odot/h$), the signal to noise is high enough and the estimated mass is stable whatever the estimator. At the low mass end, fluctuations are larger and differences can be significant. Results are usually consistent within the errors but the maximum likelihood estimator is more prone to outliers and often attracted towards the extremes of the allowed mass range. For our simulation, this problem is under control since we could fit the toy-clusters with the right NFW profile. However, the problem can be exacerbated with real clusters which can deviate from the halo modelling we assume for the fit.

The bi-weight estimator is stabler but it can be influenced by the prior. Assuming a uniform prior, masses can be biased high at the low mass end. This would not be the case assuming a prior uniform in log space, which however could be inadequate at intermediate masses. Since most of the {\it Planck} clusters are expected to be $\ga 10^{14}~M_\odot/h$, the prior uniform in mass has to be preferred.

For the simulated lenses with $M_{200}> 10^{14}~M_\odot/h$, the distribution of the relative deviations expressed as $\ln (M_\text{fit}/M_\text{true})$ has mean $0.03$ ($0.07$) and standard deviation equal to $0.67$ ($0.91$) for the bi-weight (maximum likelihood) estimator.

\end{document}